\documentclass[12pt]{article}
\newtheorem{theorem}{Theorem}
\newcommand{\bt}{\begin{theorem}}
\newcommand{\et}{\end{theorem}}
\newtheorem{definition}{Definition}
\newcommand{\bd}{\begin{definition}}
\newcommand{\ed}{\end{definition}}
\newtheorem{proposition}{Proposition}
\newcommand{\bp}{\begin{proposition}}
\newcommand{\ep}{\end{proposition}}
\newtheorem{corollary}{Corollary}
\newcommand{\bc}{\begin{corollary}}
\newcommand{\ec}{\end{corollary}}
\newtheorem{lemma}{Lemma}
\newcommand{\bl}{\begin{lemma}}
\newcommand{\el}{\end{lemma}}
\newcommand{\beq}[1]{\begin{equation} \label{#1}}
\newtheorem{example}{Example}
\newcommand{\be}{\begin{example}}
\newcommand{\ee}{\end{example}}
\newtheorem{observation}{Observation}
\newcommand{\bo}{\begin{observation}}
\newcommand{\eo}{\end{observation}}
\newtheorem{remark}{Remark}
\newcommand{\rk}{\begin{remark}}
\newcommand{\mk}{\end{remark}}
\newtheorem{Fen}{Fenchel Duality Theorem}

\usepackage[dvips]{graphicx}
\usepackage{latexsym}
  \usepackage{rotate}
\usepackage{amsmath,amssymb}
\usepackage{epsf,epsfig}
\DeclareGraphicsExtensions{.ps}

\newcommand{\qed}{ \hfill $\blacksquare$ \vspace{3mm}}

\newcommand{\Doublespace}{\renewcommand{\baselinestretch}{1.2}\small\normalsize}
\newcommand{\doublespace}{\renewcommand{\baselinestretch}{1.3}\small\normalsize}

\newcommand{\ddoublespace}{\renewcommand{\baselinestretch}{1.5}\small\normalsize}

\begin{document}

\ddoublespace

\begin{center}
\begin{tabular}[b]{c}
{\LARGE \bf Mathematics and Engineering} \\ \medskip
{\LARGE \bf Communications Laboratory}
\ \\
\ \\
{\Large \sf Technical Report}
\end{tabular}\hspace{0.4in}\epsfig{file=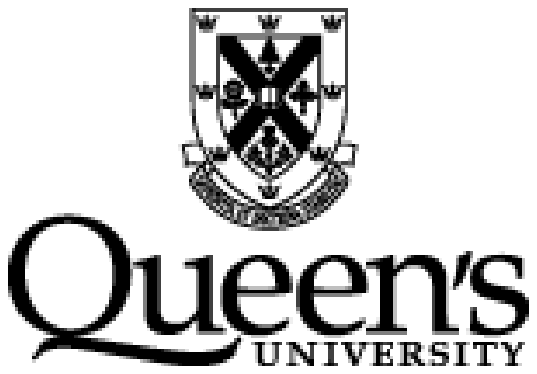,width=3.0cm}

\rule{6.0in}{0.8mm}
\end{center}

\bigskip
\bigskip
\bigskip
\bigskip
\vspace{1.0in}

\begin{center}
{\Large\bf On the Joint Source-Channel Coding Error Exponent} \\
\smallskip
{\Large\bf for Discrete Memoryless Systems: Computation} \\
\smallskip
{\Large\bf and Comparison with Separate Coding} \\
\vspace{0.6in}
{\large {\it Y. Zhong}, {\it F. Alajaji}, and {\it L.~L. Campbell}}  \\
\end{center}

\vspace{2.0in}
\begin{center}
{\large December 2005}
\end{center}

\newpage

\Doublespace

\title {\bf On the Joint Source-Channel Coding Error Exponent
for Discrete Memoryless Systems: Computation and Comparison 
with Separate Coding\footnote {This research was supported in part by the
Natural Sciences and Engineering Research Council of Canada and
the Premier's Research Excellence Award of Ontario. The authors
are with the Dept. of Mathematics \& Statistics, Queen's
University, Kingston, ON K7L 3N6, Canada. }}

\bigskip

\author{
Yangfan Zhong \and Fady Alajaji  \and L.~Lorne Campbell}


\date{}

\maketitle


\begin{center}

\vspace{-0.2in}

\end{center}

\begin{abstract}
We investigate the computation of Csisz\'ar's bounds for the joint
source-channel coding (JSCC) error exponent, $E_J$, of a
communication system consisting of a discrete memoryless source
and a discrete memoryless channel. We provide equivalent
expressions for these bounds and derive explicit formulas for the
rates where the bounds are attained. These equivalent
representations can be readily computed for arbitrary
source-channel pairs via Arimoto's algorithm. When the channel's
distribution satisfies a symmetry property, the bounds admit
closed-form parametric expressions. We then use our results to
provide a systematic comparison between the JSCC error exponent
$E_J$ and the tandem coding error exponent $E_T$, which applies if
the source and channel are separately coded. It is shown that $E_T
\leq E_J \leq 2E_T$. We establish conditions for which $E_J
> E_T$ and for which $E_J = 2E_T$. Numerical examples
indicate that $E_J$ is close to $2E_T$ for many source-channel
pairs. This gain translates into a power saving larger than 2 dB
for a binary source transmitted over additive white Gaussian noise
channels and Rayleigh fading channels with finite output
quantization. Finally, we study the computation of the lossy JSCC
error exponent under the Hamming distortion measure.

\end{abstract}


\newpage
\doublespace


\section{Introduction}
\label{intro}

Traditionally, source and channel coding have been treated
independently, resulting in what we call a {\em tandem (or
separate)} coding system. This is because Shannon in 1948
\cite{shannon} showed that separate source and channel coding
incurs no loss of optimality (in terms of reliable
transmissibility) provided that the coding blocklength goes to
infinity. In practical implementations, however, there is a price
to pay in delay and complexity, for extremely long blocklength. To
begin, we note that {\em joint source-channel coding} (JSCC) might
be expected to offer improvements for the combination of a source
with significant redundancy and a channel with significant noise,
since, for such a system, tandem coding would involve source
coding to remove redundancy and then channel coding to insert
redundancy. It is a natural conjecture that this is not the most
efficient approach (even if the blocklength is allowed to grow
without bound). Indeed, Shannon \cite{shannon} made this point as
follows:

\begin{quote}
{\it $\cdots$ However, any redundancy in the source will usually
help if it is utilized at the receiving point. In particular, if
the source already has a certain redundancy and no attempt is made
to eliminate it in matching to the channel, this redundancy will
help combat noise. For example, in a noiseless telegraph channel
one could save about $50\%$ in time by proper encoding of the
messages. This is not done and most of the redundancy of English
remains in the channel symbols. This has the advantage, however,
of allowing considerable noise in the channel. A sizable fraction
of the letters can be received incorrectly and still reconstructed
by the context. In fact this is probably not a bad approximation
to the ideal in many cases  $\cdots$ }
\end{quote}

The study of JSCC dates back to as early as the 1960's. Over the
years, many works have introduced JSCC techniques and illustrated
(analytically or numerically) their benefits (in terms of both
performance improvement and increased robustness to variations in
channel noise) over tandem coding for given source and channel
conditions and fixed complexity and/or delay constraints. In JSCC
systems, the designs of the source and channel codes are either
well coordinated or combined into a single step. Examples of (both
constructive and theoretical) previous lossless and lossy JSCC
investigations include:

\begin{enumerate}

\item JSCC theorems and the separation principle
\cite{arikan2}, \cite{balakirsky},
\cite{Chen}, \cite{Dunham}, \cite{Gallager}, \cite{Gray},
\cite{Han}, \cite{Hellman}, \cite{koshelev}, \cite{Vembu};

\item source codes that are robust against channel errors such as
optimal (or sub-optimal) quantizer design for noisy channels
\cite{COVQ}, \cite{Ayanoglu}, \cite{Farvardin}, \cite{Fine},
\cite{Gibson}, \cite{Kumazawa}--\cite{Lim}, \cite{Miller},
\cite{Nam}, \cite{Hadamard}, \cite{Skoglund}, \cite{Vaisha};

\item channel codes that exploit the source's natural redundancy
(if no source coding is applied) or its residual redundancy (if
source coding is applied) \cite{Celp}, \cite{Hagenauer},
\cite{Massey}, \cite{Sayood}, \cite{Zhu};

\item zero-redundancy channel codes with optimized codeword
assignment for the transmission of source encoder indices over
noisy channels (e.g., \cite{Farvardin}, \cite{Zeger});

\item unequal error protection source and channel codes where the
rates of the source and channel codes are adjusted to provide
various levels of protection to the source data depending on its
level of importance and the channel conditions (e.g.,
\cite{Hochwald}, \cite{Modestino});

\item uncoded source-channel matching where the source is uncoded,
directly matched to the channel and optimally decoded (e.g.,
\cite{Markov}, \cite{Gastpar}, \cite{Shamai}, \cite{Weissman}).
\end{enumerate}

The above references are far from exhaustive as the field of JSCC
has been quite active, particularly over the last 20 years.

In order to learn more about the performance of the best codes as
a function of blocklength, much research has focused on the error
exponent or reliability function for source or channel coding
(see, e.g., \cite{Blahut}, \cite{Csiszar3}, \cite{Gallager},
\cite{Jelinek}, \cite{Marton}, \cite{viterbi}). Roughly speaking,
the error exponent $E$ is a number with the property that the
probability of decoding error of a good code is approximately
$2^{-En}$ for codes of large blocklength $n$. Thus the error
exponent can be used to estimate the trade-off between error
probability and blocklength. In this paper we use the error
exponent as a tool to compare the performance of tandem coding and
JSCC. While jointly coding the source and channel offers no
advantages over tandem coding in terms of reliable
transmissibility of the source over the channel (for the case of
memoryless systems as well as the wider class of stationary information
stable \cite{Chen,Han} systems), it is possible that the same error
performance can be achieved for smaller blocklengths via optimal
JSCC coding.

The first quantitative result on error exponents for lossless JSCC
was a lower bound on the error exponent derived in 1964 by
Gallager \cite[pp. 534--535]{Gallager}. This result also indicates
that JSCC can lead to a larger exponent than the tandem coding
exponent, the exponent resulting from separately performing and
concatenating optimal source and channel coding. In 1980,
Csisz\'{a}r~\cite{Csiszar1} established a lower bound (based on
the random-coding channel error exponent) and an upper bound for
the JSCC error exponent $E_{J}(Q,W,t)$ of a communication system
with transmission rate $t$ source symbols/channel symbol and
consisting of a discrete memoryless source (DMS) with distribution
$Q$ and a discrete memoryless channel (DMC) with transition
distribution $W$. He showed that the upper bound, which is
expressed as the minimum of the sum of $te(R/t,Q)$ and $E(R,W)$
over $R$, i.e.,
\begin{equation}
\min_{R}\left[t e\left(\frac{R}{t},Q\right)+
E(R,W)\right],\label{intreq}
\end{equation}
where $e(R,Q)$ is the source error exponent~\cite{Blahut},
\cite{Csiszar1}, \cite{Jelinek} and $E(R,W)$ is the channel error
exponent~\cite{Csiszar1}, \cite{Gallager}, \cite{Jelinek}, is
tight if the latter minimum is attained for an $R$ strictly larger
than the critical rate of the channel. Another (looser) upper
bound to $E_{J}(Q,W,t)$ directly results from (\ref{intreq}) by
replacing $E(R,W)$ by the sphere-packing channel error exponent.
He extended this work in 1982 \cite{Csiszar2} to obtain a new
expurgated lower bound (based on the expurgated channel exponent)
for the above system under some conditions, and to deal with lossy
coding relative to a distortion threshold. Our first objective in
this work is to recast Csisz\'ar's results in a form more suitable
for computation and to examine the connection between Csisz\'ar's
upper and lower bounds, and also the relation between the lower
bounds of Gallager and Csisz\'ar. After this, we go on to compare
the tandem coding and joint coding error exponents in order to
discover how much potential for improvement there is via JSCC.
Since error exponents give only asymptotic expressions for system
performance, our results do not have direct application to the
construction of good codes. Rather, they point out certain systems
for which a search for good joint codes might prove fruitful.

We first investigate the analytical computation of Csisz\'{a}r's
random-coding lower bound and sphere-packing upper bound for the
JSCC error exponent. By applying Fenchel's Duality Theorem
\cite{Luenberger} regarding the optimization of the sum of two
convex functions, we provide equivalent expressions for these
bounds which involve a maximization over a non-negative parameter
of the difference between the concave hull of Gallager's channel
function and Gallager's source function \cite{Gallager}; hence,
they can be readily computed for arbitrary source-channel pairs by
applying Arimoto's algorithm \cite{Arimoto}. When the channel's
distribution is symmetric \cite{Gallager}, our bounds admit
closed-form parametric expressions. We also provide formulas of
the rates for which the bounds are attained and establish explicit
computable conditions in terms of $Q$ and $W$ under which the
upper and lower bounds coincide; in this case, $E_{J}$ can be
determined exactly. A byproduct of our results is the observation
that Csisz\'{a}r's JSCC random-coding lower bound can be larger
than Gallager's earlier lower bound obtained in \cite{Gallager}.
Using a similar approach, we obtain the equivalent expression of
Csisz\'{a}r's expurgated lower bound \cite{Csiszar2} and establish
the condition when the random-coding lower bound can be improved
by the expurgated bound. As an example, we give closed-form
parametric expressions of the improved lower bound and the
corresponding condition for equidistant DMCs.

We next employ our results to provide a systematic comparison of
the JSCC exponent $E_{J}(Q,W,t)$ and the tandem coding exponent
$E_{T}(Q,W,t)$ for a DMS-DMC pair $(Q,W)$ with the same
transmission rate $t$. Since $E_{J}\geq E_{T}$ in general (as
tandem coding is a special case of JSCC), we are particularly
interested in investigating the situation where $E_{J}> E_{T}$.
Indeed, this inequality, when it holds, provides a theoretical
underpinning and justification for JSCC design as opposed to the
widely used tandem approach, since the former method will yield a
faster exponential rate of decay for the error probability, which
may translate into substantial reductions in complexity and delay
for real-world communication systems. We establish sufficient
(computable) conditions for which $E_{J}
> E_{T}$ for any given source-channel pair $(Q,W)$, which are
satisfied for a large class of memoryless source-channel pairs.
Furthermore, we show that $E_{J}\leq 2E_T$. Numerical examples
show that $E_{J}$ can be nearly twice as large as $E_{T}$ for many
DMS-DMC pairs. Thus, for the same error probability, JSCC would
require around half the delay of tandem coding. This potential
benefit translates into more than 2~dB power gain for binary DMS
sent over binary-input quantized-output additive white Gaussian
noise and memoryless Rayleigh-fading channels.

We also partially address the computation of Csisz\'{a}r's lower
and upper bounds for the lossy JSCC exponent with distortion
threshold $\Delta$, $E^{\Delta}_{J}(Q,W,t)$. Under the case of the
Hamming distortion measure, and for a binary DMS and an arbitrary
DMC, we express the bounds for $E^{\Delta}_{J}(Q,W,t)$ and the
rates for which the bounds are attained as in the lossless case.

The rest of this paper is arranged as follows. In
Section~\ref{prelim} we describe the system, define the
terminologies and introduce some material on convexity and Fenchel
duality. Section~\ref{bounds} is devoted to study the analytical
computation of $E_{J}$ based on Csisz\'{a}r's
work~\cite{Csiszar1}, \cite{Csiszar2}. In
Section~\ref{comparison}, we assess the merits of JSCC by
comparing $E_{J}$ with $E_{T}$. The computation of the lossy JSCC
exponent is partially studied in Section~\ref{distortionsec}.
Finally, we state our conclusions in Section~\ref{Concl}.


\section{Definitions and System Description} \label{prelim}
\subsection{System}

We consider throughout this paper a communication system
consisting of a DMS $\{Q:\mathcal{S}\}$ with finite alphabet
${\cal S}$ and distribution $Q$, and a DMC
$\{W:\mathcal{X}\rightarrow \mathcal{Y}\}$ with finite input
alphabet $\cal X$, finite output alphabet ${\cal Y}$, and
transition probability $W\triangleq P_{Y|X}$. Without loss of
generality we assume that $Q(s)>0$ for each $s\in {\cal S}$. Also,
if the source distribution is uniform, optimal (lossless) JSCC
amounts to optimal channel coding which is already well-studied.
Therefore, we assume throughout that $Q$ is not the uniform
distribution on ${\cal S}$ except in Section \ref{distortionsec}
where we deal with JSCC under a fidelity criterion.

A joint source-channel (JSC) code with blocklength $n$ and
transmission rate $t>0$ (measured in source symbols/channel use)
is a pair of mappings $f_n : {\cal S}^{tn} \longrightarrow {\cal
X}^n$ and $\varphi_n : {\cal Y}^n \longrightarrow {\cal S}^{tn}$.
That is, blocks $s^{tn}\triangleq(s_1,s_2,...,s_{tn})$ of source
symbols of length $tn$ are encoded as blocks
$x^{n}\triangleq(x_1,x_2,...,x_n)=f_n (s^{tn})$ of symbols from
$\cal{X}$ of length $n$, transmitted, received as blocks
$y^{n}\triangleq(y_1,y_2,...,y_n)$ of symbols from ${\cal Y}$ of
length $n$ and decoded as blocks of source symbols $\varphi_n
(y^{n})$ of length $tn$. The probability of erroneously decoding
the block is
$$
P_e^{(n)}(Q,W,t) \triangleq \sum_{\{(s^{tn}, y^{n}):\varphi_n
(y^{n}) \neq s^{tn}\}} Q_{tn}(s^{tn}) P_{n,Y|X} \left(y^{n} | f_n
(s^{tn})\right).
$$
Here, $Q_{tn}$ and $P_{n,Y|X}$ are the $tn$-
and $n$-dimensional product distributions corresponding to $Q$ and
$P_{Y|X}$ respectively.

Throughout the paper, $\log$ will denote a base $2$ logarithm,
$|{\cal S }|$ will mean the number of elements in ${\cal S }$ and
similarly for the other alphabets, $C$ will denote the capacity of
the DMC given by
$$
C=\max_{P_X}I(P_X;W),
$$
where $I(P_X;W)$ is the mutual information between the channel
input and the channel output \cite{Gallager}. Finally, $H(\cdot)$
will denote the entropy of a discrete probability distribution.

\subsection{Error Exponents}

\begin{definition} \label{errorexp}
{\rm The JSCC error exponent $E_J (Q,W,t)$ is defined as the
largest number $E$ for which there exists a sequence of JSC codes
$(f_n, \varphi_n )$ with transmission rate $t$ and blocklength $n$
such that
$$
E \leq \liminf_{n \to \infty} -\frac{1}{n} \log P_e^{(n)}(Q,W,t).
$$}
\end{definition}
When there is no possibility of confusion, $E_J (Q,W,t)$ will be
written as $E_J$. We know from the JSCC theorem (e.g., \cite[p.
216]{Thomas}, \cite{Gallager}) that $E_J$ can be positive if and
only if $tH(Q) < C$.

For future use, we recall the source and channel functions used by
Gallager \cite{Gallager} in his treatment of the JSCC theorem. We
also introduce some useful notation and some elementary relations
among these functions. Let Gallager's source function be
\beq{source} E_s (\rho ,Q) \triangleq (1+ \rho) \log \sum_{s \in
{\cal S}}Q(s)^{\frac{1}{1+\rho }}, \qquad \rho \geq 0.
\end{equation}
Let \beq{channel2} \tilde{E}_0 (\rho,P_X, W) \triangleq -\log
\sum_{y\in {\cal Y}}\left(\sum_{x\in {\cal X}} P_X (x)
P_{Y|X}^{\frac{1}{1+\rho}} (y|x) \right)^{1+ \rho}, \qquad \rho
\geq 0,
\end{equation}
and \beq{channel3} \tilde{E}_x(\rho;P_X,W)\triangleq -\rho
\log\sum_{x\in \mathcal{X}} \sum_{x'\in
\mathcal{X}}P_{X}(x)P_{X}(x') \left(\sum_{y\in
\mathcal{Y}}\sqrt{P_{Y\mid X}(y\mid x)P_{Y\mid X}(y\mid
x')}\right)^{1/\rho}, \qquad \rho\geq 1.
\end{equation}
$P_X$ in (\ref{channel2}) and (\ref{channel3}) is an unspecified
probability distribution on ${\cal X}$. Connected with these
functions are the source error exponent, \beq{source.exp} e(R,Q)=
\sup_{0 \leq \rho < \infty}[\rho R -E_s(\rho,Q)] ,
\end{equation}
and three intermediate channel error exponents \beq{random1}
\tilde{E}_r (R,P_X,W) \triangleq \max_{0 \leq \rho \leq
1}[\tilde{E}_0 (\rho,P_X, W) - \rho R],
\end{equation}
\beq{expurgated1} \tilde{E}_{ex} (R,P_X,W) \triangleq \sup_{\rho
\geq 1}[\tilde{E}_x (\rho,P_X, W) - \rho R],
\end{equation}
and \beq{sphere1} \tilde{E}_{sp}(R,P_X,W) \triangleq \sup_{0 \leq
\rho < \infty}[\tilde{E}_0 (\rho,P_X, W) - \rho R].
\end{equation}

From these, we can form the random-coding lower bound for the
channel error exponent $E(R,W)$, \beq{random.coding} E_r (R,W)
\triangleq \max_{P_X} \tilde{E}_r (R,P_X,W),
\end{equation}
the expurgated lower bound \beq{expurgated}E_{ex} (R,W) \triangleq
\max_{P_X} \tilde{E}_{ex} (R,P_X,W) ,
\end{equation}
and the sphere-packing upper bound \beq{sphere} E_{sp} (R,W)
\triangleq \max_{P_X}\tilde{E}_{sp}(R,P_X,W).
\end{equation}
In other words, $\max\{E_r (R,W),E_{ex} (R,W)\}\leq E(R,W)\leq
E_{sp} (R,W)$. Also, we can form Gallager's channel functions
\beq{channel} E_0 (\rho,W )\triangleq \max_{P_{X}} \tilde{E}_0
(\rho,P_X, W)
\end{equation}
and \beq{channelexp} E_x (\rho,W )\triangleq \max_{P_{X}}
\tilde{E}_x (\rho,P_X, W).
\end{equation}
It should be noted that maximization over $P_X$ means maximization
over the closed bounded set $\{(p_1,\dots , p_{|{\cal X }|}) : p_i
\geq 0, \sum p_i =1 \}$. Thus, if the function involved is
continuous, the maximum is achieved for some distribution
$\overline{P}_X$.

The functions $\tilde{E}_r (R,P_X,W)$ and
$\tilde{E}_{sp}(R,P_X,W)$ in (\ref{random1}) and (\ref{sphere1})
are equal if the maximizing $\rho \leq 1$ in (\ref{sphere1}) or
equivalently, if $R \geq R_{cr}(P_X,W)$, where $R_{cr}(P_X,W)$ is
the critical rate of the channel $W$ under distribution $P_X$,
defined by \beq{crit}
 R_{cr}(P_X ,W)  \triangleq \frac{\partial \tilde{E}_0 (\rho,P_X, W)}{\partial \rho} \biggm|_{\rho =1}.
\end{equation}
For all $P_X$, $\tilde{E}_r (R,P_X,W)$ and
$\tilde{E}_{sp}(R,P_X,W)$ vanish for all $R\geq C$. Consequently,
their maxima over $P_X$, $E_r (R,W) $ and $E_{sp} (R,W)$, vanish
for $R \geq C$ and are equal on some interval $[R_{cr}(W),C]$
where $R_{cr}(W)$ is the critical rate of the channel and is
defined by \beq{rcr} R_{cr}(W) \triangleq \inf \{R : E_r (R,W)
=E_{sp} (R,W)\}.
\end{equation}
Furthermore, it is known that $E_{sp} (R,W)$ meets $E_r (R,W)$ on
its supporting line of slope $-1$ \cite[p. 171]{Csiszar3}, which
means that $E_r (R,W)$ is a straight line with slope $-1$ for
$R\leq R_{cr}(W)$ and hence
\begin{equation}
E_r(R,W)=E_0(1,W)-R, \qquad R\leq R_{cr}(W). \label{straighteq}
\end{equation}

For all $P_X$, the function $\tilde{E}_{ex} (R,P_X, W)$ is a
decreasing convex curve with a straight-line section of slope $-1$
for $R\geq R_{ex}(P_X,W)$, and $\tilde{E}_{ex}
(R,P_X,W)>\tilde{E}_{r}(R,P_X,W)$ for $R<R_{ex}(P_X,W)$, where
$R_{ex}(P_X,W)$ is the ``expurgated'' rate of the channel $W$
under distribution $P_X$, defined by \beq{expu}
 R_{ex}(P_X ,W)  \triangleq \frac{\partial \tilde{E}_x (\rho,P_X, W)}{\partial \rho} \biggm|_{\rho =1}.
\end{equation}
Since the above are satisfied for all $P_X$, we then obtain the
following relation between the two lower bounds: $E_r (R,W)<E_{ex}
(R,W)$ for $R<R_{ex}(W)$ and $E_r (R,W)\geq E_{ex} (R,W)$
otherwise, where \beq{ecr} R_{ex}(W) \triangleq \inf \{R : E_r
(R,W) =E_{ex} (R,W)\}
\end{equation}
is the expurgated rate of the channel. Furthermore, it is known
that $E_{ex} (R,W)$ and $E_r (R,W)$ meet their supporting line of
slope $-1$ (according to the fact that $E_0(1,W)=E_x(1,W)$)
\cite[p. 154]{Gallager}. This geometric relation implies that
$R_{ex}(W)\leq R_{cr}(W)$ and $E_r (R,W) =E_{ex} (R,W)$ is a
straight line in the region $[R_{ex}(W),R_{cr}(W)]$.

We remark that Csisz\'ar \cite{Csiszar1} defines $e(R,Q)$,
$\tilde{E}_r (R,P_X,W)$, and $\tilde{E}_{sp}(R,P_X,W)$ using
expressions involving constrained minima of Kullback-Leibler
divergences. He also defines $\tilde{E}_{ex}(R,P_X,W)$ in terms of
the Bhattacharya distance and the mutual information between two
channel inputs. Our expressions are equivalent, as can be shown by
the Lagrange multiplier method; see also
\cite[pp.~192--193]{Csiszar3} and \cite{Blahut}.

\subsection{Tilted Distributions} \label{Tilted}
We associate with the source distribution $Q$ a family of tilted
distributions $Q^{(\rho)}$ defined by \beq{tilt} Q^{(\rho)}(s)
\triangleq \frac{Q^{\frac{1}{1+ \rho}}(s)}{\sum_{s' \in {\cal
S}}Q^{\frac{1}{1+ \rho}} (s')}, \qquad s \in {\cal S}, \qquad \rho
\geq 0 .
\end{equation}
\begin{lemma} \label{increase}
{\rm \cite[p. 44]{Csiszar3} The entropy $H(Q^{(\rho )})$ is a
strictly increasing function of $\rho$ except in the case that
$Q(s) =1/|{\cal S}|$ for all $s \in {\cal S}$ . Moreover, for
$H(Q) \leq R \leq \log |{\cal S}|$, the equation $H(Q^{(\rho)})=R$
is satisfied by a unique value $\rho^*$ (where we define $\rho^*
\triangleq\infty$ if $R=\log |{\cal S}|$ and define
$H(Q^{(\infty)}) \triangleq\log |{\cal S}|$).}
\end{lemma}
The proof  that $H(Q^{(\rho )})$ is increasing follows easily from
differentiation with respect to $\rho$ and a use of the
Cauchy-Schwarz inequality. The remainder of the proof follows from
the facts that $H(Q^{(0)})=H(Q)$, $\lim_{\rho \to \infty}
H(Q^{(\rho)}) = \log |{\cal S}|$ and that $H(Q^{(\rho)})$ is a
continuous function of $\rho$.

It is easily seen that \beq{ES} H(Q^{(\rho)}) =\frac{\partial E_s
(\rho ,Q)}{\partial \rho} ,
\end{equation}
where $E_s(\rho ,Q)$ is defined by (\ref{source}). From this  we
see that for $R \geq H(Q)$ the supremum in (\ref{source.exp}) is
achieved at $\rho^*$.

\subsection{Fenchel Duality }
Although many of our results can be obtained by the use of the
Lagrange multiplier method, the Fenchel Duality Theorem gives more
succinct proofs and seems particularly well-adapted to the
elucidation of the connection between error exponents on the one
hand, and source and channel functions on the other.\footnote{Another
related application of Fenchel duality is carried out in \cite{arikan1}
in the context of guessing subject to distortion, where it is shown that
the guessing exponent is the Fenchel transform of the error exponent
for source coding with a fidelity criterion.}
We present
here a simplified one-dimensional version which is adequate for
our purposes. For more detailed discussion, the reader may consult
\cite[pp. 190--202]{Luenberger}, \cite[Chapter 7]{bert}, or
\cite{rock}.

For any function $f$ defined on $F \subset {\mathbb R}$, define
its convex Fenchel transform (conjugate function, Legendre
transform) $f^*$ by
 \[ f^*(y) \triangleq \sup_{x \in F} [xy - f(x)]\]
and let $F^{*}$ be the set $\{y : f^* (y) < \infty\}$. It is easy
to see from its definition that $f^*$ is a convex function on
$F^*$. Moreover, if $f$ is convex and continuous, then $(f^* )^*
=f$. More generally, $f^{**} \leq f$ and $f^{**}$ is the convex
hull of $f$, {\em i.e.} the largest convex function that is
bounded above by $f$ \cite[Section~3]{rock},
\cite[Section~7.1]{bert}.

Similarly, for any function $g$ defined on $G \subset {\mathbb
R}$, define its concave Fenchel transform  $g_*$ by
 \[ g_*(y) \triangleq \inf_{x \in G} [xy - g(x)]\]
and let $G_{*}$ be the set $\{y : g_* (y) > -\infty\}$. It is easy
to see from its definition that $g_*$ is a concave function on
$G_*$. Moreover, if $g$ is concave and continuous, then $(g_* )_*
=g$. More generally, $g_{**} \geq g$ and $g_{**}$ is the concave
hull of $g$, {\em i.e.} the smallest concave function that is
bounded below by $g$.

\begin{Fen}
{\rm \cite[p. 201]{Luenberger} Assume that $f$ and $g$ are,
respectively, convex and concave functions on the non-empty
intervals $F$ and $G$ in ${\mathbb R}$ and assume that $F\cap G$
has interior points. Suppose further that $\mu = \inf_{x \in F
\cap G}[f(x) -g(x)] $ is finite. Then
\begin{equation}
\mu = \inf_{x \in F \cap G} [f(x) -g(x)] = \max _{y \in F^* \cap
G_* } [g_{*}(y) -f^*(y)],\label{duality}
\end{equation}
where the
maximum on the right is achieved by some $y_0 \in F^* \cap G_*$.
If the infimum on the left is achieved by some $x_0 \in F \cap G$,
then
\begin{equation} \max _{x \in F} [ xy_0  -f(x)] =
x_0y_0 -f(x_0)\label{ach1}
\end{equation}
and
\begin{equation}
\min_{x \in G }[ x y_0  - g(x) ] =  x_0  y_0  -g(x_0
).\label{ach2}
\end{equation}}
\end{Fen}

\subsection{Properties of the Source and Channel Functions }

\begin{lemma} \label{sf}
{\rm The source function $E_s (\rho,Q)$  defined by
{\rm(\ref{source})} is a strictly convex  function of $\rho$.}
\end{lemma}
Convexity follows directly from (\ref{ES}) and Lemma
\ref{increase}. Strict convexity is a consequence of our general
assumption that $Q$ is not the uniform distribution. It will be
seen from (\ref{source.exp}) that $e(R,Q)$ is the convex Fenchel
transform of $E_s (\rho ,Q)$. In fact, it is easily checked that
(e.g., cf. \cite[pp.~44--45]{Csiszar3})\beq{eRQ1} e(R,Q)=
\begin{cases}
0 & \text{ if $R \leq H(Q)$},\\
D(Q^{(\rho^*)} \Vert Q) &\text{ if $H(Q) \leq R \leq \log |{\cal S}| $ },\\
\infty &\text{ if $ R> \log |{\cal S}|$ },
\end{cases} \end{equation}
where $D(\cdot \Vert \cdot )$ denotes the Kullback-Leibler
divergence and $\rho^*$ is the solution of $H(Q^{(\rho)})=R$. Note
that (\ref{eRQ1}) implies that $e(R,Q)$ is strictly convex in $R$
on $[H(Q), \log |{\cal S}|]$ when the source is nonuniform;
otherwise $H(Q)=\log |{\cal S}|$.

The relation between the Gallager's channel function $E_0 (\rho
,W)$ and the random-coding and sphere-packing bounds is more
complicated. First of all, recall that for each $P_X$,
$\tilde{E}_r (R,P_X,W)$ as defined in (\ref{random1}) is a convex
non-increasing function for all $R$, and is a linear function of
$R$ with slope $-1$  for $R \leq R_{cr}(P_X,W)$ \cite[p. 143
]{Gallager}. It will be convenient to regard this linear function
as defining $\tilde{E}_r (R,P_X,W)$ for all negative $R$. The
random coding bound $E_r (R,W)$, which is the maximum of this
family of convex functions, is  a convex strictly decreasing
function of $R$ for $R<C$, and is a linear function of $R$ with
slope $-1$ for all $R$ below the critical rate $R_{cr}(W)$. For
$R\geq C$, $E_r (R,W)=0$.  Since $E_r (R,W)$ is convex, then $-E_r
(R,W)$ is concave. Let $T_r (\rho ,W )$ be the concave transform
of $-E_r (R,W)$, {\em i.e.} \beq{Trho} T_r (\rho ,W ) \triangleq
\inf_{R \in {\mathbb R}}[\rho R + E_r (R,W)].
\end{equation} It follows from the properties of $E_r (R,W)$ noted
above that $T_r (\rho ,W ) = -\infty$ for $\rho <0$ and $\rho
>1$ and that $T_r (\rho ,W )$ is finite for $\rho \in [0,1]$.
\begin{lemma} \label{cf}
{\rm The function $T_r (\rho ,W )$ defined by {\rm (\ref{Trho})}
is the concave hull  on the interval $[0,1]$ of the channel
function $E_0 (\rho ,W)$ defined in {\rm (\ref{channel})}. Thus,
$E_0 (\rho ,W) \leq T_r (\rho ,W )$ for $0 \leq \rho \leq 1$.}
\end{lemma}
\textbf{Proof}: We form the concave transform of $E_0 (R,W)$ on
the interval $[0,1]$ to get
\[ \left(E_0 (\rho ,W)\right)_* =\inf_{0 \leq \rho \leq 1}[\rho R -E_0 (\rho , W)] = -\sup_{0 \leq \rho \leq 1} [E_0 (\rho ,W) -\rho R] .\]
Now use, in succession, (\ref{channel}), (\ref{random1}), and
(\ref{random.coding}) to get
\begin{eqnarray}
\left(E_0 (\rho ,W)\right)_* &=&-\sup_{0 \leq \rho \leq 1}\max_{P_X}[\tilde{E}_0 (\rho,P_X, W) -\rho R]\nonumber\\
&=&-\max_{P_X} \sup_{0 \leq \rho \leq 1}[\tilde{E}_0 (\rho,P_X,
W)- \rho R]\nonumber\\
&=& -\max_{P_X} \tilde{E}_r (R,P_X,W)\nonumber\\
&=& -E_r (R,W) \nonumber.
\end{eqnarray} Since $T_r (\rho ,W )$ is the concave transform of
the concave function, $-E_r (R, W)$, we have that
\[ \left(-E_r (R,W) \right)_* = T_r (\rho ,W ) \quad \text{ and so }\quad \left(E_0 (\rho ,W)\right)_{**} =T_r (\rho ,W ).\]
 Hence, $T_r (\rho ,W )$ is the concave hull on $[0,1]$ of $E_0 (\rho ,R)$. \qed

Similarly to the above, recall that $E_{sp} (R,W)$, defined in
(\ref{sphere}) is convex, zero for $R\geq C$, positive for $R <C$,
and finite if $R > R_{\infty }(W)$ \cite{Csiszar3},
\cite{Gallager}, where $R_{\infty}(W)$ is given by \beq{rinf}
 R_{\infty}(W) \triangleq \lim_{\rho \to \infty} \frac{E_0 (\rho ,W)}{\rho}.
\end{equation}
A computable expression for $R_{\infty}(W)$ is given in \cite[p.
158]{Gallager}. The normal situation is $R_{\infty}(W) =0$. (As
shown by Gallager, $R_{\infty}(W) =0$ unless each channel output
symbol is unreachable from at least one input. In the latter case,
$R_{\infty}(W) >0$.) We now let $T_{sp} (\rho ,W)$ be the concave
transform of the concave function $-E_{sp} (R,W)$, {\em i.e.}
\beq{Tsph} T_{sp} (\rho ,W) \triangleq \inf_{R_{\infty}(W)< R<
\infty} [\rho R + E_{sp}(R,W)] .
\end{equation} It follows that $T_{sp} (\rho ,W) = -\infty$ for $\rho <0$ and
that $0 \leq T_{sp} (\rho ,W) <\infty $ for $ \rho \geq 0$.
\begin{lemma} \label{cfsp}
{\rm The function $T_{sp} (\rho ,W)$ defined by {\rm (\ref{Tsph})}
is the concave hull on $[0, \infty)$ of the channel function $E_0
(\rho ,W)$ defined in {\rm (\ref{channel})}.}
\end{lemma}
\textbf{Proof}: We now form the concave transform of $E_0 (\rho
,W)$ on the interval $[0, \infty)$ to get
\[ \left(E_0 (\rho ,W)\right)_* =\inf_{0 \leq \rho < \infty}[\rho R -E_0 (\rho , W)] = -\sup_{0 \leq \rho < \infty} [E_0 (\rho ,W) -\rho R] .\]
Now use (\ref{channel}), (\ref{sphere1}), and (\ref{sphere}) to
get
\begin{eqnarray}
\left(E_0 (\rho ,W)\right)_*&=&-\sup_{0 \leq \rho < \infty}\max_{P_X}[\tilde{E}_0 (\rho,P_X, W) -\rho R]\nonumber\\
&=&-\max_{P_X} \sup_{0 \leq \rho < \infty}[\tilde{E}_0 (\rho,P_X,
W) - \rho R]\nonumber\\
&=& -\max_{P_X} \tilde{E}_{sp}(R,P_X,W)\nonumber\\
&=& -E_{sp} (R,W) \nonumber.
\end{eqnarray}
As in the previous proof, $\left(E_0 (\rho ,W)\right)_{**} =T_{sp}
(\rho ,W )$. Hence, $T_{sp} (\rho ,W )$ is the concave hull on
$[0,\infty)$ of $E_0 (\rho ,R)$. \qed

\begin{figure}[c!]
\begin{center}
\includegraphics[width=16cm]{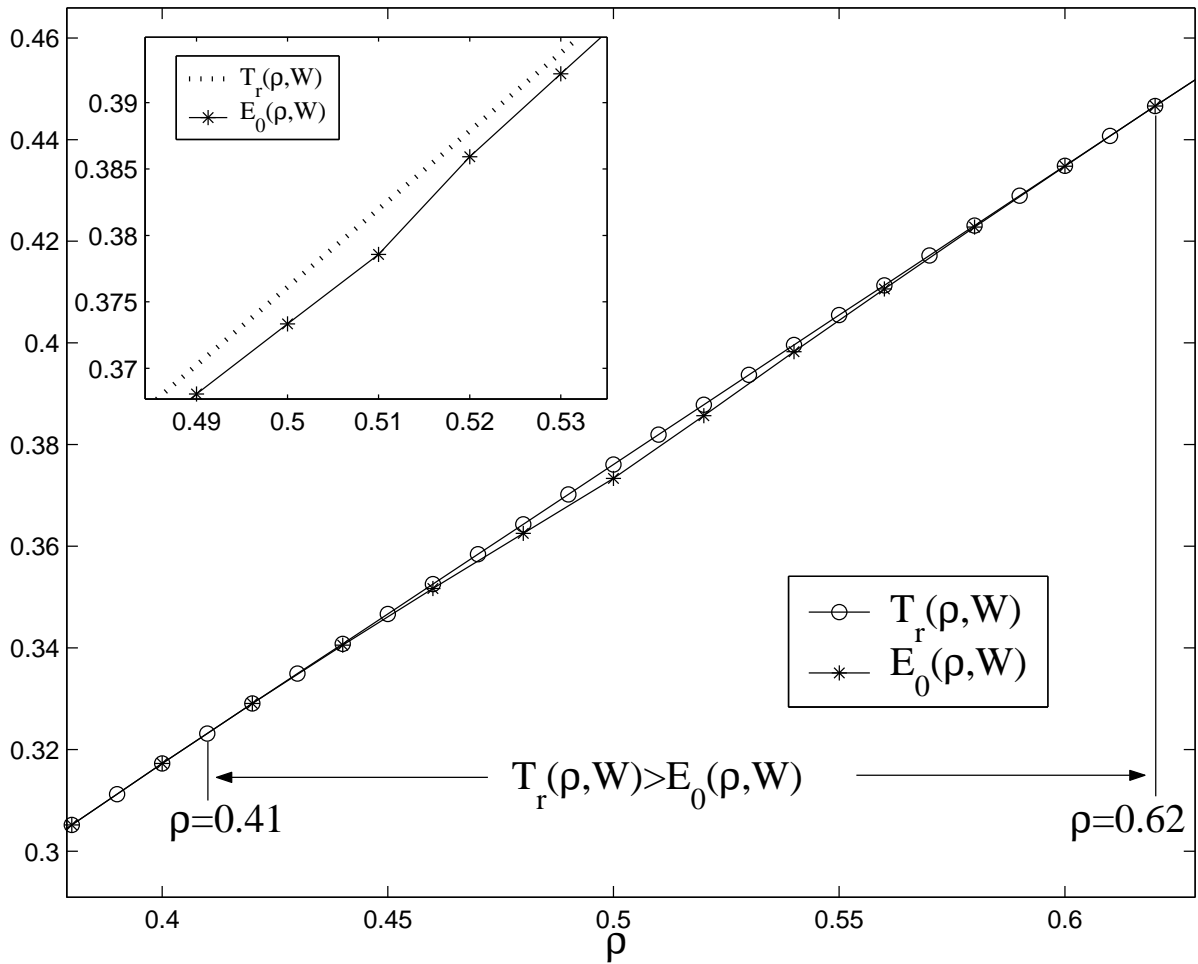}
\caption{Example of a 6-ary input, 4-ary output DMC (see
\cite[Fig.~5.6.5]{Gallager}) for which $E_0(\rho,W)$ is not
concave.} \label{concavehull}
\end{center}
\end{figure}

\begin{observation} {\rm Note that the function $\tilde{E}_0 (\rho,P_X, W)$
is concave in $\rho$ for each $P_X$ \cite[p. 142]{Gallager}.
Hence, if the maximizing $P_X$ in (\ref{channel}) is
\textit{independent} of $\rho$, $E_0 (\rho, W)$ is concave and
thus $T_r (\rho ,W )$ and $T_{sp} (\rho ,W)$ are equal to $E_0
(\rho ,W)$. This situation holds if the channel is symmetric in
the sense of Gallager \cite[p. 94]{Gallager} (also see Example
\ref{symmetric}). For this case, the maximizing distribution is
the uniform distribution $P_X (x) = 1/|{\cal X}|$ for all $x \in
{\cal X}$. However, there are channels for which $E_0 (\rho, W)$
is not concave. One example of such a channel is provided by
Gallager \cite[Fig. 5.6.5]{Gallager}. For this particular (6-ary
input, 4-ary output) channel, we plot $E_0 (\rho, W)$ against
$\rho$ in Fig. \ref{concavehull}. It is noted that the derivative
of $E_0 (\rho, W)$ has a positive jump increase at around
$\rho=0.51$ (see \cite[Fig. 5.6.5]{Gallager}), and its concave
hull $T_{r}(\rho,W)$ is strictly larger than $E_0 (\rho, W)$ in
the interval $\rho\in(0.41,0.62)$. }\label{obs1}
\end{observation}


\section{Bounds on the JSCC Error Exponent} \label{bounds}

\subsection{Csisz\'{a}r's Random-Coding and Sphere-Packing
Bounds}

Csisz\'ar \cite{Csiszar1} proved that for a DMS and a DMC the JSCC
error exponent in Definition~\ref{errorexp} satisfies
\beq{csiszar1}  \underline{E}_r(Q,W,t)\leq E_J(Q,W,t)  \leq
\overline{E}_{sp}(Q,W,t),
\end{equation}
where \beq{defEr} \underline{E}_r(Q,W,t)\triangleq \min_{t H(Q)
\leq R \leq  t\log |{\cal S}|} \left[ t e\left(\frac{R}{t},
Q\right) + E_r (R,W) \right],
\end{equation}
and \beq{defEsp} \overline{E}_{sp}(Q,W,t)\triangleq \inf_{t H(Q)
\leq R \leq t\log |{\cal S}|} \left[ t e\left(\frac{R}{t},
Q\right) + E_{sp} (R,W) \right]
\end{equation}
are called the source-channel random-coding lower bound and the
source-channel sphere-packing upper bound, since they respectively
contain $E_r(R,W)$ and $E_{sp}(R,W)$ in their expressions. These
bounds can be expressed in a form more adapted to calculation as
follows.
\begin{theorem}
\label{equiv1} {\rm Let $tH(Q) <C$ and let $t \log |{\cal S}|>
R_{\infty}(W)$. Then \beq{newform1A}
\underline{E}_r(Q,W,t)=\max_{0 \leq \rho \leq 1}[T_r (\rho ,W )
-tE_s (\rho ,Q)]
\end{equation}
and \beq{newform1B} \overline{E}_{sp}(Q,W,t)=\max_{0 \leq \rho <
\infty} [T_{sp} (\rho ,W)-tE_s (\rho ,Q)]
\end{equation}
where $T_r (\rho ,W )$ and $T_{sp} (\rho ,W)$ are the concave
hulls of $E_0 (\rho ,W)$ on $[0,1]$ and $[0, \infty)$ defined in
{\rm (\ref{Trho})} and {\rm (\ref{Tsph})}, respectively. If the
maximizing $P_X$ in {\rm (\ref{channel})} is independent of
$\rho$, $T_r (\rho ,W )$ and $T_{sp} (\rho ,W)$ can be replaced by
$E_0 (\rho ,W)$.}
\end{theorem}
\begin{remark}\label{remequi1}
{\rm When $tH(Q)\geq C$,
$\underline{E}_r(Q,W,t)=\overline{E}_{sp}(Q,W,t)=0$. }
\end{remark}

\begin{observation}
{\rm According to Lemma \ref{cf}, $E_0 (\rho ,W) \leq T_r (\rho ,W
)$. Thus the lower bound $\underline{E}_r(Q,W,t)$ can be replaced
by the {\it possibly looser} lower bound\footnote{In
\cite{ISIT04}, \cite{QBSC04}, we incorrectly stated that
Csisz\'{a}r's random-coding lower bound $\underline{E}_r(Q,W,t)$
given in (\ref{defEr}) and Gallager's lower bound given in
(\ref{dd}) are identical. This is indeed not always true; it is
true if $E_0(\rho,W)$ is a concave function of $\rho$ (e.g., for
symmetric channels) or $tH(Q^{(1)})\leq R_{cr}(W)$ (see Corollary
\ref{iden}). Thus, although both lower bounds are ``random
coding'' type bounds, Csisz\'{a}r's bound is in general tighter.}
\beq{dd} \max_{0 \leq \rho \leq 1}[E_0 (\rho,W) -t E_s (\rho, Q)].
\end{equation} This is the lower bound implied by Gallager's work
\cite[p.~535]{Gallager}. As noted earlier, if the maximizing $P_X$
in (\ref{channel}) is independent of $\rho$ (e.g., for symmetric
channels, see Example~\ref{symmetric}), the two lower bounds are
identical. }
\end{observation}

\noindent\textbf{Proof of Theorem \ref{equiv1}}: We first apply
Fenchel's Duality Theorem (\ref{duality}) to the lower bound
$\underline{E}_r(Q,W,t)$. From Lemma~\ref{sf}, (\ref{source.exp}),
and (\ref{eRQ1}), $te(R/t,Q)$ is convex on $( -\infty , t\log
|{\cal S}|]$ and has convex transform $tE_s (\rho ,Q)$ on the set
$[0, \infty)$. Also, from the discussion preceding Lemma \ref{cf},
$-E_r (R,W)$ is concave on ${\mathbb R}$ and has concave transform
$T_r (\rho ,W )$ which is bounded on $[0,1]$. Thus, by Fenchel's
Duality Theorem, \beq{aa} \inf_{-\infty \leq R \leq  t\log |{\cal
S}|} \left[ t e\left(\frac{R}{t}, Q\right) + E_r (R,W) \right] =
\max_{0 \leq \rho \leq 1}[T_r (\rho ,W ) -tE_s (\rho ,Q)].
\end{equation}
Now the convex function $te(R/t,Q)+E_r(R,W)$ is non-increasing for
$R\leq tH(Q)$ since $te(R/t,Q) =0$ in this region. This implies
that the infimum on the left side of (\ref{aa}) can be restricted
to the interval $t H(Q) \leq R \leq  t\log |{\cal S}|$. Since this
is now the infimum of a continuous function on a finite interval
this will be a minimum. Hence, (\ref{newform1A}) is an equivalent
representation of $\underline{E}_r(Q,W,t)$.

Similarly, for the upper bound, recall from the discussion
preceding Lemma \ref{cfsp} that $-E_{sp}(R,W)$ is concave and
finite for $R> R_{\infty}(W)$ and has a concave transform $T_{sp}
(\rho ,W)$, which is finite on $0 \leq \rho < \infty$. Thus, by
Fenchel's Duality Theorem, \beq{bb} \inf_{R_{\infty}(W) < R \leq
t\log |{\cal S}|}\left[ t e\left(\frac{R}{t}, Q\right) + E_{sp}
(R,W) \right] = \max_{0 \leq \rho < \infty}[T_{sp} (\rho ,W) -t
E_s(\rho ,Q)] .
\end{equation}
The assumption $R_{\infty}(W) < t \log |{\cal S}|$ ensures that
the infimum on the left of (\ref{bb}) is taken over a set with
interior points. If $R_{\infty}(W)< tH(Q)$, the infimum can be
replaced by a minimum on the interval $t H(Q) \leq R \leq t\log
|{\cal S}|$ by the same argument as for the lower bound. If
$R_{\infty}(W)\geq tH(Q)$, we no longer form the infimum of a
continuous function, but it can still be shown that there is a
minimum point which lies in the interval $t H(Q) \leq R \leq t\log
|{\cal S}|$. Hence, (\ref{bb}) is an equivalent representation of
$\overline{E}_{sp}(Q,W,t)$. \qed

\begin{observation}
{\rm The parametric form of the lower and upper bounds
(\ref{newform1A}) and (\ref{newform1B}) indeed facilitates the
computation of Csisz\'ar's bounds. In order to compute the bounds
for general non-symmetric channels (when $tH(Q)<C$ and $t\log
|{\cal S}|>R_{\infty}$), one could employ Arimoto's algorithm
\cite{Arimoto} to find the maximizing distribution and thus $E_0
(\rho ,W)$. We then can immediately obtain the concave hulls
of $E_0(\rho ,W)$, $T_r(\rho ,W)$ and $T_{sp}(\rho ,W)$,
numerically (e.g., using Matlab)
and thus the maxima of $T_r (\rho ,W ) -tE_s (\rho ,Q)$ and
$T_{sp} (\rho ,W)-tE_s (\rho ,Q)$. This significantly reduces the
computational complexity since to compute (\ref{defEr}) and
(\ref{defEsp}), we need to first compute $E_r (R,W)$ and $E_{sp}
(R,W)$ for \textit{each} $R$, which requires almost the same
complexity as above, and then we need to find the minima by
searching over \textit{all} $R$'s. For symmetric channels,
(\ref{newform1A}) and (\ref{newform1B}) are analytically solved;
see Example~\ref{symmetric}. }
\end{observation}

\begin{figure}[c!]
\begin{center}
\includegraphics[width=16cm]{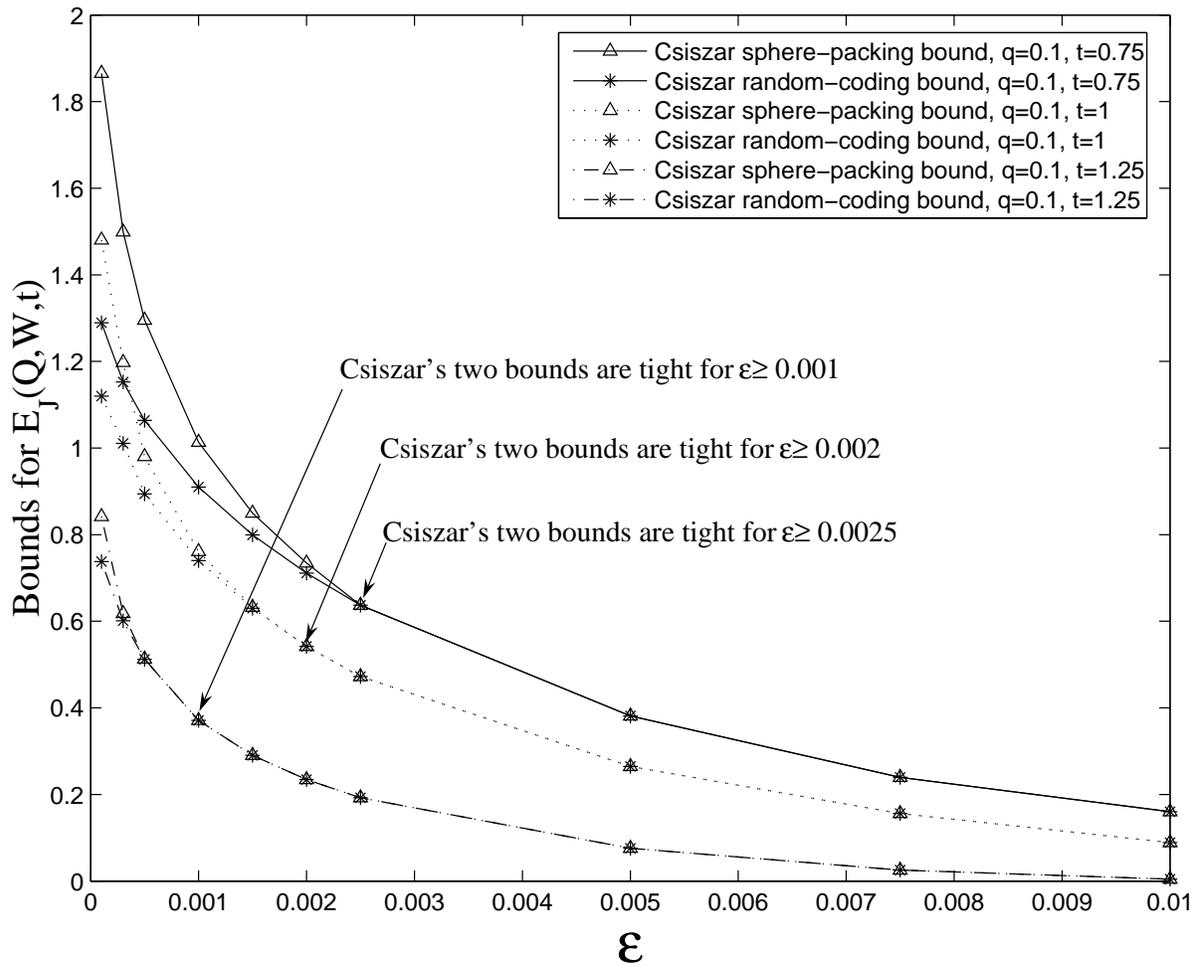}
\caption{Csisz\'{a}r's random-coding and sphere-packing bounds for
the system of
Example~\ref{computationex}.}\label{Csiszartwobounds}
\end{center}
\end{figure}

\begin{figure}[c!]
\begin{center}
\includegraphics[width=16cm]{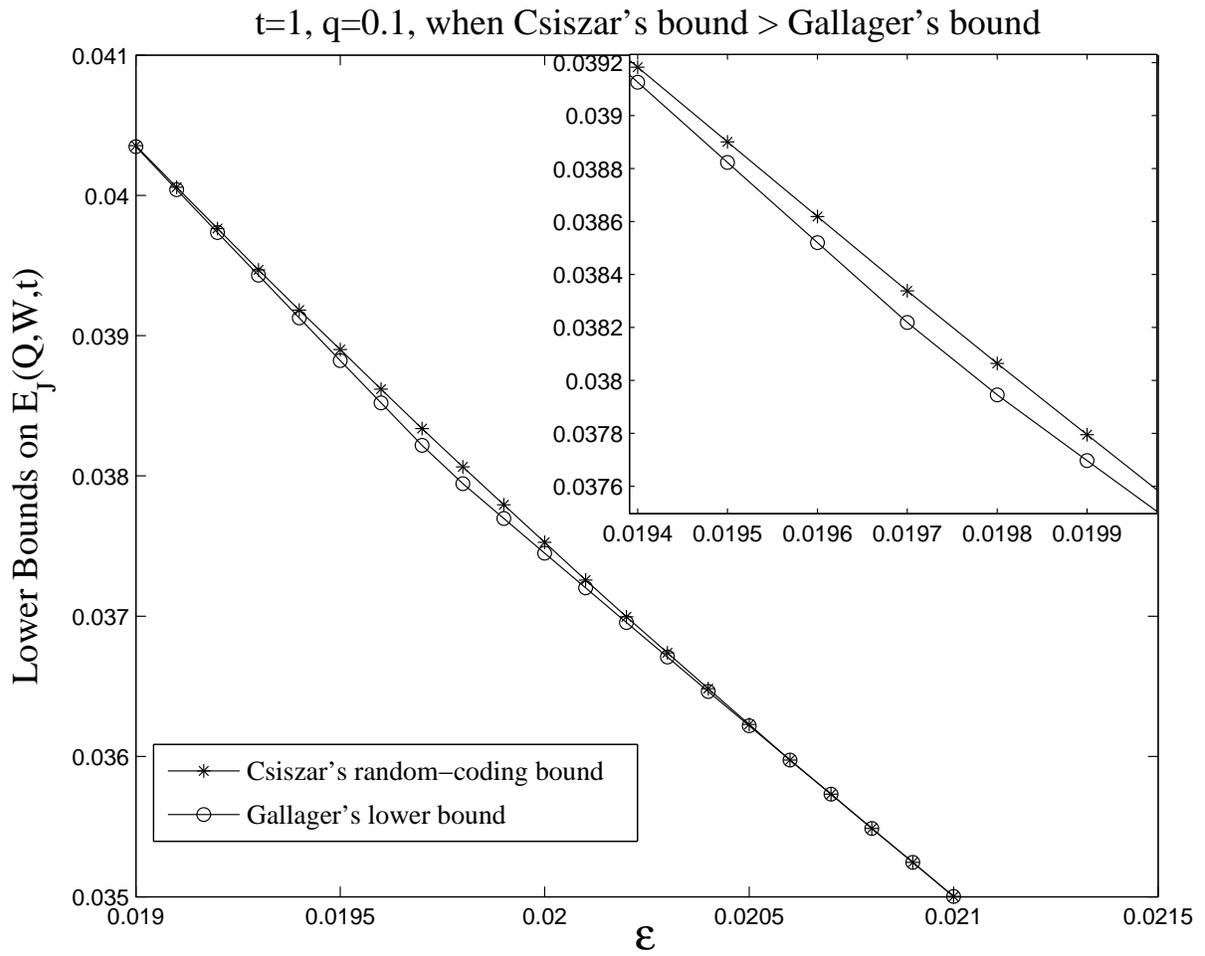}
\caption{Csisz\'{a}r's random-coding bound vs Gallager's lower
bound for the system of
Example~\ref{computationex}.}\label{CsiszarvsGallager}
\end{center}
\end{figure}

\begin{example}\label{computationex}{\rm
Consider a communication system with a binary DMS with
distribution $Q=\{q,1-q\}$ and a DMC with $|\mathcal{X}|=6$,
$|\mathcal{Y}|=4$, and transition probability matrix
\begin{eqnarray}\label{matrixp}
  W=\left[
                  \begin{array}{cccc}
                    1-18\varepsilon & 6\varepsilon     & 6\varepsilon     & 6\varepsilon\\
                    6\varepsilon    & 1-18\varepsilon  & 6\varepsilon     & 6\varepsilon\\
                    6\varepsilon    & 6\varepsilon     & 1-18\varepsilon  & 6\varepsilon\\
                    6\varepsilon    & 6\varepsilon     & 6\varepsilon     & 1-18\varepsilon\\
                    0.5-\varepsilon& 0.5-\varepsilon & \varepsilon     & \varepsilon\\
                    \varepsilon    & \varepsilon     & 0.5-\varepsilon & 0.5-\varepsilon\\
                  \end{array}
            \right], \qquad 0\leq\varepsilon\leq \frac{1}{18}.\nonumber
\end{eqnarray}

We then compute Csisz\'{a}r's random-coding and sphere-packing
bounds, $\underline{E}_r(Q,W,t)$ and $\overline{E}_{sp}(Q,W,t)$.
For fixed $Q$ and transmission rate $t$, we plot these bounds in
terms of $\varepsilon$ in Fig. \ref{Csiszartwobounds}. Our
numerical results show that $E_J$ could be determined exactly for
a large class of ($q$, $\varepsilon$, $t$) triplets: when source
$Q=\{0.1,0.9\}$ and rate $t=0.75$, $E_J$ is exactly known for
$\varepsilon\geq 0.0025$; when $Q=\{0.1,0.9\}$ and $t=1$, $E_J$ is
known for $\varepsilon\geq 0.002$; and when $Q=\{0.2,0.8\}$ and
$t=1.25$, $E_J$ is known for $\varepsilon\geq 0.001$. Since for
this channel $E_o(\rho,W)$ might not be concave (e.g., when
$\varepsilon=0.01$, $W$ reduces to the DMC discussed in
Observation \ref{obs1} at the end of Section \ref{prelim}), our
results indicate that Csisz\'{a}r's lower bound is slightly but
strictly larger (by $\approx0.0001$) than Gallager's lower bound
(\ref{dd}) for $q=0.1$, $t=1$, and $\varepsilon$ around $0.02$.
This is illustrated in Fig. \ref{CsiszarvsGallager}.}
\end{example}

\subsection{When Does \boldmath$\underline{E}_r(Q,W,t)=\overline{E}_{sp}(Q,W,t)$ ?}

One important objective in investigating the bounds for the JSCC
error exponent $E_J$ is to ascertain when the bounds are tight so
that the exact value of $E_J$ is obtained. According to
Csisz\'ar's result (\ref{csiszar1}), we note that if the minimum
in the expressions of $\underline{E}_r(Q,W,t)$ or
$\overline{E}_{sp}(Q,W,t)$ is attained for a rate (strictly)
larger than the critical rate $R_{cr}(W)$, then the two bounds
coincide and thus $E_J$ is determined exactly. This raises the
following question: how can we check whether the minimum in
$\underline{E}_r(Q,W,t)$ or $\overline{E}_{sp}(Q,W,t)$ is attained
for a rate larger than $R_{cr}(W)$? One may indeed wonder if there
exist explicit conditions for which
$\underline{E}_r(Q,W,t)=\overline{E}_{sp}(Q,W,t)$. The answer is
affirmative; furthermore, we can verify whether the two bounds are
tight in two ways: one is to compare $tH(Q^{(1)})$ with
$R_{cr}(W)$, and the other is to compare the minimizer of
$\overline{E}_{sp}(Q,W,t)$ in (\ref{newform1B}),
$\overline{\rho}^*$ say, with 1. Before we present these
conditions, we first define the following quantities which achieve
the bounds $\underline{E}_r(Q,W,t)$ and $\overline{E}_{sp}(Q,W,t)$
under the assumptions $tH(Q)<C$ and $t\log |{\cal S}|>R_{\infty}$:
\begin{eqnarray}
\underline{R}_{m} &\triangleq &\arg\min_{t H(Q) \leq R \leq  t\log
|{\cal S}|} \left[ t e\left(\frac{R}{t}, Q\right) + E_r (R,W)
\right],\label{minimizer1}\\
\overline{R}_{m} &\triangleq & \arg\min_{t H(Q) \leq R \leq t\log
|{\cal S}|} \left[ t e\left(\frac{R}{t}, Q\right) + E_{sp} (R,W)
\right],\label{minimizer2}\\
\underline{\rho}^*& \triangleq & \arg\max_{0 \leq \rho \leq
1}[T_r (\rho ,W ) -tE_s (\rho ,Q)],\label{maximizer1}\\
\overline{\rho}^* & \triangleq & \arg\max_{0 \leq
\rho<\infty}[T_{sp} (\rho ,W ) -tE_s (\rho ,Q)].\label{maximizer2}
\end{eqnarray}
Since the functions between brackets to be minimized (or
maximized) in (\ref{minimizer1})-(\ref{maximizer2}) are strictly
convex (or concave) functions of $R$ (or $\rho$),
$\underline{R}_{m}$, $\overline{R}_{m}$, $\underline{\rho}^*$ and
$\overline{\rho}^*$ are well-defined and unique. We then have the
following relations.

\begin{lemma}\label{prop}
{\rm
Let $tH(Q) <C$ and let $t \log |{\cal S}|>R_{\infty}(W)$. Then:\\
\noindent(1). $\overline{\rho}^*$ and $\underline{\rho}^*$ are
positive and finite.\\
\noindent(2). $\overline{R}_{m}=tH(Q^{(\overline{\rho}^*)})$.\\
\noindent(3). $\underline{R}_{m}=t H(Q^{(\underline{\rho}^*)})$ if
$\underline{\rho}^*<1$; $\underline{R}_{m}\geq tH(Q^{(1)})$ if
$\underline{\rho}^*=1$.}
\end{lemma}
\textbf{Proof}: We first prove (1). Since $T_{sp} (\rho ,W )$ is
the concave hull of $E_0(\rho,W)$, we have the following relation
$$
\lim_{\rho\downarrow0}\frac{T_{sp} (\rho ,W )}{\rho}\geq
\lim_{\rho\downarrow0}\frac{E_{0}(\rho,W)}{\rho}=C.
$$
where the last equality follows from \cite[Lemma 2]{Arimoto2}.
Since $\lim_{\rho\downarrow0}E_s (\rho ,Q)/\rho=H(Q)$ by
(\ref{ES}) and Lemma \ref{increase}, we have
$$
\lim_{\rho\downarrow0}\frac{T_{sp} (\rho ,W ) -tE_s (\rho
,Q)}{\rho}\geq C-tH(Q)>0.
$$
Note that the right-derivative of $T_{sp} (\rho ,W )$ (at
$\rho=0$) must exist due to its concavity
\cite[pp.~113--114]{Royden}, and hence
$\lim_{\rho\downarrow0}T_{sp} (\rho ,W )/\rho$ exists. Next we
denote $\varepsilon=t \log |{\cal S}|-R_{\infty}(W)>0$. It follows
from the definition of $T_{sp} (\rho ,W )$ that
$$
\lim_{\rho\rightarrow\infty}\frac{T_{sp} (\rho ,W)}{\rho}\leq
\lim_{\rho\rightarrow\infty}\frac{\rho
(R_{\infty}(W)+\varepsilon/2) +
E_{sp}(R_{\infty}(W)+\varepsilon/2,W)}{\rho}=R_{\infty}(W)+\varepsilon/2
$$
because of the finiteness of $E_{sp}(R,W)$ for $R>R_{\infty}(W)$.
This together with $\lim_{\rho\rightarrow\infty}E_s (\rho
,Q)/\rho=\log |{\cal S}|$ implies
$$
\lim_{\rho\rightarrow\infty}\frac{T_{sp} (\rho ,W)-tE_s (\rho
,Q)}{\rho}\leq R_{\infty}(W)+\varepsilon/2 -t \log |{\cal S}|<0.
$$
Since $T_{sp} (\rho ,W)-tE_s (\rho ,Q)$ is 0 and has a positive
right-slope at $\rho=0$ and is negative for $\rho$ sufficiently
large, by the strict concavity of $T_{sp} (\rho ,W ) -tE_s (\rho
,Q)$, the maximum in (\ref{maximizer2}) must be achieved by a
positive finite $\overline{\rho}^*$. The positivity of
$\underline{\rho}^*$ can be shown in the same way and
$\underline{\rho}^*$ is finite by its definition.

We next prove (2). If we now regard $te(R/t,Q)$ as $f^*(y)$ and
$tE_s(\rho,Q)$ as $f(x)$ (by noting that $f^{**}=f$), then
according to (\ref{ach1}) in Fenchel's Duality Theorem,
$$
\max_{0 \leq\rho<\infty}[\rho \overline{R}_{m} -tE_s (\rho
,Q)]=\overline{\rho}^* \overline{R}_{m} -tE_s (\overline{\rho}^*
,Q).
$$
Setting the derivative of $\rho \overline{R}_{m} -tE_s (\rho ,Q)$
equal to 0, we can solve for the stationary point\footnote{The
stationary points of a differentiable function $f(x)$ are the
solutions of $f'(x)=0$.} $\overline{\rho}^*$, which gives
$\overline{R}_{m}=tH(Q^{(\overline{\rho}^*)})$.

For the lower bound, using a similar argument, we obtain the
relation
$$
\max_{0 \leq \rho\leq 1}[\rho \underline{R}_{m} -tE_s (\rho
,Q)]=\underline{\rho}^* \underline{R}_{m} -tE_s
(\underline{\rho}^* ,Q).
$$
Recalling that the function between the brackets to be maximized
is strictly concave, if the above maximum is achieved by
$\underline{\rho}^*\in(0,1)$, then we can solve for the stationary
point as above and obtain $\underline{R}_{m}=t
H(Q^{(\underline{\rho}^*)})$. If the maximum is achieved at
$\underline{\rho}^*=1$, then the stationary point is beyond (at
least equal to) 1, and hence $\underline{R}_{m}\geq tH(Q^{(1)})$.
Thus (3) follows. \qed

In order to summarize the explicit conditions for the calculation
of $E_J$ it is convenient to define a critical rate for the source
by \beq{sour.crit} R_{cr}^{(s)}(Q) \triangleq \frac{\partial E_s
(\rho ,Q)}{\partial \rho}\biggr|_{\rho = 1} = H(Q^{(1)}),
\end{equation}
recalling that $Q^{(1)}(s) = \sqrt{Q(s)}/ (\sum_{s' \in
\mathcal{S}} \sqrt{Q(s')})$, $s \in \mathcal{S}$.

\begin{theorem}\label{tightcondition}
{\rm Let $tH(Q) <C$ and let $t\log |{\cal S}|>R_{\infty}(W)$. Then
\begin{itemize}
\item $tR_{cr}^{(s)}(Q)\geq R_{cr}(W)\Longleftrightarrow \overline{\rho}^*\leq 1 \Longleftrightarrow
tR_{cr}^{(s)}(Q) \geq\overline{R}_{m}=\underline{R}_{m} \geq
R_{cr}(W)$. In this case,
$$
E_{J}(Q,W,t)=T_{sp} (\overline{\rho}^* ,W ) -tE_s
(\overline{\rho}^* ,Q).
$$

\item $tR_{cr}^{(s)}(Q) < R_{cr}(W) \Longleftrightarrow\overline{\rho}^*>1 \Longleftrightarrow
R_{cr}(W) \geq \overline{R}_{m} > \underline{R}_{m} =
tR_{cr}^{(s)}(Q)$. In this case,
$$
E_0(1,W)-tE_s (1,Q)\leq E_{J}(Q,W,t) \leq T_{sp}
(\overline{\rho}^* ,W ) -tE_s (\overline{\rho}^* ,Q).
$$
\end{itemize}}
\end{theorem}

\begin{remark}
{\rm  Under the condition $tR_{cr}^{(s)}(Q)> R_{cr}(W)$,
$\overline{\rho}^*=1$ is possible. However, if $tR_{cr}^{(s)}(Q)=
R_{cr}(W)$, then we definitely have $\overline{\rho}^*=1$ and
$tR_{cr}^{(s)}(Q) =\overline{R}_{m}=\underline{R}_{m} =
R_{cr}(W)$. }
\end{remark}

\begin{remark}
{\rm It can be shown that $T_{sp}(1,W)=E_0(1,W)$ and thus when
$\overline{\rho}^*=1$, the JSCC exponent is determined by
$$
E_{J}(Q,W,t)=E_0(1,W) -tE_s (1 ,Q).
$$}
\end{remark}

\begin{corollary}\label{comp}
{\rm Let $tH(Q) <C$ and let $t\log |{\cal S}|>R_{\infty}(W)$. Then
$\underline{\rho}^*=\min\{1,\overline{\rho}^*\}$ and
$\underline{R}_{m}=tH(Q^{(\underline{\rho}^*)})$.}
\end{corollary}

The proof of Theorem \ref{tightcondition} involves a geometric
argument involving the left- and right- slopes of the convex
functions $E_{r}(R,W)$ and $E_{sp}(R,W)$ and is deferred to
Appendix \ref{pftightcon}. Corollary \ref{comp} could be regarded
as a complement of Lemma \ref{prop} (3) and it is also proved in
Appendix~\ref{pftightcon}.

\begin{corollary}\label{relateRm2}
{\rm If $\underline{R}_{m}\geq R_{cr}(W)$ or $\overline{R}_{m}>
R_{cr}(W)$, then $tR_{cr}^{(s)}(Q)\geq
\underline{R}_{m}=\overline{R}_{m}\geq R_{cr}(W)$, and the other
equivalent conditions in Theorem \ref{tightcondition} hold. }
\end{corollary}
\textbf{Proof}: If $\underline{R}_{m}\geq R_{cr}(W)$ or
$\overline{R}_{m}> R_{cr}(W)$, then
$\underline{R}_{m}=\overline{R}_{m}$ by Lemma \ref{relateRm} in
Appendix~\ref{pftightcon}. $tR_{cr}^{(s)}(Q)\geq
\underline{R}_{m}$ immediately follows from Corollary \ref{comp}.
\qed

\begin{remark}
{\rm Corollary \ref{relateRm2} states that if
$\underline{R}_{m}\geq R_{cr}(W)$ or $\overline{R}_{m}>
R_{cr}(W)$, then $E_J$ is determined exactly. Note that when
$\overline{R}_{m}=R_{cr}(W)$, the upper and lower bounds of $E_J$
may not be tight. In that case
$\underline{R}_{m}<R_{cr}(W)=\overline{R}_{m}$ is possible. The
relation between $\underline{R}_{m}$ and $\overline{R}_{m}$ is
summarized in Lemma \ref{relateRm} in Appendix~\ref{pftightcon}.}
\end{remark}

We point out that, in both the computation and analysis aspects,
the above conditions play an important role in verifying whether
$E_{J}$ can be determined exactly or not. For the class of
symmetric DMCs, we can use the conditions $tR_{cr}^{(s)}(Q) \geq
R_{cr}(W)$ and $tR_{cr}^{(s)}(Q) < R_{cr}(W)$ to derive explicit
formulas for $E_J$, see Example \ref{symmetric}. In Section
\ref{comparison}, we apply Theorem \ref{tightcondition} to
establish the conditions for which the JSCC exponent is larger
than the tandem coding exponent. Note that when $tR_{cr}^{(s)}(Q)
\leq R_{cr}(W)$, the source-channel random-coding bound admits a
simple expression \beq{Csissimple}
\underline{E}_r(Q,W,t)=E_0(1,W)-tE_s (1,Q).
\end{equation}
Consequently, we have the following statement.

\begin{corollary}\label{iden}
{\rm If $ tR_{cr}^{(s)}(Q)\leq R_{cr}(W)$, then Csisz\'{a}r's
random-coding bound and Gallager's lower bound (\ref{dd}) are
identical.}
\end{corollary}
\textbf{Proof}: Recall Gallager's lower bound to $E_J$ given by
(\ref{dd})
$$
\max_{0 \leq \rho \leq 1}[E_0 (\rho,W) -t E_s (\rho, Q)]\geq E_0
(1,W) -t E_s (1, Q).
$$
Since in general Gallager's lower bound cannot be larger than
Csisz\'{a}r's random-coding bound, they must be equal when
$tR_{cr}^{(s)}(Q)\leq R_{cr}(W)$. \qed

\begin{figure}[c!]
\begin{center}
\includegraphics[width=16cm]{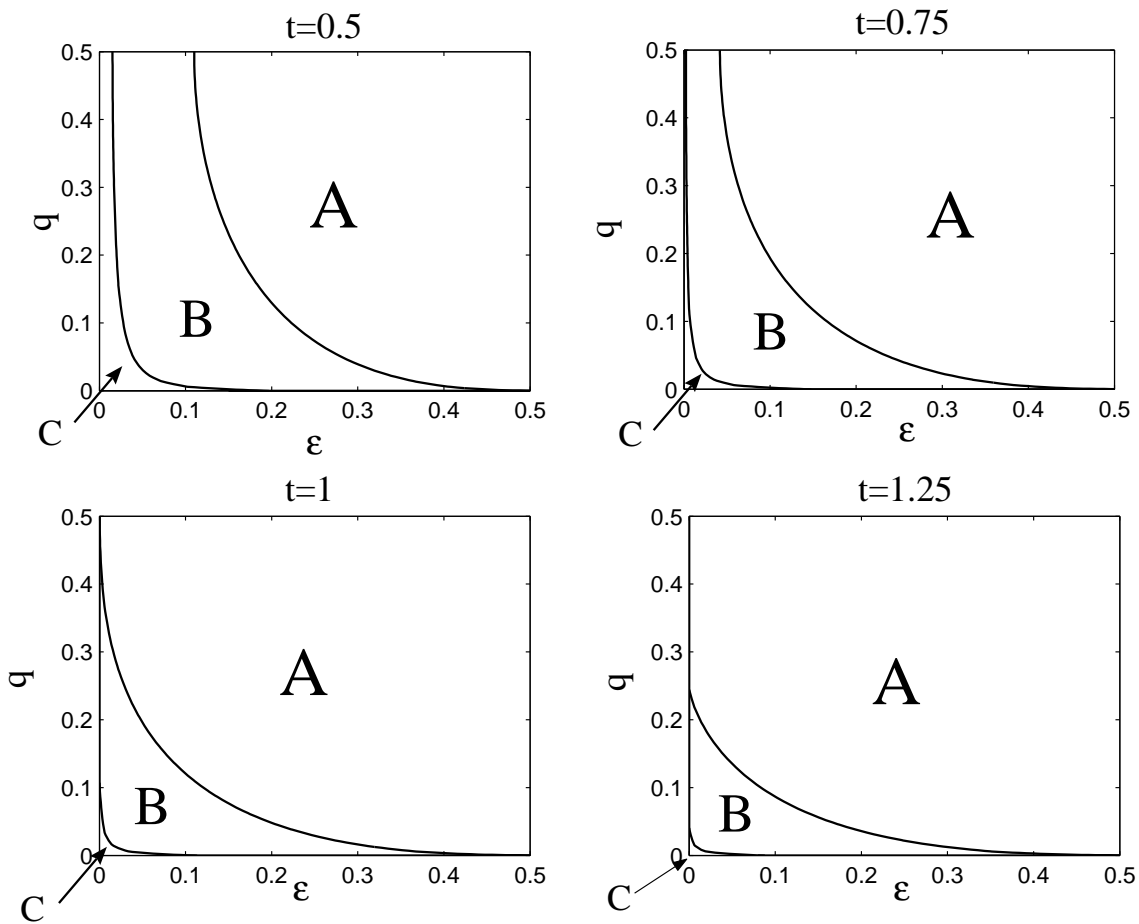}
\caption{The regions for the $(\varepsilon, q)$ pairs in the
binary DMS \{$q,1-q$\} and BSC ($\varepsilon$) system of
Example~\ref{symmetric} for different transmission rates $t$. Note
that $E_J=0$ on the boundary between $\mathbf{A}$ and
$\mathbf{B}$; $E_J$ is exactly determined on the boundary between
$\mathbf{B}$ and $\mathbf{C}$. In $\mathbf{A}$, $E_J=0$. In
$\mathbf{B}$, $E_J$ is positive and known exactly. In
$\mathbf{C}$, $E_J$ is positive and can be bounded above and
below.} \label{fig-BSC}
\end{center}
\end{figure}

\begin{example}\label{symmetric}
{\rm \textbf{(DMS and Symmetric DMC)} Consider a DMS
$\{Q:\mathcal{S}\}$ and a {\it symmetric}\footnote{Here symmetry
is defined in the Gallager sense~\cite[p.~94]{Gallager}; it is a
generalization of the standard notion of symmetry \cite{Thomas}
(which corresponds to $s=1$ above).} DMC \{$W:
\mathcal{X}\rightarrow \mathcal{Y}$\} with rate $t$, where the
channel transition matrix $W$ can be partitioned along its columns
into sub-matrices $W_{1},W_{2},\cdots,W_{s}$, such that in each
$W_{i}$ with size $|\mathcal{X}|\times |\mathcal{Y}_{i}|$, each
row is a permutation of each other row and each column is a
permutation of each other column. Denote the transition
probabilities in any column of sub-matrix $W_{i}$,
$i=1,2,\cdots,s$, by
$\left\{p_{i1},p_{i2},...,p_{i|\mathcal{X}|}\right\}$. Then both
$E_0(\rho,W)$ and the channel capacity are achieved by the uniform
distribution $P_{X}=1/|\mathcal{X}|$ and have the form
\begin{equation}
E_{0}(\rho,W)=(1+\rho)\log|\mathcal{X}|-\log\left\{\sum^s_{i=1}|\mathcal{Y}_{i}|\left(\sum^{|\mathcal{X}|}_{j=1}p^{\frac{1}{1+\rho}}_{ij}\right)^{1+\rho}\right\}\label{Esym}
\end{equation}
and
$$
C=\log|\mathcal{X}|-\frac{1}{|\mathcal{X}|}\sum^s_{i=1}|\mathcal{Y}_{i}|
\left(\sum^{|\mathcal{X}|}_{j=1}p_{ij}\right)H(P_i^{(0)}),
$$
where the tilted distribution $P^{(\alpha)}_i$, $\alpha \ge 0$,
for each $i=1,2,\cdots,s$, is defined on
$I_{\mathcal{X}} \triangleq \{1,2,\cdots,|\mathcal{X}| \}$ by
$$
P^{(\alpha)}_i(j) \triangleq
\frac{p^{\frac{1}{1+\alpha}}_{ij}}{(\sum_{j=1}^{|\mathcal{X}|}
p^{\frac{1}{1+\alpha}}_{ij})}, \quad j \in I_{\mathcal{X}}.
$$
Since now $E_{0}(\rho,W)$ is a concave and differentiable function
of $\rho$, the bounds $\underline{E}_r(Q,W,t)$ and
$\overline{E}_{sp}(Q,W,t)$ can be analytically obtained. If
\begin{equation}
\frac{1}{|\mathcal{X}|}\sum^s_{i=1}|\mathcal{Y}_{i}|
\left(\sum^{|\mathcal{X}|}_{j=1}p_{ij}\right)H(P_i^{(0)})+tH(Q)<
\log|\mathcal{X}|\label{co1}
\end{equation}
and
\begin{equation}
 \frac{\sum^s_{i=1}|\mathcal{Y}_{i}|\left(\sum^{|\mathcal{X}|}_{j=1}\sqrt{p_{ij}}\right)^{2}H(P^{(1)}_i)}{\sum^s_{i=1}|\mathcal{Y}_{i}|\left(\sum^{|\mathcal{X}|}_{j=1}\sqrt{p_{ij}}\right)^{2}}
+tH(Q^{(1)}) \geq\log|\mathcal{X}|,\label{co2}
\end{equation}
then the source-channel exponent is positive and is exactly
determined by
\begin{equation}
E_{J}(Q,W,t)=(1+\overline{\rho}^*)\log|\mathcal{X}|-\log\left\{\left[\sum^s_{i=1}|\mathcal{Y}_{i}|\left(\sum^{|\mathcal{X}|}_{j=1}p^{\frac{1}{1+\overline{\rho}^*}}_{ij}\right)^{1+\overline{\rho}^*}\right]\left(\sum_{s
\in \mathcal{S}}
Q^{\frac{1}{1+\overline{\rho}^*}}(s)\right)^{t(1+\overline{\rho}^*)}\right\},
\label{cor1a}
\end{equation}
where $\overline{\rho}^*$ is the unique root of the equation
\begin{eqnarray}
\frac{\sum^s_{i=1}|\mathcal{Y}_{i}|\left(\sum^{|\mathcal{X}|}_{j=1}p^{\frac{1}{1+\rho}}_{ij}\right)^{1+\rho}
H(P_i^{(\rho)})}
{\sum^s_{i=1}|\mathcal{Y}_{i}|\left(\sum^{|\mathcal{X}|}_{j=1}p^{\frac{1}{1+\rho}}_{ij}\right)^{1+\rho}}+
tH(Q^{(\rho)})
=\log|\mathcal{X}|.\label{root}
\end{eqnarray}
In the case when (\ref{co1}) does not hold, which means $tH(Q)\geq
C$, $E_{J}(Q,W,t)=0$. When (\ref{co1}) holds but (\ref{co2}) does
not hold, the right-hand side of (\ref{cor1a}) becomes the upper
bound $\overline{E}_{sp}(Q,W,t)$ and meanwhile, $E_{J}$ is lower
bounded by $E_0(1,W)-tE_s(1,Q)$, where $E_0(\rho,W)$ is given by
(\ref{Esym}).

Now we apply the conditions (\ref{co1}) and (\ref{co2}) to a
communication system with a binary source with distribution
\{$q,1-q$\}, a binary symmetric channel (BSC) with crossover
probability $\varepsilon$ and transmission rates $t=$0.5, 0.75, 1,
and 1.25. Note that
$$
R_{cr}(W)=1-h_{b}\left(\frac{\sqrt{\varepsilon}}{\sqrt{\varepsilon}+\sqrt{1-\varepsilon}}\right)
$$
and
$$
R_{cr}^{(s)}(Q)=h_{b}\left(\frac{\sqrt{q}}{\sqrt{q}+\sqrt{1-q}}\right),
$$
where $h_b(\cdot)$ is the binary entropy function. In
Fig.~\ref{fig-BSC}, we partition the set of possible points for
the $(\varepsilon,q)$ pairs into three regions: $\mathbf{A}$,
$\mathbf{B}$ and $\mathbf{C}$. If $(\varepsilon,q) \in
\mathbf{B}$, where conditions (\ref{co1}) and (\ref{co2}) hold,
i.e., $tH(Q)<C$ and $tR_{cr}^{(s)}(Q) \geq R_{cr}(W)$, then the
corresponding $E_{J}$ is positive and exactly known.\footnote{
In light of the recent work in \cite{barg},
where the random coding exponent $E_r (R,W)$ of the BSC is shown to be indeed
the true value of the channel error exponent $E(R,W)$ for code rates $R$
in some interval directly below the channel critical rate
(in other words, it is shown that for the BSC with its $\varepsilon$ above
a certain threshold,
$E_r (R,W) = E(R,W)$ for $R_1 \leq R \leq C$ where $R_1$ can be {\it less} 
than $R_{cr}(W)$ \cite{barg}), we note
via (\ref{intreq}) and the lower bound in (\ref{csiszar1})-(\ref{defEr})
that region $\mathbf{B}$ where $E_{J}$ is exactly known
can be enlarged.}
Furthermore,
if $(\varepsilon,q) \in \mathbf{C}$, then $E_{J}$ is bounded above
(below, respectively) by the right-hand side of (\ref{cor1a})
($E_0(1,W)-tE_s (1,Q)$, respectively). When $(\varepsilon,q)\in
\mathbf{A}$, where $tH(Q)>C$, $E_{J}$ is zero, and the error
probability of this communication system converges to 1 for $n$
sufficiently large. So we are only interested in the cases when
$(\varepsilon,q)\in \mathbf{B}\cup \mathbf{C}$. }
\end{example}

\subsection{Csisz\'{a}r's Expurgated Lower Bound}
In \cite{Csiszar2}, Csisz\'{a}r extended his work and obtained
another lower bound to $E_J$ for a class of source-channel pairs:
for a DMS and a DMC with zero-error capacity equal to 0,
\beq{csiszar2} E_{J}(Q,W,t)\geq \underline{E}_{ex}(Q,W,t)
\end{equation}
if $E_{ex}(R,W)=\max_{P_X}\tilde{E}_{ex}(R,P_X,W)$ is attained for
a $P_X$ not depending on $R$, where
\begin{eqnarray}
\underline{E}_{ex}(Q,W,t)\triangleq\min_{tH(Q)\leq R\leq
t\log|\mathcal{S}|}\left[t e\left(\frac{R}{t},Q\right)+
E_{ex}(R,W)\right]
\end{eqnarray}
is called the source-channel expurgated lower bound since it
contains $E_{ex}(R,W)$ in its expression. We then use Fenchel's
Duality Theorem to derive an equivalent expression of
$\underline{E}_{ex}(R,W,t)$.

\begin{theorem} \label{equiv2}
{\rm For a DMS and a DMC with zero-error capacity equal to 0, if
$E_{ex}(R,W)=\max_{P_X}\tilde{E}_{ex}(R,P_X,W)$ is attained for a
$P_X$ not depending on $R$, then \beq{newform2}
\underline{E}_{ex}(Q,W,t)= \sup_{\rho \geq 1} [E_{x} (\rho
,W)-tE_s (\rho ,Q)].
\end{equation}}
\end{theorem}
\textbf{Proof}: Recall that $\tilde{E}_{x}(\rho, P_{X},W)$ is
concave in $\rho$ on the interval $G= [1,+\infty)$
\cite[pp.~153--154]{Gallager}. Note that
$$
-\tilde{E}_{ex}(R,P_{X},W)\triangleq -\sup_{\rho\in G}[E_{x}(\rho,
P_{X},W)-\rho R]=\inf_{\rho\in G}[\rho R-\tilde{E}_{x}(\rho;
P_{X},W)]
$$
is the concave transform of $\tilde{E}_{x}(\rho, P_{X},W)$ on
$R\in G^*=\{R: -\tilde{E}_{ex}(R,P_{X},W)>-\infty\}=[0,+\infty)$
for DMCs with zero-error capacity equal to 0. Also recall that
$tE_{s}(\rho,Q)$ is strictly convex in $\rho$ on the interval $F=
[0,+\infty)$. Its convex transform
$$
\sup_{\rho\in F}[\rho R-t E_{s}(\rho,Q)]=t
e\left(\frac{R}{t},Q\right)
$$
is a function of $R$ on $F^*= \{R: t
e(R/t,Q)<+\infty\}=(-\infty,t\log|\mathcal{S}|]$. Fenchel's
Duality Theorem states that
$$
\inf_{\rho\in F\cap G }[t
E_{s}(\rho,Q)-\tilde{E}_{x}(\rho,P_{X},W)]=\max_{R\in F^*\cap G^*
}\left[-\tilde{E}_{ex}(R,P_{X},W)-t
e\left(\frac{R}{t},Q\right)\right]
$$
or
$$
\sup_{\rho\geq 1}[\tilde{E}_{x}(\rho, P_{X},W)-t
E_{s}(\rho,Q)]=\min_{0<R\leq t\log|\mathcal{S}|}\left[t
e\left(\frac{R}{t},Q\right)+\tilde{E}_{ex}(R,P_{X},W)\right].
$$
We can now maximize over $P_{X}$ and get the two equivalent lower
bounds:
\begin{eqnarray}
\sup_{\rho\geq
1}[E_{x}(\rho,W)-tE_{s}(\rho,Q)]&=&\max_{P_X}\min_{0<R\leq t
\log|\mathcal{S}|}\left[t
e\left(\frac{R}{t},Q\right)+\tilde{E}_{ex}(R,P_{X},W)\right]\nonumber\\
&\stackrel{(a)}{=}&\min_{0<R\leq t \log|\mathcal{S}|}\left[t
e\left(\frac{R}{t},Q\right)+\max_{P_X}\tilde{E}_{ex}(R,P_{X},W)\right]\nonumber\\
&\stackrel{(b)}{=}& \min_{tH(Q)\leq R\leq t
\log|\mathcal{S}|}\left[t
e\left(\frac{R}{t},Q\right)+E_{ex}(R,W)\right]\nonumber\\
&=&\underline{E}_{ex}(Q,W,t),\nonumber
\end{eqnarray}
where (a) follows by assumption that the maximizing $P_X$ does not
depend on $R$ and (b) holds since the convex function $t
e(R/t,Q)+E_{ex}(R,W)$ is either infinity or strictly decreasing
for $R<tH(Q)$. \qed

In the following lemma we note that the supremum in
(\ref{newform2}) can be replaced by a maximum, and the relation
between the maximizer $\underline{\rho}_x$ and its dual minimizer
$\underline{R}_{xm}$ is given.

\begin{lemma}\label{propexpur}
{\rm For DMC with zero-error capacity equal to 0, the function
$E_{x}(\rho,W)-tE_{s}(\rho,Q)$ has a global maximum at a finite
$\rho\geq 1$. Let
\begin{equation}
\underline{\rho}_x\triangleq\arg \max_{\rho\geq
1}[E_{x}(\rho,W)-tE_{s}(\rho,Q)]\label{maximizer3}
\end{equation}
and
\begin{equation}
\underline{R}_{xm}\triangleq\arg\min_{tH(Q)\leq R\leq t
\log|\mathcal{S}|}\left[t
e\left(\frac{R}{t},Q\right)+E_{ex}(R,W)\right].\label{minimizer3}
\end{equation}
Then $\underline{R}_{xm}=tH(Q^{(\underline{\rho}_x)})$ if
$\underline{\rho}_x>1$; $\underline{R}_{xm}\leq tR_{cr}^{(s)}(Q)$
if $\underline{\rho}_x=1$.}
\end{lemma}
\begin{remark}
{\rm Since the function between brackets to be optimized in
(\ref{maximizer3}) (or (\ref{minimizer3})) is strictly concave (or
convex), $\underline{\rho}_x$ and $\underline{R}_{xm}$ are
well-defined and unique. }
\end{remark}
\textbf{Proof}: We first show that $\underline{\rho}_x$ is finite.
Recall that for any $P_X$, Gallager's source and channel functions
$E_s(\rho,Q)$ and $\tilde{E}_x(\rho;P_X,W)$ given in
(\ref{channel3}) at $\rho=1$ reduce to
$$
E_s(1,Q)=\log \left(\sum_{s \in {\cal S}}\sqrt{Q(s)}\right)^2
$$
and
$$
\tilde{E}_x(1;P_X,W)=-\log \sum_{y\in {\cal Y}}\left(\sum_{x\in
{\cal X}} P_X (x) \sqrt{P_{Y|X}(y|x)} \right)^{2}.
$$
Using Jensen's inequality \cite{Thomas} on the convex function
$x^2$, we obtain
$$
E_s(1,Q)\leq \log \sum_{s \in {\cal
S}}(Q(s)Q(s)^{-1})=\log|\mathcal{S}|
$$
with equality if and only if $Q$ is uniform, and
$$
\tilde{E}_x(1;P_X,W)\geq -\log \sum_{y\in {\cal Y}}\sum_{x\in
{\cal X}}P_X (x)P_{Y|X}(y|x)=0.
$$
Therefore,
$$
E_{x}(1,W)-tE_{s}(1,Q)> -\log|\mathcal{S}|
$$
because of the nonuniform source assumption. On the other hand,
because the zero-error capacity is 0 we know that
$\lim_{\rho\rightarrow\infty}\frac{E_x(\rho,W)}{\rho}=0$ (from
\cite[p.~155]{Gallager}) and hence
$$
\lim_{\rho\rightarrow\infty}\frac{E_x(\rho,W)-tE_{s}(\rho,Q)}{\rho}\leq
-t\log_2|\mathcal{S}|.
$$
Clearly, since the concave function $E_{x}(\rho,W)-tE_{s}(\rho,Q)$
is finite (bounded below) at $\rho=1$, and approaches to $-\infty$
as $\rho\rightarrow \infty$, there exists a global maximum at a
finite $\underline{\rho}_x$. We next show the relation between
$\underline{\rho}_x$ and $\underline{R}_{xm}$. Following the proof
of Theorem \ref{equiv2}, let $f^*(y)$ be $te(R/t,Q)$ and let
$f(x)$ be $E_s(\rho,Q)$. Fenchel's Duality Theorem (\ref{ach1})
says that $\underline{\rho}_x$ and $\underline{R}_{xm}$ should
satisfy
$$
\max_{\rho\geq 1}[\rho \underline{R}_{xm} -tE_s (\rho
,Q)]=\underline{\rho}_x \underline{R}_{xm} -tE_s (\rho ,Q).
$$
If $\underline{\rho}_x>1$, then $\underline{\rho}_x$ is the
stationary point of the concave function $\rho \underline{R}_{xm}
-tE_s (\rho ,Q)$, and hence
$$
\underline{R}_{xm}=tH(Q^{(\underline{\rho}_x)}).
$$
Otherwise (if $\underline{\rho}_x=1$), which means that the
stationary point is less than or equal to 1,
$\underline{R}_{xm}\leq tR_{cr}^{(s)}(Q)$.

\qed

Analogously to Theorem~\ref{tightcondition}, we have the following
explicit conditions regarding the expurgated lower bound to the
JSCC exponent.
\begin{theorem}\label{rootcon2}
{\rm For the expurgated lower bound in Theorem \ref{equiv2}, the
following conditions are equivalent.
\begin{itemize}
\item $tR_{cr}^{(s)}(Q)<
R_{ex}(W)\Longleftrightarrow\underline{\rho}_x> 1
\Longleftrightarrow tR_{cr}^{(s)}(Q)<\underline{R}_{xm} \leq
R_{ex}(W)$. Thus,
$$
E_{J}(Q,W,t)\geq
E_x(\underline{\rho}_x,W)-tE_{s}(\underline{\rho}_x,Q).
$$

\item $tR_{cr}^{(s)}(Q)\geq R_{ex}(W)\Longleftrightarrow\underline{\rho}_x=1 \Longleftrightarrow
\underline{R}_{xm} =tR_{cr}^{(s)}(Q)\geq R_{ex}(W)$. Thus,
$$
E_{J}(Q,W,t) \geq E_x(1,W)-tE_{s}(1,Q).
$$
\end{itemize}}
\end{theorem}

The proof of Theorem \ref{rootcon2} is similar to that of Theorem
\ref{tightcondition} and is hence omitted. We next use
Theorems~\ref{tightcondition} and~\ref{rootcon2} to compare
Csisz\'{a}r's random-coding and expurgated lower bounds. Of clear
interest is the case when the expurgated bound improves upon the
random-coding bound.

\begin{corollary}\label{twolower}
{\rm The source-channel random-coding bound is improved by the
expurgated bound (i.e.,
$\underline{E}_r(Q,W,t)<\underline{E}_{ex}(Q,W,t)$) if and only if
$tR_{cr}^{(s)}(Q)< R_{ex}(W)$.}
\end{corollary}
\textbf{Proof}: When $tR_{cr}^{(s)}(Q)< R_{ex}(W)$, we must have
that $tR_{cr}^{(s)}(Q)< R_{cr}(W)$, since $R_{ex}(W)$ is never
larger than $R_{cr}(W)$. It follows from
Theorem~\ref{tightcondition} that the random-coding lower bound is
attained at $\underline{R}_{m} = tR_{cr}^{(s)}(Q)$. By Theorem
\ref{rootcon2} the expurgated lower bound is attained at
$R_{ex}(W)\geq \underline{R}_{xm}> tR_{cr}^{(s)}(Q)$. On account
of Lemma \ref{propexpur}, this must happen if
$\underline{R}_{xm}=tH(Q^{(\underline{\rho}_x)})$ with
$\underline{\rho}_x>1$. Thus,
$\underline{R}_{xm}>\underline{R}_{m}$ and
\begin{eqnarray}
\underline{E}_r(Q,W,t)&=&E_{r}(\underline{R}_{m},W)+te\left(\frac{\underline{R}_{m}}{t},Q\right)\nonumber\\\nonumber\\
&<& E_{r}(\underline{R}_{xm},W)+te\left(\frac{\underline{R}_{xm}}{t},Q\right)\nonumber\\
&\leq &
E_{ex}(\underline{R}_{xm},W)+t e\left(\frac{\underline{R}_{xm}}{t},Q\right)\nonumber\\
&=& \underline{E}_{ex}(Q,W,t).\nonumber
\end{eqnarray}
In this case, the source-channel expurgated lower bound is tighter
than the random-coding lower bound. We then show that
$\underline{E}_r(Q,W,t)\geq \underline{E}_{ex}(Q,W,t)$ if
$tR_{cr}^{(s)}(Q)\geq R_{ex}(W)$.

When $R_{ex}(W)\leq tR_{cr}^{(s)}(Q)\leq R_{cr}(W)$, it follows
from Theorems~\ref{tightcondition} and \ref{rootcon2} that
\begin{eqnarray}
\underline{E}_r(Q,W,t)&=&
E_0(1,W)-tE_s (1,Q)\nonumber\\
&=&
E_x(1,W)-tE_{s}(1,Q)\nonumber\\
&=& \underline{E}_{ex}(Q,W,t),\nonumber
\end{eqnarray}
where the second equality follows from the fact that, for any
$P_X$, Gallager's channel functions $\tilde{E}_0(1,P_X,W)$ and
$\tilde{E}_x(1,P_X,W)$ are equal \cite{Gallager}, and hence their
maxima are equal. In this case, the source-channel random-coding
lower bound is identical to the expurgated lower bound.

When $tR_{cr}^{(s)}(Q)> R_{cr}(W)$, we must have
$tR_{cr}^{(s)}(Q)> R_{ex}(W)$. Then the expurgated lower bound is
attained at $\underline{R}_{xm}=tR_{cr}^{(s)}(Q)$ by
Theorem~\ref{rootcon2}. On account of
Theorems~\ref{tightcondition} and Corollary \ref{comp}, the
random-coding lower bound is attained at
$\underline{R}_{m}=tH(Q^{(\underline{\rho}^*)})\geq R_{cr}(W)$
with $\underline{\rho}^*\leq 1$. Consequently,
\begin{eqnarray}
\underline{E}_r(Q,W,t) &=& E_{r}(\underline{R}_{m},W)+te\left(\frac{\underline{R}_{m}}{t},Q\right)\nonumber\\
&\geq & E_{ex}(\underline{R}_{m},W)+te\left(\frac{\underline{R}_{m}}{t},Q\right)\nonumber\\
&\geq &
E_{ex}(\underline{R}_{xm},W)+t e\left(\frac{\underline{R}_{xm}}{t},Q\right)\nonumber\\
&= & \underline{E}_{ex}(Q,W,t).\nonumber
\end{eqnarray}
In this case, the source-channel random-coding lower bound is
tighter than or equal to the expurgated lower bound. \qed

\begin{figure}[c!]
\begin{center}
\includegraphics[width=14cm]{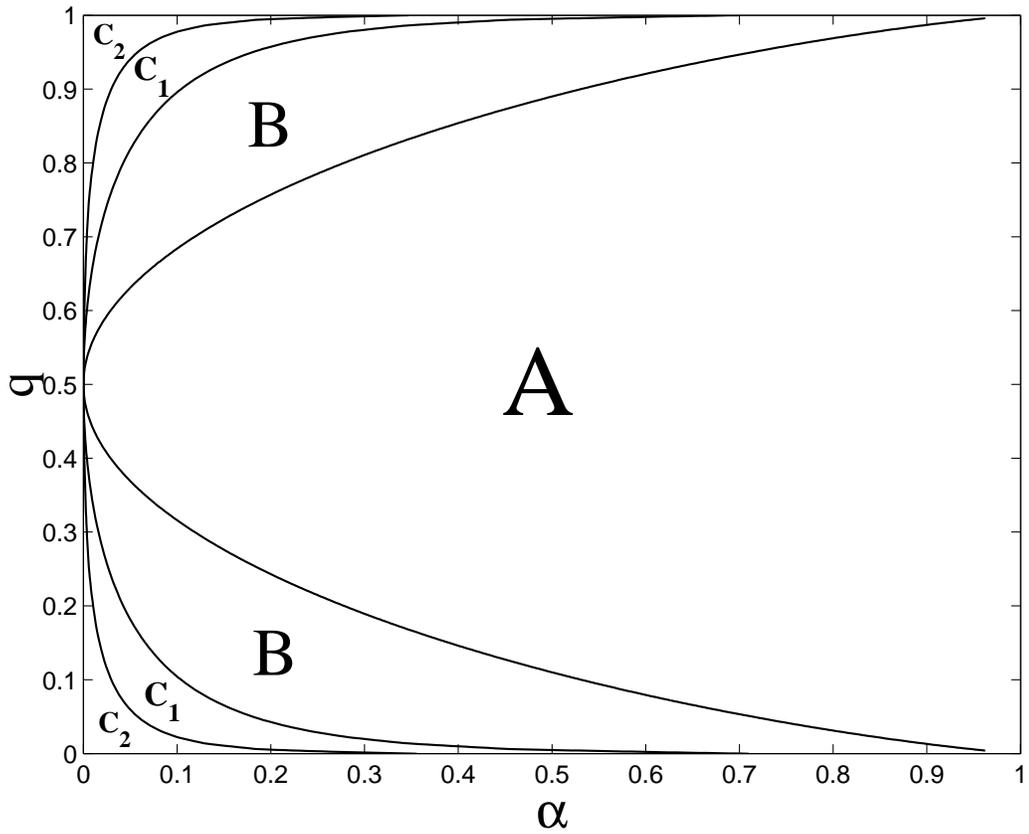}
\caption{The regions for the $(\alpha, q)$ pairs in the binary DMS
\{$q,1-q$\} and BEC ($\alpha$) system of
Example~\ref{equidischannel} with $t=1$. Note that $E_J=0$ on the
boundary between $\mathbf{A}$ and $\mathbf{B}$; $E_J$ is
determined on the boundary between $\mathbf{B}$ and
$\mathbf{C}_1$; The random-coding bound and expurgated bound to
$E_J$ are equal on the boundary between $\mathbf{C}_1$ and
$\mathbf{C}_2$.}\label{expurgatedBEC}
\end{center}
\end{figure}

\begin{figure}[c!]
\begin{center}
\includegraphics[width=16cm]{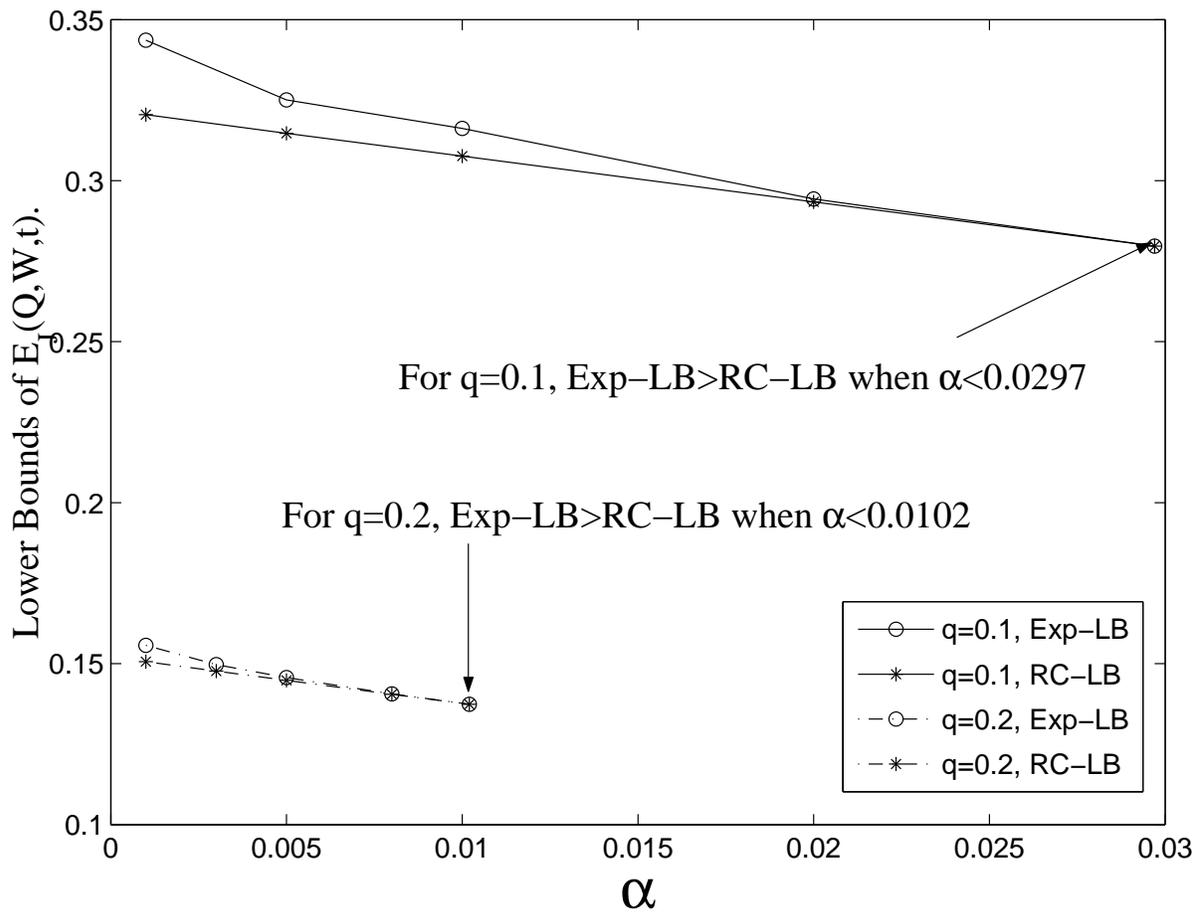}
\caption{Improvement due to the expurgated lower bound for the
binary DMS ($\alpha,q$) and BEC ($\alpha$) system with $t=1$.
Exp-LB and RC-LB stand for the expurgated and random-coding lower
bounds, respectively.}\label{expurgatedlowerbound}
\end{center}
\end{figure}

\begin{example}\label{equidischannel}
{\rm \textbf{(DMS and Equidistant DMC)} A DMC $W= P_{Y|X}$ is
called equidistant if there exists a number $\beta> 0$ such that
for all pairs of inputs $x\neq \widetilde{x}$,
$$
\sum_{y}\sqrt{P_{Y|X}(y|x)P_{Y|X}(y|\widetilde{x})}=\beta.
$$
Note that equidistant DMCs have 0 zero-error capacity, and every
DMC with binary input alphabet is equidistant. It is shown in
\cite{Jelinek} that for an equidistant channel, $E_{x}(\rho, W)$
is achieved in the range $\rho\geq 1$ by a uniform input
distribution $P_{X}(x)=1/|\mathcal{X}|$. Therefore, we can write
$E_{x}(\rho,W)$ as
$$
E_{x}(\rho,W)=-\rho\log\left(\frac{|\mathcal{X}|-1}{|\mathcal{X}|}\beta
^{\frac{1}{\rho}}+\frac{1}{|\mathcal{X}|}\right)\qquad \mbox{for}
\qquad \rho\geq 1.
$$
Now we apply Theorems~\ref{equiv2} and \ref{twolower} to DMS $Q$
and equidistant DMC $W$ with transmission rate $t$. We then see
that if
\begin{equation}
tH(Q^{(1)})+\log\left(\frac{|\mathcal{X}|-1}{|\mathcal{X}|}\beta+\frac{1}{|\mathcal{X}|}\right)\leq
\frac{\beta\log\beta}{\beta+\frac{1}{|\mathcal{X}|-1}},\label{co3}
\end{equation}
the expurgated JSCC lower bound is tighter than the random-coding
lower bound and is given by
\begin{equation}\label{expurgatedbound}
E_{J}(Q,W,t)\geq
-\underline{\rho}_{x}\log\left(\frac{|\mathcal{X}|-1}{|\mathcal{X}|}\beta
^{\frac{1}{\underline{\rho}_{x}}}+\frac{1}{|\mathcal{X}|}\right)-t(1+\underline{\rho}_{x})\log\sum_{s\in\mathcal{S}}Q^{\frac{1}{1+\underline{\rho}_{x}}}(s),
\end{equation}
where $\underline{\rho}_{x}$ is the unique root of the equation
$$
tH(Q^{(\rho)})+\log\left(\frac{|\mathcal{X}|-1}{|\mathcal{X}|}
\beta^{\frac{1}{\rho}}+\frac{1}{|\mathcal{X}|}\right)=
\frac{\rho^{-1}\beta^{\frac{1}{\rho}}\log\beta}{\beta^{\frac{1}{\rho}}
+\frac{1}{|\mathcal{X}|-1}}.
$$

Consider a communication system with a binary source with
distribution \{$q,1-q$\}, a binary erasure channel (BEC) with
erasure probability $\alpha$ and transmission rate $t=1$ (similar
results hold for other cases, as in the last example). Using the
conditions (\ref{co1}), (\ref{co2}) in Example \ref{symmetric},
and together with (\ref{co3}), we present in
Fig.~\ref{expurgatedBEC} the set of ($\alpha,q$) points,
partitioned into four regions. If the pair $(\alpha,q)$ is located
in region $\mathbf{B}$, then the system $E_{J}$ is positive and
exactly known. If $(\alpha,q) \in \mathbf{C}=\mathbf{C}_{1}\cup
\mathbf{C}_{2}$, then upper and lower bounds for $E_{J}$ are
known. Here, region $\mathbf{C}_2$ consists of the values of
$(\alpha,q)$ for which the source-channel expurgated lower bound
given in (\ref{expurgatedbound}) is tighter than the
source-channel random-coding lower bound. Finally, when
$(\alpha,q)\in \mathbf{A}$, $E_J(Q,W,t)=0$. In
Fig.~\ref{expurgatedlowerbound}, we plot the random-coding and
expurgated lower bounds for different source and BEC pairs. We
observe that when the source distribution is $Q$=\{0.1,0.9\}
(respectively $Q$=\{0.2,0.8\}), the expurgated lower bound for
$E_{J}$ is tighter than the random-coding lower bound if
$\alpha<0.0297$ (respectively if $\alpha<0.0102$). }
\end{example}


\section{When is JSCC Worthwhile: JSCC vs Tandem Coding Exponents}\label{comparison}

\subsection{Tandem Coding Error Exponent}
A tandem code $(f^*_n,\varphi^*_{n})\triangleq(f_{cn}\circ
f_{sn},\varphi_{sn}\circ\varphi_{cn})$ for a DMS
$\{Q:\mathcal{S}\}$ and a DMC $\{W:\mathcal{X}\rightarrow
\mathcal{Y}\}$ with blocklength $n$ and transmission rate $t$
(source symbols/channel use) is composed independently by a
$(tn,M)$ block source code $(f_{sn},\varphi_{sn})$ defined by
$f_{sn}:{\mathcal{S}}^{tn}\longrightarrow \{1,2,...,M\}$ and $
\varphi_{sn}:\{1,2,...,M\}\longrightarrow {\mathcal{S}}^{tn}$ with
source code rate
$$
R_s\triangleq \frac{\log M}{{tn}}\qquad\mbox{source code
bits/source symbol},
$$
and an $(n,M)$ block channel code $(f_{cn},\varphi_{cn})$ defined
by $f_{cn}:\{1,2,...,M\}\longrightarrow {\mathcal{X}}^n$ and $
\varphi_{cn}:{\mathcal{Y}}^n\longrightarrow \{1,2,...,M\}$ with
channel code rate
$$
R_c\triangleq \frac{\log M}{n}\qquad\mbox{source code bits/channel
use},
$$
where ``$\circ$'' means composition and $R_s$ and $R_c$ are
independent of $n$. That is, blocks $s^{tn}$ of source symbols of
length $tn$ are encoded as integers (indices) $f_{sn}(s^{tn})$
from $\{1,2,...,M\}$, and these integers are further encoded as
blocks $x^{n}=f_{cn}\left[ f_{sn}(s^{tn})\right]$ of symbols from
$\cal{X}$ of length $n$, transmitted, received as blocks $y^{n}$
of symbols from ${\cal Y}$ of length $n$.  These received blocks
$y^{n}$ are decoded as integers $\varphi_{cn} (y^{n})$ from
$\{1,2,...,M\}$, and finally, these integers are decoded as blocks
of source symbols $\varphi^*_n
(y^{n})=\varphi_{sn}\left[\varphi_{cn} (y^{n})\right]$ of length
$tn$. Thus, the probability of erroneously decoding the block is
$$
P^{(n)}_{e^*}(Q,W,t)\triangleq
\sum_{\{(s^{tn},y^n):\varphi_{sn}\left[\varphi_{cn}
(y^{n})\right]\neq s^{tn}\}}Q_{tn}(s^{tn})P_{n,Y\mid X}\left(y^n
\left| f_{cn}\left[ f_{sn}(s^{tn})\right]\right)\right.,
$$
where $Q_{tn}$ and $P_{n,Y|X}$ are the $tn$- and $n$-dimensional
product distributions corresponding to $Q$ and $P_{Y|X}$.
respectively.
\begin{definition}
{\rm The tandem coding error exponent $E_{T}(Q,W,t)$ is defined as
the largest number $\widehat{E}$ for which there exists a sequence
of tandem codes $(f^*_n,\varphi^*_{n})=(f_{cn}\circ
f_{sn},\varphi_{sn}\circ\varphi_{cn})$ with transmission rate $t$
and block length $n$ such that
$$
\widehat{E}\leq \liminf_{n\rightarrow \infty}-\frac{1}{n}\log
P^{(n)}_{e^*}(Q,W,t).
$$}
\end{definition}
When there is no possibility of confusion, $E_T (Q,W,t)$ will
often be written as $E_T$. In general, we know that $E_{J}\geq
E_{T}$ since by definition tandem coding is a special case of
JSCC. We are hence interested in determining the conditions for
which $E_{J}> E_{T}$ for the same transmission rate $t$.
Meanwhile, it immediately follows (from the JSCC theorem) that
$E_{T}$ can be positive if and only if $tH(Q)<C$; otherwise, both
$E_J$ and $E_{T}$ are zero.

By definition, the tandem coding exponent results from separately
performing and concatenating optimal source and channel coding,
which can be expressed by (e.g., see \cite{Csiszar1})
\begin{eqnarray}
E_{T}(Q,W,t) &=& \sup_{R_s,R_c:R_c=tR_s} \min
\left\{te(R_s,Q),E(R_c,W)\right\}\nonumber\\
&=&
\sup_{R}\min\left\{te\left(\frac{R}{t},Q\right),E(R,W)\right\},\label{Tandem}
\end{eqnarray}
where $e(R,Q)$ and $E(R,W)$ are the source and channel error
exponents, respectively. Note that
$$
\sup_{R\leq t\log|\mathcal{S}|}te\left(\frac{R}{t},Q\right)=
te(\log|\mathcal{S}|,Q) =-t\log(|\mathcal{S}|\overline{Q(s)}),
$$
where $\overline{Q(s)}$ is the geometric mean of the source
probabilities, i.e.
$\overline{Q(s)}\triangleq\left(\prod_{s\in\mathcal{S}}Q(s)\right)^{1/|\mathcal{S}|}\leq
{1}/{|\mathcal{S}|}$. If $-t\log(|\mathcal{S}|\overline{Q(s)})\geq
E(t\log|\mathcal{S}|,W)$, then the graphs of $te(R/t,Q)$ and
$E(R,W)$ must have exactly one intersection $R_{o}$ and by
(\ref{Tandem}) \beq{int}
E_{T}(Q,W,t)=te\left(\frac{R_{o}}{t},Q\right)=E(R_{o},W),
\end{equation}
since $te(R/t,Q)$ is strictly increasing in
$R\in[tH(Q),t\log|\mathcal{S}|]$ and $E(R,W)$ is non-increasing in
$R$. If $-t\log(|\mathcal{S}|\overline{Q(s)})<
E(t\log|\mathcal{S}|,W)$, then there is no intersection between
$te(R/t,Q)$ and $E(R,W)$. Recall (\ref{eRQ1}) that $te(R/t,Q)$ is
infinite in the open interval $(t\log|\mathcal{S}|,\infty)$. In
this case, we have that \beq{noint}
E_{T}(Q,W,t)=E(t\log|\mathcal{S}|,W)
\end{equation}
by (\ref{Tandem}). Without loss of generality, we denote
\begin{eqnarray}
            R_{o}\triangleq\left\{\begin{array}{l}
            \mbox{the rate satisfying}\qquad
            te(\frac{R_{o}}{t},Q)=E(R_{o},W)\\
            \qquad\qquad\mbox{if $-t\log(|\mathcal{S}|\overline{Q(s)})\geq
E(t\log|\mathcal{S}|,W)$},\\
            t\log|\mathcal{S}|\\
            \qquad\qquad\mbox{if
            $-t\log(|\mathcal{S}|\overline{Q(s)})<
E(t\log|\mathcal{S}|,W)$},
            \end{array}\right.\label{defofRo}
\end{eqnarray}
so that we can always write that $E_{T}(Q,W,t)=E(R_{o},W)$.

When the DMS is uniform, the optimal source coding operation
reduces to the trivial enumerating (identity) function with
$M=|S|^{tn}$ as the source is incompressible. Hence only channel
coding is performed in both JSCC and tandem coding and
$E_J(Q,W,t)=E_T(Q,W,t)=E(t\log|\mathcal{S}|,W)$. Thus, our
comparison of the two exponents is nontrivial only if the source
is nonuniform and $tH(Q)<C$. Even though we know that $E_{J}$ is
never worse than $E_{T}$, the following theorem gives a limit on
how much $E_{J}$ can outperform $E_{T}$.
\begin{theorem}\label{generalbound}
{\rm JSCC exponent can at most be equal to double the tandem
coding exponent, i.e.,
$$
E_J(Q,W,t)\leq 2E_T(Q,W,t),
$$
with equality if $tR_{cr}^{(s)}(Q)\geq R_{cr}(W)$ and $T_{sp}
(\overline{\rho}^* ,W )=tE_s (\overline{\rho}^*
,Q)+2tD(Q^{(\overline{\rho}^*)}\parallel Q)$.}
\end{theorem}

\begin{remark}
{\rm Equivalently, this upper bound also implies that $E_J$ can at
most exceed $E_T$ by $E_J/2$, i.e., \beq{defupper}
E_J(Q,W,t)-E_T(Q,W,t)\leq \frac{1}{2}E_J(Q,W,t).\end{equation}}
\end{remark}
\textbf{Proof}: We first refer to the upper bound of $E_J(Q,W,t)$
given by Csisz\'{a}r \cite[Lemma 2]{Csiszar1} \beq{tupper}
E_J(Q,W,t)\leq \min_{tH(Q)\leq R\leq t\log|\mathcal{S}|}\left[t
e\left(\frac{R}{t},Q\right)+ E(R,W)\right],
\end{equation}
where $te(R/t,W)$ is the source error exponent, which is strictly
convex and increasing in $[tH(Q),t\log|\mathcal{S}|]$, and
$E(R,W)$ is the channel error exponent, which is a positive and
non-increasing in $[0,C)$. Unlike the source exponent, the
behavior of $E(R,W)$ is unknown for $R<R_{cr}(W)$. Let $C_0$ be
the zero-error capacity of the channel $W$, i.e., $E(R,W)=\infty$
if and only if $R<C_0$ \cite{Gallager}. If $C_0>
t\log|\mathcal{S}|$, obviously, we have
$$
E_J(Q,W,t)=E_T(Q,W,t)=+\infty.
$$
If $C_0\leq t\log|\mathcal{S}|$, the upper bound in (\ref{tupper})
is finite and the minimum must be achieved by some rate, say
$R_m$, in the interval $[C_0,t\log|\mathcal{S}|]$. Then
\begin{eqnarray} E_J(Q,W,t)&\stackrel{(a)}{\leq} &t
e\left(\frac{R_m}{t},Q\right)+
E(R_m,W)\nonumber\\
&\stackrel{(b)}{\leq} &t e\left(\frac{R_o}{t},Q\right)+
E(R_o,W)\nonumber\\
&\stackrel{(c)}{\leq} & 2E(R_o,W)\nonumber\\
&=& 2E_T(Q,W,t).\nonumber
\end{eqnarray}
Here, the equality in (a) holds if our computable upper and lower
bounds, $\overline{E}_{sp}(Q,W,t)$ and $\underline{E}_{r}(Q,W,t)$,
are equal. To ensure this, we need the condition
$tR_{cr}^{(s)}(Q)\geq R_{cr}(W)$ by Theorem~\ref{tightcondition}.
The equality in (b) holds if $R_m=R_o$ by definition of $R_m$. The
equality (c) holds if and only if there is an intersection between
$te(R/t,W)$ and $E(R,W)$, i.e., $t e(R_o/t,Q)=E(R_o,W)$. Now
taking these considerations together, and applying Theorem
\ref{tightcondition} again, we conclude that $E_J=2E_T$ if
$tR_{cr}^{(s)}(Q)\geq R_{cr}(W)$ and $T_{sp} (\overline{\rho}^* ,W
)-tE_s (\overline{\rho}^*
,Q)=2te(\overline{R}_m/t,Q)=2tD(Q^{(\overline{\rho}^*)}\parallel
Q)$. \qed

\begin{observation}
{\rm The condition for the equality states that, if the minimum in
the expression of $\underline{E}_{r}(Q,W,t)$ given in
(\ref{defEr}) is attained at the intersection of
$te(\frac{R}{t},W)$ and $E_{r}(R,W)$ which is no less than the
critical rate of the channel, then the JSCC exponent is
\textit{twice} as large as the tandem coding exponent. In that
case, the rate of decay of the error probability for the JSCC
system is \textit{double} that for the tandem coding system. In
other words, for the same probability of error $P_e$, the delay of
(optimal) JSCC is approximately \textit{half} of the delay of
(optimal) tandem coding,
$$
P_e\approx 2^{-nE_T(Q,W,t)}=2^{-\frac{n}{2}E_J(Q,W,t)}
\qquad\mbox{for $n$ sufficiently large}.
$$
}
\end{observation}

\subsection{Sufficient Conditions for which \boldmath$E_J>E_T$}

In the following we will use our previous results to derive
computable sufficient conditions for which $E_J>E_T$. We first
define
\begin{eqnarray}
            \gamma\triangleq\left\{\begin{array}{ll}
            \mbox{the root of}\qquad
            tH(Q^{(\gamma)})=R_{cr}(W)&\mbox{if $tH(Q)\leq R_{cr}(W)\leq t\log|\mathcal{S}|$},\\
            0&\mbox{if $tH(Q)> R_{cr}(W)$}.
            \end{array}\right. \label{defgam}
\end{eqnarray}
such that the source error exponent $te(R/t,Q)$ has a parametric
expression at $R_{cr}(W)$ \beq{pa}
te\left(\frac{R_{cr}(W)}{t},Q\right)=tD(Q^{(\gamma)}\parallel Q).
\end{equation}
Note that $\gamma$ is well defined only if $R_{cr}(W)\leq
t\log|\mathcal{S}|$. Denote
\begin{equation}
T(\overline{\rho}^*)\triangleq T_{sp} (\overline{\rho}^* ,W )
-tE_s (\overline{\rho}^* ,Q).
\end{equation}

\begin{theorem}\label{TandemTheorem1}
{\rm Let $R_{cr}(W)\leq t\log|\mathcal{S}|$. If
\begin{equation}
\max\left\{tR_{cr}^{(s)}(Q),E_{o}(1,W)-tD(Q^{(\gamma)}\parallel
Q)\right\}\geq R_{cr}(W),\label{maincondition}
\end{equation}
then
$$
E_{J}(Q,W,t)>E_{T}(Q,W,t).
$$
More precisely, we have the
following bounds. \\
\noindent(a) If
$\min\left\{tR_{cr}^{(s)}(Q),E_{o}(1,W)-tD(Q^{(\gamma)}\parallel
Q)\right\}\geq R_{cr}(W)$, then \beq{deffbou1}
E_{J}(Q,W,t)-E_{T}(Q,W,t)\geq
\frac{1}{2}T(\overline{\rho}^*)-\left|\frac{1}{2}T(\overline{\rho}^*)-tD(Q^{(\overline{\rho}^*)}\parallel
Q)\right|\geq 0,
\end{equation}
where the two equalities in (\ref{deffbou1}) cannot hold
simultaneously.

\noindent(b) If $tR_{cr}^{(s)}(Q)\geq R_{cr}(W)>
E_{o}(1,W)-tD(Q^{(\gamma)}\parallel Q)$, then \beq{deffbou2}
E_{J}(Q,W,t)-E_{T}(Q,W,t)>
T(\overline{\rho}^*)-tD(Q^{(\gamma)}\parallel Q)\geq 0.
\end{equation}
\noindent(c) If $E_{o}(1,W)-tD(Q^{(\gamma)}\parallel Q)\geq
R_{cr}(W)> tR_{cr}^{(s)}(Q)$, then \beq{deffbou3}
E_J(Q,W,t)-E_T(Q,W,t)\geq R_{cr}(W)-tE_s(1,Q)>0.
\end{equation}}
\end{theorem}
\textbf{Proof}: We shall show that, in each of the three cases, (a), (b), and (c), we have $E_J>E_T$. \\

\noindent(a). Assume $tR_{cr}^{(s)}(Q)\geq R_{cr}(W)$ and
$E_{o}(1,W)-tD(Q^{(\gamma)}\parallel Q)\geq R_{cr}(W)$. By
definition of $\gamma$, we have $tD(Q^{(\gamma)}\parallel
Q)=te(R_{cr}(W)/t,Q)$, see (\ref{eRQ1}) and (\ref{pa}). Thus, the
latter condition is equivalent to $E(R_{cr}(W),W)\geq
te(R_{cr}(W)/t,Q)$ and by (\ref{straighteq}) and the related
discussion it guarantees that $R_o\geq R_{cr}(W)$, where $R_o$ is
defined in (\ref{defofRo}). According to
Theorem~\ref{tightcondition}, when $tR_{cr}^{(s)}(Q)\geq
R_{cr}(W)$, $\overline{E}_{sp}(Q,W,t)$ is attained by
$\overline{R}_{m}\geq R_{cr}(W)$ and $E_J$ is determined by
$$
E_J(Q,W,t)=te\left(\frac{\overline{R}_{m}}{t},Q\right)+E_{sp}(\overline{R}_{m},W).
$$
Since $R_o\geq R_{cr}(W)$, $E_T$ is determined by $E_{sp}(R_o,W)$.
If $R_o\neq\overline{R}_{m}$, we must have
$$
E_T(Q,W,t)<
\max\left\{te\left(\frac{\overline{R}_{m}}{t},Q\right),E_{sp}(\overline{R}_{m},W)\right\},
$$
because $te(R/t,Q)$ is strictly increasing and $E_{sp}(R,W)$ is
strictly decreasing at $\overline{R}_{m}$. Thus, \beq{bt1}
E_{J}(Q,W,t)-E_{T}(Q,W,t)> \min
\left\{te\left(\frac{\overline{R}_{m}}{t},Q\right),E_{r}(\overline{R}_{m},W)\right\}\geq
0,
\end{equation}
where equality holds if $\overline{R}_{m}=C$. If
$R_o=\overline{R}_{m}$, then immediately, \beq{bt2}
E_J(Q,W,t)-E_T(Q,W,t)=te\left(\frac{\overline{R}_{m}}{t},Q\right)=tD(Q^{(\overline{\rho}^*)}\parallel
Q),
\end{equation}
where the above is positive since $\overline{\rho}^*>0$ by Lemma
\ref{prop} (1). Note also that in this case
$te(\overline{R}_{m}/t,Q)= E_{r}(\overline{R}_{m},W)$, so
(\ref{bt1}) and (\ref{bt2}) can be
summarized by (\ref{deffbou1}).\\

\noindent(b). In this case, we have $\overline{R}_{m}\geq
R_{cr}(W)>R_o$. We can upper bound $E_T$ by
$$
E_{T}(Q,W,t)=te\left(\frac{R_{o}}{t},Q\right)<te\left(\frac{R_{cr}(W)}{t},Q\right)=tD(Q^{(\gamma)}\parallel
Q)
$$
and hence
$$
E_{J}(Q,W,t)-E_{T}(Q,W,t)> T_{sp} (\overline{\rho}^* ,W ) -tE_s
(\overline{\rho}^* ,Q)-tD(Q^{(\gamma)}\parallel Q).
$$
The above lower bound must be nonnegative since
\begin{eqnarray}
T_{sp} (\overline{\rho}^* ,W ) -tE_s (\overline{\rho}^*
,Q)-tD(Q^{(\gamma)}\parallel Q)&=&
E_{r}(\overline{R}_m,W)+t\left[e\left(\frac{\overline{R}_m}{t},Q\right)-
e\left(\frac{R_{cr}(W)}{t},Q\right)\right]\nonumber\\
&\geq &E_{r}(\overline{R}_m,W)\nonumber\\
&\geq &0\nonumber,
\end{eqnarray}
and it is equal to 0 if $R_{cr}(W)=\overline{R}_m=C$.\\

\noindent(c). In this case, we have $R_o\geq R_{cr}(W)>
\underline{R}_{m}$ and from (\ref{Csissimple}) $E_J$ is bounded by
$$
E_J(Q,W,t)\geq E_0(1,W)-tE_s(1,Q).
$$
On the other hand, by the monotonicity of $E_r(R,W)$, we can upper
bound $E_T$ by
$$
E_T(Q,W,t)=E_r(R_o,W)\leq E_r(R_{cr}(W),W)=E_0(1,W)-R_{cr}(W).
$$
Thus we obtain
$$
E_J(Q,W,t)-E_T(Q,W,t)\geq R_{cr}(W)-tE_s(1,Q).
$$
The above is positive since
\begin{eqnarray}
E_0(1,W)-tE_s(1,Q)&= &
te\left(\frac{\underline{R}_m}{t},Q\right)+E_r(\underline{R}_m,W)\nonumber\\
&>& E_r(\underline{R}_m,W)\nonumber\\
&>& E_r(R_{cr}(W),W)\nonumber\\
&=& E_0(1,W)-R_{cr}(W),\nonumber
\end{eqnarray}
where the first inequality follows from the fact that
$\underline{R}_m>tH(Q)$ by Lemma \ref{prop} and
Corollary~\ref{comp}.

\qed

As pointed out in the proof, the condition $tR_{cr}^{(s)}(Q)\geq
R_{cr}(W)$ means that the JSCC exponent $E_J$ is achieved at a
rate no less than $R_{cr}(W)$. The second condition,
$E_{o}(1,W)-tD(Q^{(\gamma)}\parallel Q)\geq R_{cr}(W)$ means that
the tandem coding exponent $E_T$ is achieved at a rate no less
than $R_{cr}(W)$. Hence (\ref{maincondition}) in
Theorem~\ref{TandemTheorem1} states that $E_J$ would be strictly
larger than $E_T$ if either $E_J$ or $E_T$ is determined exactly.
Conversely, if the conditions in Theorem \ref{TandemTheorem1} are
not satisfied, then neither $E_{J}$ nor $E_{T}$ are exactly known.
Nevertheless, if the lower bound of $E_{J}$ is strictly larger
than the upper bound of $E_{T}$, then we must have $E_{J}>E_{T}$.
Hence we obtain the following sufficient conditions.

\begin{theorem}\label{TandemTheorem2}
{\rm Let $E_{ex}(0,W)<\infty$ and let $t\log|\mathcal{S}|\geq
R_{cr}(W)$, where $E_{ex}(R,W)$ is the expurgated channel error
exponent \cite{Gallager}. If
$$
E_0(1,W)-tE_s(1,Q)\geq E_{R_{l}}\triangleq
\frac{k_{1}k_{2}t\log|\mathcal{S}|+
k_2t\log(|\mathcal{S}|\overline{Q(s)})+k_{1}E_{ex}(0,W)}{k_{1}-k_{2}},
$$
where
$$
k_{1}=\frac{D\left(Q^{(1)}\parallel
Q\right)+\log(|\mathcal{S}|\overline{Q(s)})}{H\left(Q^{(1)}\right)
-\log|\mathcal{S}|} \qquad\mbox{and}\qquad
k_{2}=\frac{E_{0}(1,W)-E_{ex}(0,W)}{R_{cr}(W)}-1,
$$
then $E_{J}(Q,W,t)>E_{T}(Q,W,t)$. }
\end{theorem}

\begin{theorem}\label{TandemTheorem3}
{\rm Let $t\log|\mathcal{S}|\geq R_{cr}(W)$. If
$E_0(1,W)-tE_s(1,Q)\geq tD\left(Q^{(\gamma)}\parallel Q\right)$,
where $\gamma$ is defined in (\ref{defgam}), then
$E_{J}(Q,W,t)>E_{T}(Q,W,t)$.}
\end{theorem}

In Theorems \ref{TandemTheorem2} and \ref{TandemTheorem3}, we
establish the sufficient conditions by comparing the
source-channel random-coding bound derived in Theorem
\ref{tightcondition}, with the upper bound of tandem coding
exponent obtained by using the geometric characteristics of
$e(R,W)$ and $E(R,W)$. The proofs of Theorems \ref{TandemTheorem2}
and \ref{TandemTheorem3} are given in Appendices
\ref{pftandemtheorem2} and \ref{pftandemtheorem3}, respectively.
These conditions can be readily computed since it only requires
the knowledge of $R_{cr}(W)$ and $E_{ex}(0,W)$. Note that the
condition $E_{ex}(0,W)<\infty$ in Theorem~\ref{TandemTheorem2} is
satisfied by the DMCs with zero-error capacity equal to $0$, see
\cite[p.~187]{Csiszar3}. Thus, Theorem~\ref{TandemTheorem2}
applies to equidistant channels, in particular, to every channel
with binary input alphabet. An expression of $E_{ex}(0,W)$ for the
DMC with $0$ zero-error capacity is given in
\cite[Problem~5.24]{Gallager}.

\begin{figure}[c!]
\begin{center}
\includegraphics[width=16cm]{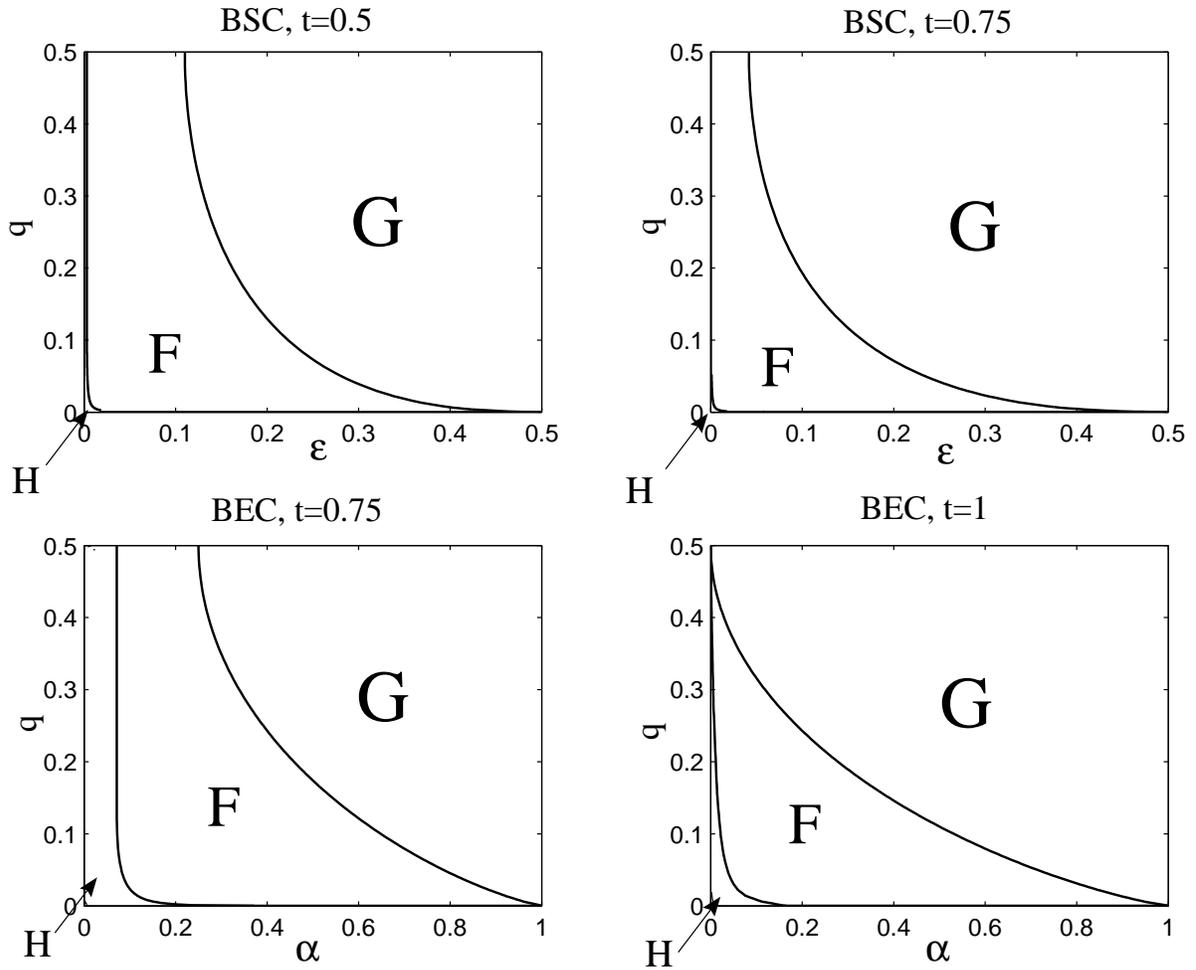}
\caption{The regions for binary DMS-BSC ($q,\varepsilon$) pairs
and binary DMS-BEC ($q,\alpha$) pairs under different transmission
rates $t$. In region $\mathbf{F}$ (including the boundary between
$\mathbf{F}$ and $\mathbf{H}$), $E_J>E_T>0$; in region
$\mathbf{G}$ (including the boundary between $\mathbf{G}$ and
$\mathbf{F}$), $E_J=E_T=0$; and in region $\mathbf{H}$, $E_J\geq
E_T>0$.}\label{ETEJregion}
\end{center}
\end{figure}

\begin{example}\label{exwhen}
{\rm \textbf{(When Does the JSCC Exponent Outperform the Tandem
Coding Exponent?)} We apply Theorems~\ref{TandemTheorem1},
\ref{TandemTheorem2} and \ref{TandemTheorem3} to the binary DMS
with distribution $\{q,1-q\}$ and BSC with crossover probability
$\varepsilon$, and the binary DMS $\{q,1-q\}$ and the binary
erasure channel (BEC) with erasure
probability $\alpha$, under different transmission rates $t$. If
any one of the conditions in these theorems holds, then
$E_{J}>E_{T}$. The above conditions are summarized by Region
$\mathbf{F}$ in Fig.~\ref{ETEJregion}. Indeed, Region $\mathbf{F}$
shows that $E_{J}>E_{T}$ for a wide range of $(\varepsilon,q)$ or
$(\alpha,q)$ pairs. Region $\mathbf{G}$ consists of the pairs
$(\varepsilon,q)$ or $(\alpha,q)$ such that $tH(Q) \ge C$; in this
case, $E_{J}=E_{T}=0$. Finally, when $(\varepsilon,q)$ or
$(\alpha, q)$ falls in Region $\mathbf{H}$, we are not sure
whether $E_{J}$ is still strictly larger than $E_{T}$.}
\end{example}

\begin{example}\label{lbr}
{\rm \textbf{(By How Much Can the JSCC Exponent Be Larger Than the
Tandem Coding Exponent?)} In the last example we have seen that
$E_J>E_T$ holds for a wide large class of source-channel pairs.
Now we evaluate the performance of $E_J$ over $E_T$ by looking at
the ratio of the two quantities. Recall that when
Theorem~\ref{TandemTheorem1} (a) is satisfied, both $E_J$ and
$E_T$ are exactly determined. In this case we can directly compute
$E_J$ (using the results of Section \ref{bounds}) and $E_T$ (using
(\ref{int}) and (\ref{noint})). When $E_J$ (respectively, $E_T$)
is not known, i.e., when $tR_{cr}^{(s)}(Q)<R_{cr}(W)$
(respectively, $E_{o}(1,W)-tD(Q^{(\gamma)}\parallel Q)<R_{cr}(W)$),
we can calculate the lower bound of $E_J$ (respectively, the
upper bound of $E_T$) instead and thus obtain a
lower bound for $E_J/E_T$. For general DMCs, we lower bound $E_J$
by its random-coding lower bound $\underline{E}_{r}(Q,W,t)$. For
equidistant DMCs, particularly for binary DMCs, when
$tR_{cr}^{(s)}(Q)< R_{ex}(W)$, we use the expurgated lower bound
$\underline{E}_{ex}(Q,W,t)$; when $tR_{cr}^{(s)}(Q)\geq
R_{ex}(W)$, we use the random-coding lower bound
$\underline{E}_{r}(Q,W,t)$. To calculate the upper bound of $E_T$,
when $ E_{o}(1,W)-tD(Q^{(\gamma)}\parallel Q)<R_{cr}(W)\leq
R_{cr}^{(s)}(Q)$, or equivalently when $R_o<R_{cr}(W)\leq
\overline{R}_m$, we can bound $E_T$ by
$$
E_{T}(Q,W,t)\leq\min\left\{tD\left(Q^{(\gamma)}\parallel Q\right),
E_{sp}(R_s,W)\right\},
$$
where $R_s$ is the intersection of $E_{sp}(R,W)$ and $te(R/t,Q)$
if any; otherwise $R_s=t\log|S|$. When
$E_{o}(1,W)-tD(Q^{(\gamma)}\parallel Q)<R_{cr}(W)$ and
$R_{cr}^{(s)}(Q)<R_{cr}(W)$, we bound $E_T$ by
$$
E_{T}(Q,W,t)\leq E_{sp}(R_s,W).
$$
Table~\ref{ratiobinaryBSC} exhibits $E_J/E_T$ (or its lower bound,
which must be no less than 1) for the binary DMS $\{q,1-q\}$ and
BSC ($\varepsilon$) system under transmission rates $t=0.5$,
$0.75$ and $1$. It is seen that the ratio $E_J/E_T$ can be very
close to 2 (its upper bound) for many $(q,\varepsilon)$ pairs. For
other systems, we have similar results: $E_J$ substantially
outperforms $E_T$. For instance, for binary DMS $\{q, 1-q\}$ and
BEC ($\alpha$) with $t=1$, we note that $E_J/E_T\geq 1.4$ for a
wide range of ($q,\alpha$)'s; for ternary DMS and BSC or for DMS
and ternary symmetric channel, if transmission rate $t$ is chosen
suitably (such that $tH(Q)<C$), we obtain that $E_J/E_T\geq 1.5$
for many source-channel pairs.}
\end{example}

\subsection{Power Gain Due to JSCC for DMS over Binary-input AWGN and Rayleigh-Fading Channels with Finite
Output Quantization}

\begin{figure}[c!]
\begin{center}
\includegraphics[width=16cm]{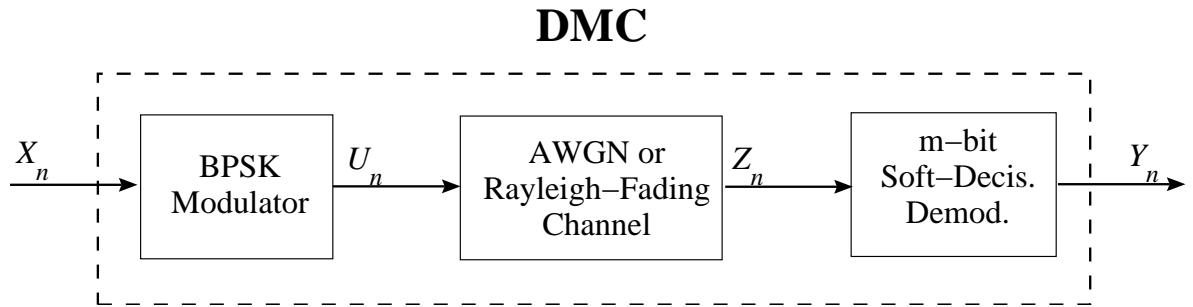}
\caption{Binary-input AWGN or Rayleigh-fading channel with finite
output quantization.}\label{AWGNRF}
\end{center}
\end{figure}

It is well known that $M$-ary modulated additive white Gaussian
noise (AWGN) and memoryless Rayleigh-fading channels can be
converted to a DMC when finite quantization is applied at their
output. For example, as illustrated in \cite{COVQ}, \cite{Nam}, we
know that the concatenation of a binary phase-shift keying (BPSK)
modulated AWGN or Rayleigh-fading channel with $m$-bit
soft-decision demodulation is equivalent to a binary-input,
$2^m$-output DMC (cf. Fig. \ref{AWGNRF}). We next study the JSCC
and tandem coding exponent for a system involving such channels to
assess the potential benefits of JSCC over tandem coding in terms
of power or channel signal-to-noise ratio (SNR) gains.

We assume that the BPSK signal $U_n\in \{-1,+1\}$ corresponding to
the signal input $X_n$ is of unit energy, and $V_n$ is a zero-mean
independent and identically distributed (i.i.d.) Gaussian random
process with variance $N_o/2$. The channel SNR is defined by
$\mbox{SNR}\triangleq E[U_n^2]/E[V_n^2]=2/N_o$ and the received
signal is
$$
Z_n=A_nU_n+V_n, \qquad n=1,2,...,
$$
where $A_n$ is 1 for the AWGN channel (no fading), and for the
Rayleigh-fading channel, $\{A_n\}$ is the amplitude fading process
assumed to be i.i.d. with probability density function (pdf)
$$
  f_A(a)=\left\{
            \begin{array}{ll}
              2ae^{-a^2}, & \mbox{if $a>0$},\\
              0, & \mbox{otherwise},
            \end{array}
      \right.
$$
such that $E[A_n^2]=1$. We also assume for the Rayleigh-fading
channel that $A_n$, $U_n$ and $V_n$ are independent of each other,
and the values of $A_n$ are not available at the receiver. At the
receiver, as shown in Fig.~\ref{AWGNRF}, each $Z_n\in \mathbb{R}$
is demodulated via an $m$-bit uniform scalar quantizer with
quantization step $\Delta$ to yield $Y_n\in\{0,1\}^{m}$. If the
channel input alphabet is $\mathcal{X}=\{0,1\}$ and the channel
output alphabet is $\mathcal{Y}=\{0,1,2,...,2^m-1\}$, then the
transition probability matrix $\Pi$ is given by
$$
\Pi=[\pi_{ij}], \qquad i\in\mathcal{X},\qquad j\in\mathcal{Y},
$$
where
$$
\pi_{ij}\triangleq
P(Y=j|X=i)={\mathcal{Q}}\left((T_{j-1}-(2i-1))\sqrt{\mbox{SNR}}\right)-{\mathcal{Q}}\left((T_j-(2i-1))\sqrt{\mbox{SNR}}\right)
$$
for the AWGN channel \cite{Nam}, and
$$
\pi_{ij}\triangleq P(Y=j|X=i)=F_{Z|X}(T_j|i)-F_{Z|X}(T_{j-1}|i)
$$
for the Rayleigh-fading channel \cite{COVQ}. Here
$F_{Z|X}(z|i)=Pr\{Z\leq z|Z=i\}$ is given by \cite{COVQ},
\cite{taricco}
$$
F_{Z|X}(z|1)=1-F_{Z|X}(-z|0)=1-{\mathcal{Q}}\left(\frac{z}{\sqrt{N_o/2}}\right)-\frac{e^{-(z^2/(N_o+1))}}{\sqrt{N_o+1}}\times\left[1-{\mathcal{Q}}\left(\frac{z}{\sqrt{N_o(N_o+1)/2}}\right)\right],
$$
where ${\mathcal{Q}}(x)$ is the complementary error function
$$
{\mathcal{Q}}(x)=\frac{1}{\sqrt{2\pi}}\int_x^\infty
\exp\left\{-t^2/2\right\}dt,
$$
and $\{T_j\}$ are the thresholds of the receiver's soft-decision
quantizer given by
\begin{eqnarray}
  T_j=\left\{
            \begin{array}{ll}
              -\infty, & \mbox{if $j=-1$},\\
              (j+1-2^{m-1})\Delta, & \mbox{if $j=0,1,...,2^m-2$},\\
              +\infty, & \mbox{if $j=2^m-1$}
            \end{array}
      \right.\label{thres}
\end{eqnarray}
with uniform step-size $\Delta$. For each channel SNR, the
suitable quantization step $\Delta$ is chosen as in \cite{Nam},
\cite{COVQ} to yield the maximum capacity of the binary-input
$2^m$-output DMC.

\begin{figure}[c!]
\begin{center}
\includegraphics[width=16cm]{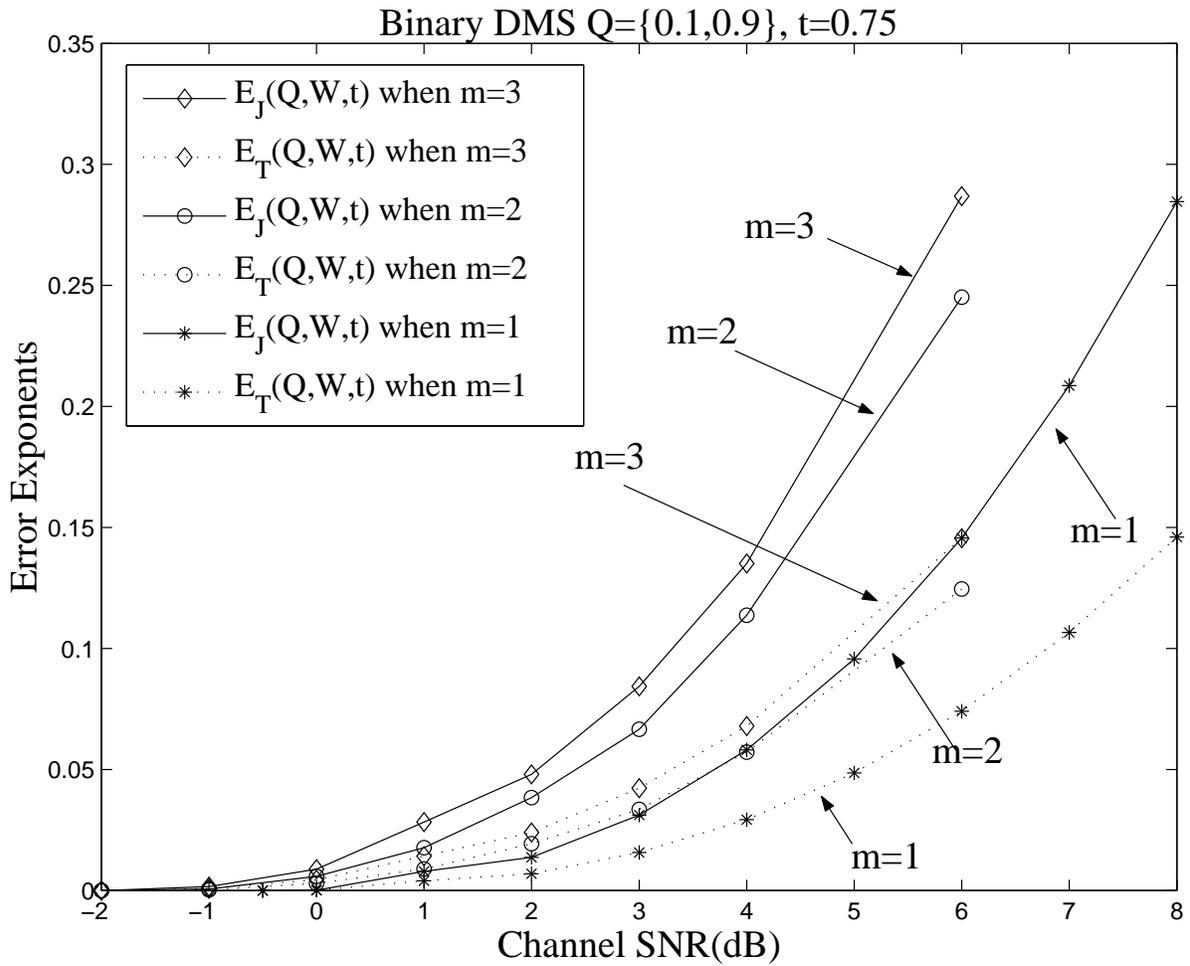}
\caption{The power gain due to JSCC for binary DMS and
binary-input $2^m$-output DMC (AWGN channel) with $t=0.75$.}
\label{exponentforCOVQAWGN075}
\end{center}
\end{figure}

\begin{figure}[c!]
\begin{center}
\includegraphics[width=16cm]{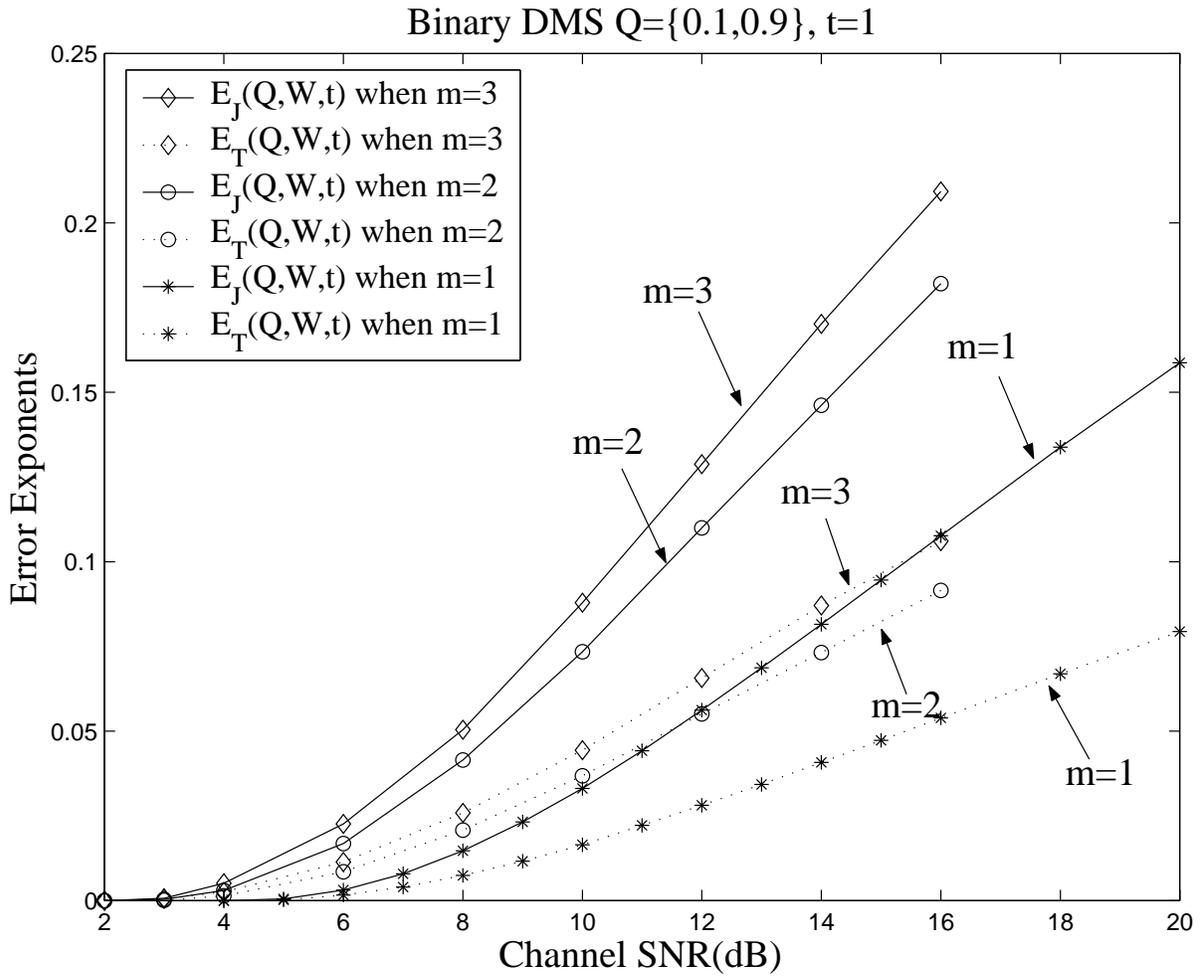}
\caption{The power gain due to JSCC for binary DMS and
binary-input $2^m$-output DMC (Rayleigh-fading channel) with
$t=1$.} \label{exponentforCOVQRF}
\end{center}
\end{figure}

We compute the JSCC and tandem coding exponents for the binary
source and the binary-input $2^m$-output DMC converted from the
AWGN (Rayleigh-fading, respectively) channel under transmission
rate $t=0.75$ ($t=1$, respectively), and illustrate the power gain
due to JSCC. In Figs. \ref{exponentforCOVQAWGN075} and
\ref{exponentforCOVQRF}, we plot $E_J$ and $E_T$ for binary DMS
$Q=\{0.1,0.9\}$ and $m=1,2,3$ by varying the channel SNR (in dB).
We point out that in both the two figures, when $\mbox{SNR}\leq
6$~dB for $m=2,3$ and when $\mbox{SNR}\leq 8$~dB for $m=1$, $E_J$
and $E_T$ are determined exactly. We observe that for the same
SNR, $E_J$ is almost twice as large as $E_T$. Furthermore, for the
same exponent and the same (asymptotic) encoding length, JSCC
would yield the same probability of error as tandem coding with a
power gain of more than 2~dB. A similar behavior was noted for
other values of transmission rate $t$.


\section{JSCC Error Exponent with Hamming Distortion Measure}\label{distortionsec}
Let $\mathcal{S}$ be a finite set and $d(\cdot,\cdot)$ be a
distortion measure, i.e., a nonnegative valued function $d$
defined on $\mathcal{S}\times \mathcal{S}$ and extended to
${\mathcal{S}}^n\times {\mathcal{S}}^n$ by setting
$$
d(s^n,{\widetilde{s}}^n)\triangleq
\frac{1}{n}\sum^n_{i=1}d(s_{i},\widetilde{s}_{i}).
$$

A JSC code with blocklength $n$ and transmission rate $t>0$ for a
$tn$-length DMS $\{Q:\mathcal{S}\}$ and a DMC
$\{W:\mathcal{X}\rightarrow \mathcal{Y}\}$ with a threshold
$\Delta$ of tolerated distortion is a pair of mappings
$f_{n}:{\mathcal{S}}^{tn}\longrightarrow {\mathcal{X}}^n $ and $
\varphi_{n}:{\mathcal{Y}}^n\longrightarrow
 {\mathcal{S}}^{tn}$. The probability of the code exceeding the threshold
$\Delta$ is given by
$$
P^{(n)}_{\Delta}(Q,W,t)\triangleq \sum_{\{(s^{tn},
y^n):d(s^{tn},\varphi_n
(y^n))>\Delta\}}Q_{tn}(s^{tn})P_{n,Y|X}(y^n\mid f_n(s^{tn})),
$$
where $Q_{tn}$ and $P_{n,Y|X}$ are the $tn$- and $n$-dimensional
product distributions corresponding to $Q$ and $P_{Y|X}$
respectively. $P^{(n)}_{\Delta}(Q,W,t)$ is also called the
probability of excess distortion. We remark that for the JSCC with
a distortion threshold, we allow that the source has a uniform
distribution.
\begin{definition}
{\rm The JSCC error exponent $E^{\Delta}_{J}(Q,W,t)$ is defined as
the largest number $E^{\Delta}$ for which there exists a sequence
of JSC codes $(f_n,\varphi_n)$ with blocklength $n$ and
transmission rate $t$ such that
$$
E^{\Delta}\leq\liminf_{n\rightarrow \infty}-\frac{1}{n}\log
P^{(n)}_{\Delta}(Q,W,t).
$$}
\end{definition}
When there is no possibility of confusion, $E^{\Delta}_J (Q,W,t)$
will often be written $E^{\Delta}_J$. In \cite{Csiszar2},
Csisz\'{a}r proved that for a DMS $Q$ and a DMC $W$, the JSCC
error exponent under distortion threshold $\Delta$ satisfies
\beq{cisadisto1} \underline{E}^{\Delta}_{r}(Q,W,t)\leq
E^{\Delta}_{J}(Q,W,t)\leq \overline{E}^{\Delta}_{sp}(Q,W,t),
\end{equation}
where \beq{p4} \underline{E}^{\Delta}_{r}(Q,W,t)\triangleq
\inf_{R> 0
}\left[tF\left(\frac{R}{t},Q,\Delta\right)+E_{r}(R,W)\right]
\end{equation}
and \beq{p5} \overline{E}^{\Delta}_{sp}(Q,W,t)\triangleq \inf_{R>
0 }\left[tF\left(\frac{R}{t},Q,\Delta\right)+E_{sp}(R,W)\right].
\end{equation}
In the above,
\begin{equation}
F(R,Q,\Delta)= \inf_{P: R(P,\Delta)> R}D(P\parallel
Q)\label{sourcedis}
\end{equation}
is the source error exponent with a fidelity criterion \cite{Marton}
and $R(P,\Delta)$ is the rate distortion function (e.g.,
\cite{Thomas}, \cite{Csiszar3}). $E_r(R,W)$ and $E_{sp}(R,W)$ are
the random-coding and sphere-packing bounds to the channel error
exponent. Likewise, if the infimum in (\ref{p4}) or (\ref{p5}) is
attained for a rate larger than the channel critical rate, then
the lower and upper bounds coincide, and we can determine
$E^{\Delta}_J$ exactly. Of course, the two bounds are nontrivial
if and only if $tR(Q,\Delta)<C$ by the JSCC theorem.

It can be shown that $F(R,Q,\Delta)$ is a nondecreasing function
in $R$. However, unlike $e(R,Q)$, $F(R,Q,\Delta)$ is not
necessarily convex or even continuous in $R$ \cite{Ahlswede},
\cite{Marton}. Therefore, it is hard to analytically compute the
JSCC exponent $E^{\Delta}_{J}$ in general. In this section we only
address the computation of $E^{\Delta}_{J}$ for a binary DMS and
an arbitrary DMC under the Hamming distortion measure
$d_{H}(\cdot,\cdot)$, given by \beq{hamming}
  d_{H}(s,\widetilde{s})=\left\{
            \begin{array}{ll}
              1, & \mbox{if $s\neq \widetilde{s}$},\\
              0, & \mbox{if $s=\widetilde{s}$}.
            \end{array}
      \right.
\end{equation}
We first need to derive a parametric form of $F(R,Q,\Delta)$.
Define \beq{Esdel} E^{\Delta}_{s}(\rho,Q)\triangleq
(1+\rho)\log\left(q^{\frac{1}{1+\rho}}+(1-q)^{\frac{1}{1+\rho}}\right)-\rho
 h_{b}(\Delta).
\end{equation}

\begin{lemma} \label{sourceexponentdist}
{\rm For binary DMS $Q\triangleq\{q,1-q\}$ $(q\leq 1/2)$ under the
Hamming distortion measure (\ref{hamming}) and distortion
threshold $\Delta$ such that $\Delta\leq 1/2$, the following hold.
\begin{equation}
  F(R,Q,\Delta)=\left\{
            \begin{array}{ll}
              +\infty, & R>1-h_{b}(\Delta),\\
              \sup_{\rho\geq \rho_0}[\rho R-E^{\Delta}_{s}(\rho,Q)],&R(Q,\Delta)< R\leq
              1-h_{b}(\Delta),\\
              0, & R\leq R(Q,\Delta),
            \end{array}\label{symm}
      \right.
\end{equation}
where the rate-distortion function
$R(Q,\Delta)=h_{b}(q)-h_{b}(\Delta)$ and $\rho_0=0$ if $q\geq
\Delta$; otherwise $R(Q,\Delta)=0$ and $\rho_0$ is the unique root
of equation $H(Q^{(\rho)})=h_{b}(\Delta)$ such that $\rho_0>0$.}
\end{lemma}

The proof of this lemma is given in Appendix \ref{pfsourceexpdis}.
It can be easily verified that $F(R,Q,\Delta)$ is continuous and
convex in $R\in (-\infty, 1-h_{b}(\Delta)]$ if $q\geq \Delta$ and
$F(R,Q,\Delta)$ is continuous and convex in $R\in
(0,1-h_{b}(\Delta)]$ and has a jump at $R=R(Q,\Delta)=0$ if $q<
\Delta$. According to Lemma {\ref{sourceexponentdist}}, the source
error exponent $tF(R/t,Q,\Delta)$ is the convex transform of
$tE^{\Delta}_{s}(\rho,Q)$ in $[\rho_0,+\infty)$. Define the binary
divergence by \beq{tildeD} \widetilde{D}(\Delta\parallel
q)\triangleq
\Delta\log\frac{\Delta}{q}+(1-\Delta)\log\frac{1-\Delta}{1-q}.
\end{equation}
Adopting the approach of Section~\ref{bounds}, we can apply
Fenchel's Duality Theorem to $\underline{E}^{\Delta}_{r}(Q,W,t)$
and $ \overline{E}^{\Delta}_{sp}(Q,W,t)$ and obtain equivalent
computable bounds.

\begin{theorem}\label{Disthe}
{\rm Given a binary DMS $(q\leq 1/2)$ and a DMC $W$ under the
Hamming distortion measure and distortion threshold $\Delta$
$(\Delta\leq 1/2)$, the JSCC exponent satisfies the following.

\smallskip\noindent
1) {\it Lower Bound:} If $0\leq
\Delta<\sqrt{q}/(\sqrt{q}+\sqrt{1-q})$, then $\rho_0< 1$ and
\begin{equation}
\underline{E}^{\Delta}_{r}(Q,W,t)= \max_{\rho_0\leq\rho\leq
1}[T_{r}(\rho,W)-tE^{\Delta}_{s}(\rho,Q)],\label{dist1}
\end{equation}
Otherwise, if $\Delta\geq\sqrt{q}/(\sqrt{q}+\sqrt{1-q})$, then
\begin{equation}
\underline{E}^{\Delta}_{r}(Q,W,t)= t\widetilde{D}(\Delta\parallel
q)+E_{0}(1,W).\label{dist2}
\end{equation}
\smallskip\noindent
2) {\it Upper Bound:}
\begin{equation} \overline{E}^{\Delta}_{sp}(Q,W,t)= \sup_{\rho\geq \rho_0
}[T_{sp}(\rho,W)-tE^{\Delta}_{s}(\rho,Q)].\label{dist3}
\end{equation}}
\end{theorem}

Since the above result is a simple extension of the results in
Section \ref{bounds}, the proof is omitted and we hereby only
provide the following remarks.
\begin{enumerate}
\item Similar to the lossless case, if $t(h_{b}(q)-h_{b}(\Delta))\geq C$, then
$\underline{E}^{\Delta}_{r}(Q,W,t)=\overline{E}^{\Delta}_{sp}(Q,W,t)=0$.
If $R_{\infty}(W)> t(1-h_{b}(\Delta))$, then
$\overline{E}^{\Delta}_{sp}(Q,W,t)=+\infty$.

\item Note that when
$ \Delta\geq\sqrt{q}/(\sqrt{q}+\sqrt{1-q})$,
$\underline{E}^{\Delta}_{r}(Q,W,t)$ in (\ref{p4}) is achieved at
$R\downarrow 0^+$, and
\begin{eqnarray}
\underline{E}^{\Delta}_{r}(Q,W,t)&=&\lim_{R\downarrow0^+}\left[tF(\frac{R}{t},Q,\Delta)+E_{r}(R,W)\right]\nonumber\\
&=&\lim_{R\downarrow0^+}\left[t\inf_{P: R(P,\Delta)>
\frac{R}{t}}D(P\parallel
Q)+E_{0}(1,W)-R\right]\nonumber\\
&=& t\widetilde{D}(\Delta\parallel q)+E_{0}(1,W).\nonumber
\end{eqnarray}

\item In the special case where the binary source
is uniform, i.e., $q=1/2$, Theorem~\ref{Disthe} reduces to
$$
\max_{0\leq \rho\leq 1}\left[-\rho
t(1-h_b(\Delta))+T_{r}(\rho,W)\right]\leq E_J^{\Delta}(Q,W,t)\leq
\sup_{\rho\geq 0}\left[-\rho
t(1-h_b(\Delta))+T_{sp}(\rho,W)\right].
$$
This is clearly equivalent to
\begin{equation}
E_r\left(t(1-h_b(\Delta)),W\right)\leq E_J^{\Delta}(Q,W,t)\leq
E_{sp}\left(t(1-h_b(\Delta)),W\right)\label{distspecial1}
\end{equation}
by the definition of $T_{r}(\rho,W)$ and $T_{sp}(\rho,W)$. In other words,
$E_J^{\Delta}$ is bounded by the channel random-coding and
sphere-packing bounds at rate $t(1-h_b(\Delta))$.  If
$t(1-h_b(\Delta))\geq R_{cr}(W)$, then $E_J^{\Delta}$ is exactly
determined.

\item When the source is nonuniform, $E^{\Delta}_{s}(\rho,Q)=
E_{s}(\rho,Q)-\rho t h_{b}(\Delta)$ is strictly concave in $\rho$.
In this case, the maximizer
$$
\overline{\rho}^{\Delta}\triangleq\arg\sup_{\rho\geq
\rho_0}[T_{sp}(\rho,W)-tE^{\Delta}_{s}(\rho,Q)]
$$
is strictly larger than $\rho_0$ if $t(h_{b}(q)-h_{b}(\Delta))< C$
and $R_{\infty}(W)\leq t (1-h_{b}(\Delta))$. Particularly,
$\overline{\rho}^{\Delta}<\infty$ if $R_{\infty}(W)<t
(1-h_{b}(\Delta))$. As counterparts of Lemma \ref{prop} and
Corollary \ref{comp}, it can be shown that the upper bound
$\overline{E}^{\Delta}_{sp}(Q,W,t)$ in (\ref{p5}) is attained at
$\overline{R}^{\Delta}_{m}=H(Q^{(\overline{\rho}^{\Delta})})-h_{b}(\Delta)$
and the lower bound in (\ref{p4}) is attained at
$\underline{R}^{\Delta}_{m}=H(Q^{(\underline{\rho}^{\Delta})})-h_{b}(\Delta)$,
where
$\underline{\rho}^{\Delta}=\min\{\overline{\rho}^{\Delta},1\}$.
Consequently, other similar results to the lossless case regarding
these optimizers can be obtained.

\end{enumerate}

\begin{figure}[c!]
\begin{center}
\includegraphics[width=16cm]{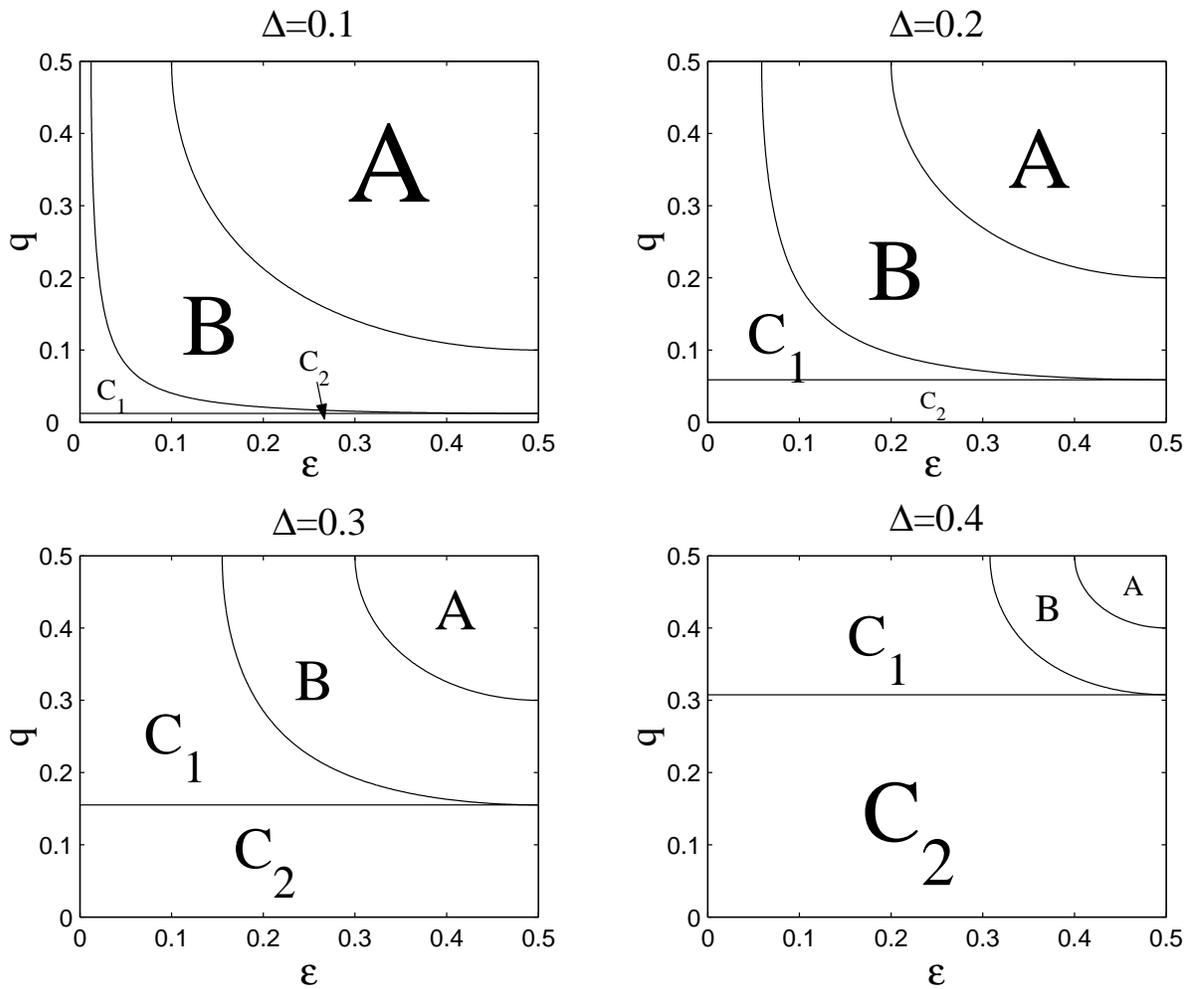}
\caption{The regions for the $(\varepsilon, q)$ pairs in the
binary DMS \{$q,1-q$\} and BSC ($\varepsilon$) system of Example
\ref{distortionex} with Hamming distortion for different values of
the distortion threshold $\Delta$ with $t=1$. Note that
$E_J^\Delta=0$ on the boundary between $\mathbf{A}$ and
$\mathbf{B}$, and $E_J^\Delta>0$ is determined on the boundary
between $\mathbf{B}$ and $\mathbf{C}_1$.}
\label{BSCwithdistortion}
\end{center}
\end{figure}

\begin{figure}[c!]
\begin{center}
\includegraphics[width=16cm]{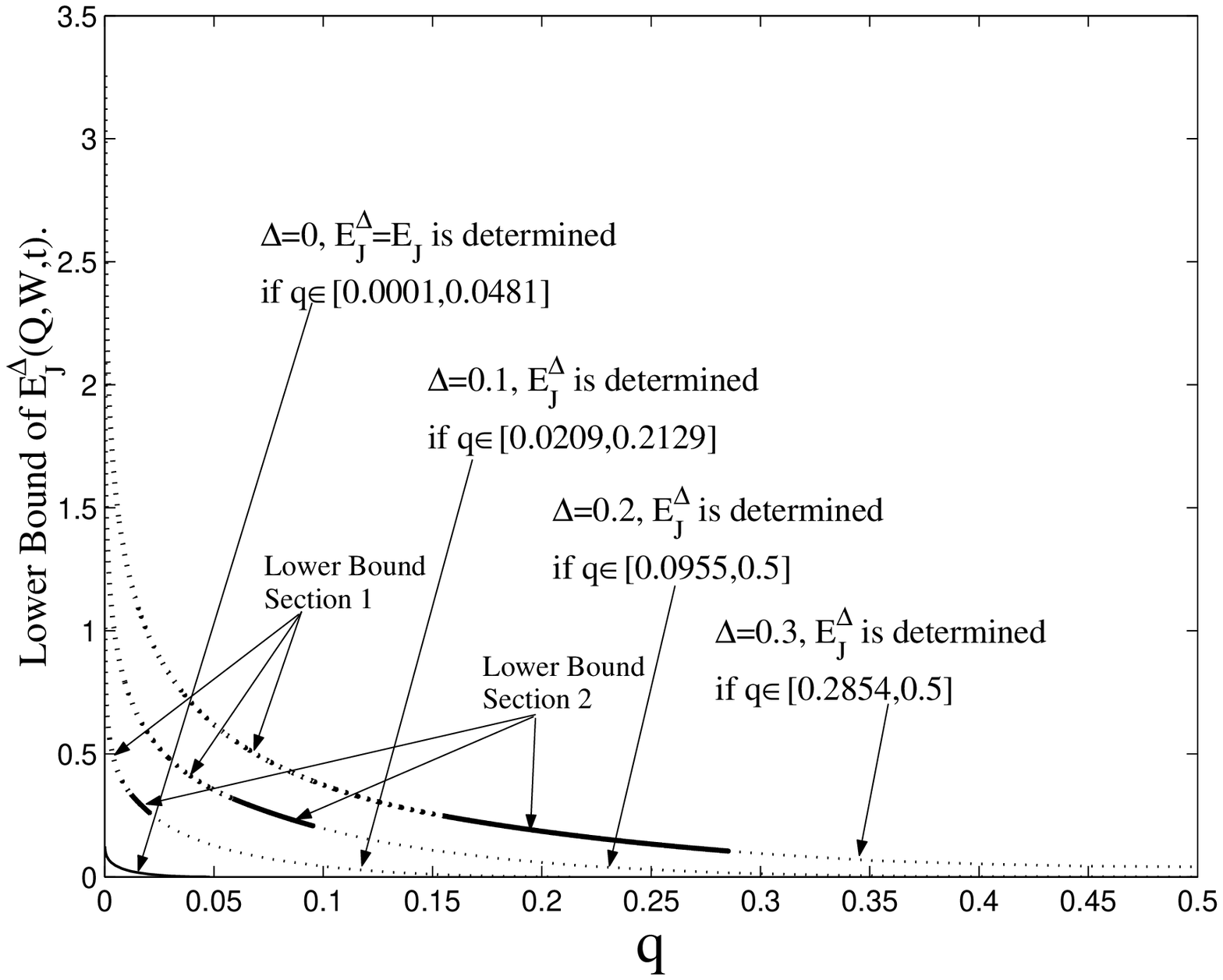}
\caption{Fix $\varepsilon=0.2$. The JSCC exponent lower bound of
the binary DMS $\{q,1-q\}$ ($q\leq 0.5$) and BSC ($\varepsilon$)
pairs under Hamming distortion with $t=1$. For $\Delta=0$,
$E_J^\Delta$ is determined if $q\in [0.0001,0.0481]$, which is the
same as the random-coding lower bound for the lossless JSCC error
exponent. For $\Delta=0.1$, $E_J^\Delta$ is determined if $q\in
[0.0209,0.2129]$. For $\Delta=0.2$, $E_J^\Delta$ is determined if
$q\in [0.0955,0.5]$. For $\Delta=0.3$, $E_J^\Delta$ is determined
if $q\in [0.2854,0.5]$.} \label{reliabilityunderdistortion}
\end{center}
\end{figure}

\begin{example}\label{distortionex}
{\rm For a binary DMS $\{q,1-q\}$ $(q\leq 0.5)$ and a BSC
($\varepsilon$) under transmission rate $t=1$, we compute the JSCC
error exponent under the Hamming distortion measure with
distortion threshold $\Delta$ $(\Delta<\frac{1}{2})$. In
Fig.~\ref{BSCwithdistortion}, if the pair $(\varepsilon,q)$ is
located in region $\mathbf{B}$, then the corresponding JSCC
exponent can be determined exactly (the lower and upper bounds are
equal). If $(\varepsilon,q)$ is located in region
$\mathbf{C}_{1}$, then $E^{\Delta}_{J}$ is bounded by
(\ref{dist1}) and (\ref{dist3}). If $(\varepsilon,q)$ is located
in region $\mathbf{C}_{2}$, then $E^{\Delta}_{J}$ is bounded by
(\ref{dist2}) and (\ref{dist3}). When $(\varepsilon,q)\in
\mathbf{A}$, $E^{\Delta}_{J}$ is zero, and the error probability
of this communication system converges to 1 for $n$ sufficiently
large. So we are only interested in the cases when
$(\varepsilon,q)\in \mathbf{B}\cup \mathbf{C}_{1}\cup
\mathbf{C}_{2}$.

Fig.~\ref{reliabilityunderdistortion} shows the JSCC error
exponent lower bound of the binary DMS $\{q,1-q\}$ $(q \leq 0.5)$ and
BSC ($\varepsilon$) pairs under different distortion thresholds.
We fix the BSC parameter $\varepsilon=0.2$, and vary $q$ from 0 to
0.5. In Fig.~\ref{reliabilityunderdistortion}, Segment~1 is
determined by (\ref{dist2}), and Segments~2 and~3 are determined
by (\ref{dist1}). Furthermore, the lower bound coincides with the
upper bound (\ref{dist3}) in Segment~3; i.e., the JSCC exponent is
exactly determined in Segment~3. }
\end{example}


\section{Conclusions}\label{Concl}

In this work, we establish equivalent parametric representations
of Csisz\'{a}r's lower and upper bounds for the JSCC exponent
$E_J$ of a communication system with a DMS and a DMC, and we
obtain explicit conditions for which the JSCC exponent is exactly
determined. As a result, the computation of the bounds for $E_J$
is facilitated for arbitrary DMS-DMC pairs. Furthermore, the
bounds enjoy closed-form expressions when the channel is
symmetric. A byproduct of our result is the fact that
Csisz\'{a}r's random-coding lower bound for $E_J$ is in general
larger than Gallager's lower bound \cite{Gallager}.

We also provide a systematic comparison between $E_J$ and $E_T$,
the tandem coding error exponent. We show that JSCC can at most
double the error exponent vis-a-vis tandem coding by proving that
$E_J \leq 2 E_T$ and we provide the condition for achieving this
doubling effect. In the case where this upper bound is not tight,
we also establish sufficient explicit conditions under which $E_J
> E_T$. Numerical results indicate that $E_J \approx 2 E_T$ for a
large class of DMS-DMC pairs, hence illustrating the substantial
potential benefit of JSCC over tandem coding. This benefit is also
shown to result into a power saving gain of more than 2~dB for a
binary DMS and a BPSK-modulated AWGN/Rayleigh channel with finite
output quantization. Finally, we partially investigate the
computation of Csisz\'{a}r's lower and upper bounds for the lossy
JSCC exponent under the Hamming distortion measure, and obtain
equivalent representations for these bounds using the same
approach as for the lossless JSCC exponent.

\appendix

\section{Proof of Theorem \ref{tightcondition} and Corollary \ref{comp}}\label{pftightcon}

Theorem \ref{tightcondition} can be shown by a left- and right-
derivatives argument combined with the results of Lemma
\ref{prop}. Let $s_l(R)$ and $s_r(R)$ be the left and right-slopes
(or left- and right-derivatives) of $E_{sp}(R,W)$ at each $R>
R_\infty(W)$. Let $r_l(R)$ and $r_r(R)$ be the left and right
slopes of $E_{r}(R,W)$ at each $R\geq 0$. Let $\rho(R)$ be the
slope of $te(R/t,Q)$ for any $R\in [tH(Q), t\log|\mathcal{S}|]$.
It is easy to verify that these slopes have the following
properties (cf.
\cite{Blahut}, \cite{Gallager}, \cite{Royden}):\\

\noindent(a) $s_l(R)$ and $s_r(R)$ exist for every
$R>R_{\infty}(W)$ and are nondecreasing in $R$.\\

\noindent(b) $r_l(R)$ and $r_r(R)$ exist for every $R\geq 0$ and
are nondecreasing in $R$.\\

\noindent(c) $s_l(R)\leq s_r(R)< -1$ for $R<R_{cr}(W)$, $-1\leq
s_l(R)\leq s_r(R)\leq 0$ for $R_{cr}(W)<R<C$, and
$s_l(R)=s_r(R)=0$ for $R> C$. $s_l(R_{cr}(W))\leq-1\leq
s_r(R_{cr}(W))$ and $s_l(C)\leq 0= s_r(C)$.\\

\noindent(d) $r_l(R)= r_r(R)= -1$ for $R<R_{cr}(W)$,
$r_l(R)=s_l(R)$ for $R>R_{cr}(W)$, and $r_r(R)=s_r(R)$ for $R\geq
R_{cr}(W)$.
$r_l(R_{cr}(W))=-1\leq r_r(R_{cr}(W))$.\\

\noindent(e) $\rho(R)$ is a strictly increasing function of $R$
and is determined by $R=tH\left(Q^{(\rho(R))}\right)$ for
$tH(Q)\leq R\leq t\log|\mathcal{S}|$. Specifically,
$\rho(tH(Q))=0$ and $\rho(t\log|\mathcal{S}|)=\infty$.\\

\noindent(f) $\overline{\rho}^*=\rho(\overline{R}_m)$, where $\overline{\rho}^*$ and $\overline{R}_m$ are defined in (\ref{minimizer2}) and (\ref{maximizer2}), respectively.\\

(a) and (b) follows from the convexity of $E_{sp}(R,W)$ for
$R>R_{\infty}(W)$ and $E_{r}(R,W)$ for $R\geq 0$, see \cite[pp.
113--114]{Royden}. Recalling that $E_{r}(R,W)$ involves a
straight-line section with slope $-1$ for $R\in[0,R_{cr}(W)]$ and
$E_{r}(R,W)=E_{sp}(R,W)$ only for $R\geq R_{cr}(W)$, where they
both are equal to 0 for $R\geq C$, we obtain (c) and (d) from (a)
and (b). From (\ref{eRQ1}), we know that
$te(R/t,Q)=tD\left(Q^{(\rho^*)}\parallel Q\right)$ for $tH(Q)\leq
R\leq t\log|\mathcal{S}|$, where $\rho^*$ is the unique root of
$tH(Q^{(\rho)})=R$. Also, it is easy to verify \cite{Blahut} that
such $\rho^*$ is exactly the slope of $te(R/t,Q)$ at $R$, i.e.,
$$
\frac{\partial te(R/t,Q)}{\partial R}=\rho^*.
$$
Thus (e) follows. Recalling also that in Lemma \ref{prop} we have
shown the relation $\overline{R}_m=tH(Q^{(\overline{\rho}^*)})$,
since there is unique $\rho$ satisfying this equation, we obtain (f).\\

Based on the above setup, the following lemma illustrates the
geometric conditions for which $\underline{E}_r(Q,W,t)$ and
$\overline{E}_{sp}(Q,W,t)$ are attained.

\begin{lemma}\label{geocon}
{\rm Let $tH(Q)<C$ and let $R_\infty(W)< t\log|\mathcal{S}|$. The
minimum in (\ref{defEsp}) is attained at $\overline{R}_m$ if and
only if $-s_l(\overline{R}_m)\geq \rho(\overline{R}_m)\geq
-s_r(\overline{R}_m)$, and the minimum in (\ref{defEr}) is
attained at $\underline{R}_m$ if and only if
$-r_l(\underline{R}_m)\geq \rho(\underline{R}_m)\geq
-r_r(\underline{R}_m)$.}
\end{lemma}
\textbf{Proof}:\\
\noindent1. {\em Forward part}: We only show the case for the
upper bound $\overline{E}_{sp}(Q,W,t)$, since the case for the
lower bound can be shown in a similar manner. We first show that a
rate $R_1\in [tH(Q),t\log|\mathcal{S}|]$ satisfying $-s_l(R_1)\geq
\rho(R_1)\geq -s_r(R_1)$ must achieve the minimum in
$\overline{E}_{sp}(Q,W,t)$. Define functions
\begin{eqnarray}
            f_1(R)\triangleq\left\{\begin{array}{ll}
            E_{sp}(R,W) & \mbox{if}\quad R\leq R_1,\nonumber\\
            E_{sp}(R_1,W)-\frac{|s_l(R_1)|+|\rho(R_1)|}{2}(R-R_1)&\mbox{if}\quad  R\geq R_1.\nonumber
            \end{array}\right.
\end{eqnarray}
and
\begin{eqnarray}
            g_1(R)\triangleq\left\{\begin{array}{ll}
            te\left(\frac{R}{t},Q\right)  & \mbox{if}\quad R\leq R_1,\nonumber\\
            te\left(\frac{R_1}{t},Q\right)+\frac{|\rho(R_1)|+|s_l(R_1)|}{2}(R-R_1) &\mbox{if}\quad  R\geq R_1.\nonumber
            \end{array}\right.
\end{eqnarray}
Since $-s_l(R_1)\geq \rho(R_1)$ implies $s_l(R_1)\leq
-(|s_l(R_1)|+|\rho(R_1)|)/2$ and $\rho(R_1)\leq
(|\rho(R_1)|+|s_l(R_1)|)/2$, we claim that $f_1(R)$ and $g_1(R)$
are both convex functions and hence their sum is convex,
\begin{eqnarray}
            f_1(R)+g_1(R)=\left\{\begin{array}{ll}
            te\left(\frac{R}{t},Q\right)+E_{sp}(R,W)  & \mbox{if}\quad R\leq R_1,\nonumber\\
            te\left(\frac{R_1}{t},Q\right)+E_{sp}(R_1,W) &\mbox{if}\quad  R\geq R_1.\nonumber
            \end{array}\right.
\end{eqnarray}
Since the convex function $f_1(R)+g_1(R)$ is constant for $R\geq
R_1$ (noting that the convexity is strict in the interval
$[tH(Q),R_1]$), we may write
$$
\min_{tH(Q)\leq R\leq
R_1}\left[te\left(\frac{R}{t},Q\right)+E_{sp}(R,W)\right]=te\left(\frac{R_1}{t},Q\right)+E_{sp}(R_1,W).
$$
Similarly, using the relation $\rho(R_1)\geq -s_r(R_1)$ we can
construct convex functions
\begin{eqnarray}
            f_2(R)\triangleq\left\{\begin{array}{ll}
            E_{sp}(R,W) &\mbox{if}\quad R\geq R_1,\nonumber\\
            E_{sp}(R_1,W)+\frac{s_r(R_1)-\rho(R_1)}{2}(R-R_1)&\mbox{if}\quad R\leq R_1.\nonumber
            \end{array}\right.
\end{eqnarray}
and
\begin{eqnarray}
            g_2(R)\triangleq\left\{\begin{array}{ll}
            te\left(\frac{R}{t},Q\right) &\mbox{if}\quad R\geq R_1,\nonumber\\
            te\left(\frac{R_1}{t},Q\right)+\frac{\rho(R_1)-s_r(R_1)}{2}(R-R_1)&\mbox{if}\quad R\leq R_1,\nonumber
            \end{array}\right.
\end{eqnarray}
and use them to show that the minimum
$$
\min_{R_1\leq R\leq t\log|\mathcal{S}|
}\left[te\left(\frac{R}{t},Q\right)+E_{sp}(R,W)\right]
$$
is attained at $R_1$. Thus, $R_1$ is the minimizer of
$\overline{E}_{sp}(Q,W,t)$, i.e.,
$$
\min_{tH(Q)\leq R\leq
t\log|\mathcal{S}|}\left[te\left(\frac{R}{t},Q\right)+E_{sp}(R,W)\right]=te\left(\frac{R_1}{t},Q\right)+E_{sp}(R_1,W).
$$

\noindent 2. {\it Converse part:} We assume
$\overline{R}_m\in(R_{\infty}(W),t\log|\mathcal{S}|)$ achieves the
minimum in (\ref{defEsp}) but $\rho(\overline{R}_m)<
-s_r(\overline{R}_m)$. Note that $\rho(t\log|\mathcal{S}|)=\infty>
-s_r(t\log|\mathcal{S}|)$ provided that
$t\log|\mathcal{S}|>R_{\infty}(W)$. Now let $R_1$ be the smallest
rate in $[R_{\infty}(W),t\log|\mathcal{S}|]$ satisfying
$\rho(R_1)\geq -s_r(R_1)$. According to our assumption together
with (a) and (e), $R_1>\overline{R}_m$. However, using our
previous method, we can construct two convex functions $f_1(R)$
and $g_1(R)$ associated with $R_1$ to show
$$
\min_{tH(Q)\leq R\leq
R_1}\left[te\left(\frac{R}{t},Q\right)+E_{sp}(R,W)\right]=te\left(\frac{R_1}{t},Q\right)+E_{sp}(R_1,W).
$$
This is clearly contradicted with the assumption that the minimum
is attained at $\overline{R}_m$, a rate smaller than $R_1$, since
there is unique minima due to the strict convexity. Thus, at
$\overline{R}_m$ we must have $\rho(\overline{R}_m)\geq
-s_r(\overline{R}_m)$. Consequently, we can show in a similar
manner that $\rho(\overline{R}_m)\leq -s_l(\overline{R}_m)$. \qed

The following facts immediately follow from Lemma \ref{geocon}.
\begin{lemma}\label{relateRm}
{\rm We have the following relations between $\overline{R}_m$ and
$\underline{R}_m$:\\
(1). If $\overline{R}_m>R_{cr}(W)$ or $\underline{R}_m>R_{cr}(W)$,
then
$\underline{R}_m=\overline{R}_m>R_{cr}(W)$ and $\overline{E}_{sp}(Q,W,t)=\underline{E}_{r}(Q,W,t)$.\\
(2). If $\overline{R}_m=R_{cr}(W)$, then $\underline{R}_m\leq
R_{cr}(W)$.\\
(3). $\overline{R}_m\geq \underline{R}_m$. }
\end{lemma}
\textbf{Proof}: (1) is trivial since $E_r(R,W)=E_{sp}(R,W)$ for
$R\geq R_{cr}(W)$. If $\overline{R}_m=R_{cr}(W)$, then by Lemma
\ref{geocon} and (d), $\rho(R_{cr}(W))\geq
-s_r(R_{cr}(W))=-r_r(R_{cr}(W))$. Using Lemma \ref{geocon} again
we obtain (2). To show (3), we only need to show the case when
$\overline{R}_m<R_{cr}(W)$. According to Lemma \ref{geocon}
together with (c) and (d), we see $\rho(\overline{R}_m)>1$ and
$\rho(\underline{R}_m)=1$. It follows from (e) that
$\overline{R}_m>\underline{R}_m$.

\qed


This lemma emphasizes that when the JSCC error exponent upper
bound is achieved at a rate equal to the channel critical rate
$R_{cr}(W)$, the lower bound could be achieved at a rate smaller
than $R_{cr}(W)$.

In the sequel we shall use properties (c)-(f), and Lemmas
\ref{prop}, \ref{geocon} and \ref{relateRm} to prove Theorem
\ref{tightcondition}. To show $A\Longleftrightarrow
B\Longleftrightarrow C$, we only need to show: $A\Longrightarrow
B$ (Forward) and
$B\Longrightarrow C\Longrightarrow A$ (Converse). \\
1. {\em Converse Part}. We start from
\begin{eqnarray}
\overline{\rho}^*< 1 \Longrightarrow & \rho(\overline{R}_m)< 1 & \mbox{(by (f))} \nonumber\\
\Longrightarrow & \overline{R}_m< tR_{cr}^{(s)}(Q)
&\mbox{(by (e))}\nonumber\\
\mbox{and}& s_r(\overline{R}_m)> -1 &
\mbox{(by Lemma \ref{geocon})}\nonumber\\
\Longrightarrow &\overline{R}_m\geq R_{cr}(W)
& \mbox{(by (c))}\nonumber\\
\Longrightarrow & tR_{cr}^{(s)}(Q)>
\underline{R}_m=\overline{R}_m> R_{cr}(W)& \mbox{(by Lemma
\ref{relateRm} (1))}\label{c1}\\
\mbox{or} & tR_{cr}^{(s)}(Q)> \overline{R}_m= R_{cr}(W)\geq
\underline{R}_m& \mbox{(by Lemma
\ref{relateRm} (2))}\label{c2}\\
\Longrightarrow & 0<\underline{\rho}^*=\overline{\rho}^*<1
\label{c3}\\
\mbox{and}& tR_{cr}^{(s)}(Q)> \overline{R}_m= \underline{R}_m\geq
R_{cr}(W)\label{c4},
\end{eqnarray}
where (\ref{c3}) and (\ref{c4}) are explained as follows. We first
claim $\underline{\rho}^*<1$, because $\underline{\rho}^*=1$ would
yield $\underline{R}_m\geq tR_{cr}^{(s)}(Q)$ by Lemma~\ref{prop}
(3), which is contradicted with (\ref{c1}) and (\ref{c2}). Since
now $\underline{\rho}^*<1$, from Lemma~\ref{geocon} and (d) we
know $\underline{R}_m\geq R_{cr}(W)$. Thus in (\ref{c2}) we must
have $\underline{R}_m=R_{cr}(W)$ and consequently (\ref{c1}) and
(\ref{c2}) can both be summarized by (\ref{c4}). Meanwhile,
$\underline{\rho}^*=\overline{\rho}^*$ follows by Lemma
\ref{prop}. If now
\begin{eqnarray}
\overline{\rho}^*= 1 \Longrightarrow & \rho(\overline{R}_m)= 1 & \mbox{(by (f))} \nonumber\\
\Longrightarrow & \overline{R}_m= tR_{cr}^{(s)}(Q)
&\mbox{(by (e))}\nonumber\\
\mbox{and}& s_l(\overline{R}_m)\leq -1\leq s_r(\overline{R}_m) &
\mbox{(by Lemma \ref{geocon})}\nonumber\\
\Longrightarrow &\overline{R}_m\geq R_{cr}(W)
& \mbox{(by (c))}\nonumber\\
\Longrightarrow & tR_{cr}^{(s)}(Q)=
\underline{R}_m=\overline{R}_m> R_{cr}(W)& \mbox{(by Lemma
\ref{relateRm} (1))}\label{d1}\\
\mbox{or} & tR_{cr}^{(s)}(Q)= \overline{R}_m= R_{cr}(W)\geq
\underline{R}_m& \mbox{(by Lemma
\ref{relateRm} (2))}\label{d2}\\
\Longrightarrow & \underline{\rho}^*=\overline{\rho}^*=1\label{d3}\\
\mbox{and} & tR_{cr}^{(s)}(Q)= \underline{R}_m=\overline{R}_m\geq
R_{cr}(W),\label{d4}
\end{eqnarray}
where (\ref{d3}) and (\ref{d4}) are explained as follows. We first
claim that $\underline{\rho}^*=1$. If $\underline{\rho}^*<1$, then
by Lemma \ref{prop} (3) we have
$\underline{R}_m<tR_{cr}^{(s)}(Q)$. In (\ref{d1}), we see
$\underline{R}_m=tR_{cr}^{(s)}(Q)$, contradicted. In (\ref{d2}),
it is still impossible that
$\underline{R}_m<tR_{cr}^{(s)}(Q)=R_{cr}(W)$, because in that case
we have $\rho(\underline{R}_m)<\rho(tR_{cr}^{(s)}(Q))=1$ by (e),
which violates Lemma \ref{geocon} since
$\underline{R}_m<R_{cr}(W)$ implies $\rho(\underline{R}_m)=1$.
Thus we must have $\underline{\rho}^*=1$ and (\ref{d3}) follows.
According to Lemma \ref{prop} (3) again, $\underline{\rho}^*=1$
implies $\underline{R}_m\geq tR_{cr}^{(s)}(Q)$. Hence in
(\ref{d2}) we must have $\underline{R}_m= tR_{cr}^{(s)}(Q)$.
(\ref{d1}) and (\ref{d2}) can both be summarized by (\ref{d4}).
Next if
\begin{eqnarray}
\overline{\rho}^*> 1
\Longrightarrow &\rho(\overline{R}_m)>1 & \mbox{(by (f))} \nonumber\\
\Longrightarrow &\overline{R}_m>tR_{cr}^{(s)}(Q)
&\mbox{(by (e))}\label{e1}\\
\mbox{and}& s_l(\overline{R}_m)< -1 &
\mbox{(by Lemma \ref{geocon})}\nonumber\\
\Longrightarrow &\overline{R}_m\leq R_{cr}(W)
& \mbox{(by (c))}\nonumber\\
\Longrightarrow & \underline{R}_m\leq \overline{R}_m\leq
R_{cr}(W)& \mbox{(by Lemma
\ref{relateRm} (1) and (3))}\nonumber\\
\Longrightarrow & \underline{R}_m<R_{cr}(W) & \label{e3}\\
\Longrightarrow & r_l(\underline{R}_m)=-1= r_r(\underline{R}_m) &\mbox{(by (d))}\nonumber\\
\Longrightarrow & \rho(\underline{R}_m)=1 &
\mbox{(by Lemma \ref{geocon})}\nonumber\\
\Longrightarrow & \underline{R}_m=tR_{cr}^{(s)}(Q) & \mbox{(by (e))}\label{e2}\\
\Longrightarrow & \underline{\rho}^*=1 & \mbox{(by Lemma
\ref{prop} (3))}\nonumber\\
\mbox{and}& \overline{R}_m>\underline{R}_m.& \mbox{(by (\ref{e1})
and (\ref{e2}))}\nonumber.
\end{eqnarray}
To see (\ref{e3}), we let
$\underline{R}_m=\overline{R}_m=R_{cr}(W)$. Then using (d) and
Lemma \ref{geocon} yields $\rho(\underline{R}_m)\leq 1$, which is
contradicted with the assumption
$\rho(\underline{R}_m)=\rho(\overline{R}_m)>1$. To show the last
step, we assume $\underline{\rho}^*<1$, then Lemma \ref{prop} (3)
ensures
$\underline{R}_m=tH(Q^{(\underline{\rho}^*)})<tR_{cr}^{(s)}(Q)$,
which is contradicted with the last second step.\\

\noindent2. \textit{Forward Part}. First recall that
$\rho(tR_{cr}^{(s)}(Q))=1$ by (e). Now if $tR_{cr}^{(s)}(Q)\geq
R_{cr}(W)$, then $\overline{R}_m$ cannot be strictly larger than
$tR_{cr}^{(s)}(Q)$ because in that case
$\rho(\overline{R}_m)>\rho(tR_{cr}^{(s)}(Q))=1$,
$-s_l(\overline{R}_m)\leq 1$ by (c), which violates Lemma
\ref{geocon}. It then follows $\overline{R}_m\leq
tR_{cr}^{(s)}(Q)$ and hence $\overline{\rho}^*\leq 1$ by (e).
Conversely, if $tR_{cr}^{(s)}(Q)< R_{cr}(W)$, then
$\overline{R}_m$ cannot be less than (or equal to)
$tR_{cr}^{(s)}(Q)$ because in that case
$\rho(\overline{R}_m)\leq\rho(tR_{cr}^{(s)}(Q))=1$,
$-s_r(\overline{R}_m)> 1$ by (c), which violates Lemma
\ref{geocon}. It then follows $\overline{R}_m>tR_{cr}^{(s)}(Q)$
and hence $\overline{\rho}^*> 1$ by (e).

Finally, we should note that when $tR_{cr}^{(s)}(Q)<R_{cr}(W)$, or
$\overline{\rho}^*>1$, the lower bound is achieved by
$\underline{R}_m= tR_{cr}^{(s)}(Q)< R_{cr}(W)$ and
$\underline{\rho}^*=1$. Thus
\begin{eqnarray}
\underline{E}_{r}(Q,W,t)&=&
t e\left(\frac{\underline{R}_m}{t},Q\right)+E_r(\underline{R}_m,W)\nonumber\\
&=&
\left[\underline{\rho}^*\underline{R}_m-tE_s(\underline{\rho}^*,Q)\right]+\left[E_0(1,W)-\underline{\rho}^*\underline{R}_m\right]\nonumber\\
&=& E_0(1,W)-tE_s(1,Q).\nonumber
\end{eqnarray}
Meanwhile, Corollary \ref{comp} immediately follows by the above
argument. \qed

\section{Proof of Theorem \ref{TandemTheorem2}}\label{pftandemtheorem2}
We first recall that if $-t\log(|\mathcal{S}|\overline{Q(s)})<
E(t\log|\mathcal{S}|,W)$, then there is no intersection between
$te(R/t,Q)$ and $E(R,W)$. Clearly, the tandem coding exponent
satisfies
\begin{eqnarray}
E_T(Q,W,t)&=& E(t\log|\mathcal{S}|,W)\nonumber\\
&=& E_r(t\log|\mathcal{S}|,W)\label{addineq1}\\
&<& E_r(\underline{R}_m,W)\label{addineq2}\\
&\leq & E_J(Q,W,t),\nonumber
\end{eqnarray}
Here, (\ref{addineq1}) follows by hypothesis $R_{cr}(W)\leq
t\log|\mathcal{S}|$. (\ref{addineq2}) holds since
$\underline{R}_{m}$ must be a quantity smaller than
$t\log|\mathcal{S}|$ by Corollary \ref{comp}.

We hence assume that $-t\log(|\mathcal{S}|\overline{Q(s)}\geq
E(t\log|\mathcal{S}|,W)$, i.e., we assume that $te(R/t,Q)$ and
$E(R,W)$ intersect at rate $R_o$. If $R_o\geq R_{cr}(W)$, which
means that $E_{o}(1,W)-R_{cr}(W)\geq te(R_{cr}(W)/t,Q)$, then Theorem
\ref{TandemTheorem1} guarantees that $E_J>E_T$. If
$\underline{R}_m\geq R_{cr}(W)$, which implies
$tR_{cr}^{(s)}(Q)\geq R_{cr}(W)$ by Corollary \ref{relateRm2}.
This ensures $E_J>E_T$ by Theorem \ref{TandemTheorem1}.
Furthermore, if $R_{cr}(W)>\underline{R}_m \geq R_{o}$, then
\begin{eqnarray}
E_{J}(Q,W,t)& \geq & t e\left(\frac{\underline{R}_m}{t},Q\right)
+E_{r}(\underline{R}_m,W)\nonumber\\
&>& t e\left(\frac{\underline{R}_m}{t},Q\right)\nonumber\\
&\geq & t e\left(\frac{R_{o}}{t},Q\right)\nonumber\\
&=&E_{T}(Q,W,t).\nonumber
\end{eqnarray}

In the remaining, we assume that $te(R/t,Q)$ and $E(R,W)$
intersect at rate $R_o$ and that $\underline{R}_m < R_{o}<R_{cr}$.

For a DMC with $E_{ex}(0,W)<\infty$, we may define the upper bound
of the channel error exponent by
\begin{eqnarray}
E_{s}(R,W)\triangleq\left\{\begin{array}{ll}
             E_{sl}(R,W), \qquad 0\leq R\leq R_{s},\nonumber\\
             E_{sp}(R,W), \qquad R_{s}\leq R\leq C,\nonumber
            \end{array}\right.
\end{eqnarray}
where $E_{sl}(R,W)$ is the straight-line upper bound for the
channel error exponent, and $R_{s}$ is the rate where the
straight-line upper bound is tangent to the sphere-packing bound
and $R_{s}\leq R_{cr}(W)$ \cite{Csiszar3}, \cite{Gallager}.
Clearly, $E_{s}(R,W)$ is also convex in $0\leq R\leq C$, and it is
shown in \cite{Csiszar3}, \cite{Gallager} that
$$
E_{s}(0,W)=E_{sl}(0,W)=E_{ex}(0,W).
$$
Now connect $(0, E_{s}(0,W))$ and $(R_{cr}(W),
E_{s}(R_{cr}(W),W))$ with a straight line, denoted by $l_{1}$,
where
$$
E_{s}(R_{cr}(W),W)=E_{r}(R_{cr}(W),W)=E_{0}(1,W)-R_{cr}(W).
$$
Again, connect $(\underline{R}_{m}, te(\underline{R}_{m}/t,Q))$
and $(t\log|\mathcal{S}|$, $te(\log|\mathcal{S}|,Q))$ with a
straight line, denoted by $l_{2}$, where
$$
te\left(\frac{\underline{R}_{m}}{t},Q\right)=tD(Q^{(1)}\parallel
Q),
$$
and
$$
te(\log|\mathcal{S}|,Q)=-t\log(|\mathcal{S}|\overline{Q(s)}).
$$
Suppose that the intersection of $E_{s}(R,W)$ and $te(R/t,Q)$ is
$(R_{1}, te(R_{1}/t,Q))$, and that the intersection of $l_{1}$ and
$l_{2}$ is $(R_{l}, E_{R_{l}})$. By assumption, $R_o$, the
intersection of $te(R/t,W)$ and $E(R,W)$, is strictly larger than
$\underline{R}_m$ and strictly less than $R_{cr}(W)$; hence by
definition, $R_{1}$, the intersection of $te(R/t,W)$ and
$E_s(R,W)$, must be strictly larger than $\underline{R}_m$ and
strictly less than $R_{cr}(W)$, i.e., $\underline{R}_m<R_1\leq
R_o<R_{cr}(W)$. Likewise, it is easily seen that
$\underline{R}_{m}<R_{l}<R_{cr}(W)$. Furthermore, because of the
convexity of $te(R/t,Q)$ and $E_{s}(R,W)$ in the region
$[\underline{R}_{m}, R_{cr}(W)]$, $E_{R_{l}}$ must be strictly
larger than $te(R_{1}/t,Q)$ (as $te(R/t,W)$ is strictly convex in
this interval). It follows that
$$
E_{J}(Q,W,t)\geq E_0(1,W)-tE_s(1,Q)\geq
E_{R_{l}}>te\left(\frac{R_1}{t},Q\right)\geq
te\left(\frac{R_o}{t},Q\right)=E_T(Q,W,t).
$$
\qed

\section{Proof of Theorem \ref{TandemTheorem3}}\label{pftandemtheorem3}
As in the previous proof, we only consider the case
$-t\log_2(|\mathcal{S}|\overline{Q(s)})\geq
E(t\log_2|\mathcal{S}|,W)$ and $\underline{R}_{m}< R_o<R_{cr}(W)$.
Thus, we can upper bound $E_T$ by
\begin{eqnarray}
E_{T}(Q,W,t)&=& te(\frac{R_{o}}{t},Q)\nonumber\\
&<&
te\left(\frac{R_{cr}(W)}{t},Q\right)\nonumber\\
&=& tD\left(Q^{(\gamma)}\parallel Q)\right)\nonumber
\end{eqnarray}
by the strict monotonicity of the source error exponent. On the
other hand, Theorem \ref{tightcondition} gives that
$$
E_{J}(Q,W,t)\geq E_0(1,W)-tE_s(1,Q).
$$
By assumption, if $E_0(1,W)-tE_s(1,Q)\geq
tD\left(Q^{(\gamma)}\parallel Q)\right)$, then $E_J>E_T$. \qed

\section{Proof of Lemma \ref{sourceexponentdist}}\label{pfsourceexpdis}
Recall that the rate-distortion function $R(Q,\Delta)$ for a
binary DMS $Q=\{q,1-q\}$ under the Hamming distortion measure is
given by (e.g., \cite{Thomas})

\begin{eqnarray}
  R(Q,\Delta)=\left\{
            \begin{array}{ll}
              h_{b}(q)-h_{b}(\Delta), & 0\leq \Delta\leq q,\\
              0, & \Delta> q.
            \end{array}
      \right.\label{ratedistortion}
\end{eqnarray}

\noindent Clearly, $F(R,Q,\Delta)=0$ for $R\leq 0$ since the
infimum in (\ref{sourcedis}) is attained at $P=Q$. Similarly,
since $R(P,\Delta)\leq 1-h_b(\Delta)$ for all $P$,
$F(R,Q,\Delta)=\infty$ for $R>1-h_b(\Delta)$. For the remainder of
the proof, we assume $0<R\leq 1-h_b(\Delta)$.\\

\noindent(1){\em Case of $0\leq\Delta\leq q$.} For $R\leq
R(Q,\Delta)=h_{b}(q)-h_{b}(\Delta)$, we have
$$
F(R,Q,\Delta)=\left.\inf_{P: R(P,\Delta)> R}D(P\parallel
Q)=D(P\parallel Q)\right|_{P=Q}=0.
$$
For $h_{b}(q)-h_{b}(\Delta)< R\leq 1-h_{b}(\Delta)$, we have
\begin{eqnarray}
F(R,Q,\Delta)&=&\inf_{P: R(P,\Delta)> R}D(P\parallel
Q)\nonumber\\
&=& \min_{P\triangleq\{p,1-p\}: R(P,\Delta)=R}D(P\parallel Q)\label{lem12exp1}\\
&=& \min_{p:
h_{b}(p)-h_{b}(\Delta)=R}D(P\parallel Q)\nonumber\\
&=& e(R+h_{b}(\Delta),Q), \qquad \qquad \text{for} \ H(Q)\leq
R+h_{b}(\Delta)\leq
\log|\mathcal{S}|\label{lem12exp3}\\
&=& \sup_{\rho\geq
0}[\rho(R+h_{b}(\Delta))-E_{s}(\rho)]\label{lem12exp4}\\
&=& \sup_{\rho\geq 0}[\rho R-E^{\Delta}_{s}(\rho,Q)].
\nonumber
\end{eqnarray}
Here (\ref{lem12exp1}) follows from the facts that the continuous
function $\theta(p)\triangleq
p\log\frac{p}{q}+(1-p)\log\frac{1-p}{1-q}$ is increasing for
$p\geq q$ and $R(P,\Delta)$ given in (\ref{ratedistortion}) is
continuous and increasing in $p$ for $\Delta\leq p\leq
\frac{1}{2}$. In (\ref{lem12exp3}), we note that $H(Q)=h_b(q)$ and
that $\log|\mathcal{S}|=1$ as the source is binary.
(\ref{lem12exp4}) follows by the well known parametric form of
source exponent function introduced by Blahut \cite{Blahut} and
noting that $R'\triangleq R+h_{b}(\Delta)\in
[H(Q),\log|\mathcal{S}|]$.\\

\noindent(2) {\em Case of $\Delta>q$.} For $0< R\leq
1-h_{b}(\Delta)$, similarly as (\ref{lem12exp3}), we have
$$
F(R,Q,\Delta) = e(R',Q)=\sup_{\rho\in A}[\rho R'-E_s(\rho)],
$$
where $R'=R+h_{b}(\Delta)$ such that $H(Q)<h_{b}(\Delta)<R'\leq
1=\log|\mathcal{S}|$ and
\begin{eqnarray}
A&=&\left\{\rho^*: \left.\frac{\partial [\rho
R'-E_s(\rho)]}{\partial \rho}\right|_{\rho=\rho^*}=0,\qquad
h_{b}(\Delta)\leq R'\leq 1\right\}\nonumber\\
&=&\left\{\rho^*: h_{b}(\Delta)\leq R'=H(Q^{(\rho^*)})\leq
1\right\}\nonumber\\
&=& \{\rho^*: \rho_0\leq \rho^*<\infty\},\label{lem12exp5}
\end{eqnarray}
where $\rho_0$ is the unique root of equation
$H(Q^{(\rho)})=h_{b}(\Delta)$ and $\rho_0>0$. Here
(\ref{lem12exp5}) follows from the monotone property of
$H(Q^{(\rho)})$. Therefore, we write
$$
F(R,Q,\Delta)=\sup_{\rho\geq \rho_0}[\rho R-E^{\Delta}_{s}(\rho,Q)].
$$
In fact, it can be shown that $\rho_0$ is the right slope of
$F(R,Q,\Delta)$ at $R=R(Q,\Delta)$. \qed

\renewcommand{\baselinestretch}{1.12}

\newpage


\begin{table}[c!]
\begin{center}
\begin{tabular}{||c|c|c|c|c||}\hline\hline
$E_J/E_T$              & t=0.5, q=0.1 & t=0.75, q=0.1 & t=0.75, q=0.15 & t=1, q=0.05  \\
\hline $\varepsilon=0.0005$ & $1.0^\dag$         &$1.60^\dag$    & $1.58^\dag$    &  $1.87^\dag$\\
\hline $\varepsilon=0.001$  & $1.0^\dag$         &$1.70^\dag$    & $1.68^\dag$    &  $1.93^\dag$\\
\hline $\varepsilon=0.005$ &  $1.36^\dag$          &$1.94^\dag$    & 1.89           &    1.99\\
\hline $\varepsilon=0.01$  & $1.70^\dag $         &1.95           & 1.91           &    2.0\\
\hline $\varepsilon=0.04$  & 1.85         &1.97           & 1.95           &    2.0\\
\hline $\varepsilon=0.08$  & 1.91         &1.99           & 1.96           &    2.0\\
\hline $\varepsilon=0.12$  & 1.95         &1.97           & 2.0            &    2.0\\
\hline $\varepsilon=0.16$  & 1.96         &1.95           & N/A            &    2.0\\
\hline $\varepsilon=0.2$   & 1.86         &N/A            & N/A            &      N/A\\
\hline\hline
\end{tabular}
\caption{$E_J/E_T$ for the binary DMS and BSC pairs of Example
\ref{lbr}. ``N/A'' means that $tH(Q)>C$ such that $E_J=E_T=0$.
``$\dag$'' means that this quantity is only a lower bound for
$E_J/E_T$.}\label{ratiobinaryBSC}
\end{center}
\end{table}

\clearpage



\small\normalsize

\begin{thebibliography}{40}
\bibitem{Ahlswede} R.~Ahlswede, ``Extremal properties of
rate-distortion functions,'' {\em IEEE Trans. Inform. Theory},
vol.~36, pp.~166--171, Jan.~1990.

\bibitem{Markov} F. Alajaji, N. Phamdo, N. Farvardin, and T. Fuja, ``Detection of binary Markov sources over channels with additive Markov noise,'' {\em IEEE Trans. Inform. Theory}, vol.~42, No.~1, pp.~230--239, Jan.~1996.

\bibitem{Celp} F.~Alajaji, N.~Phamdo, and T.~Fuja, ``Channel codes that exploit
the residual redundancy in CELP-encoded speech,'' {\em IEEE Trans.
Speech and Audio Processing}, vol.~4, no.~5, pp.~325--336,
Sept.~1996.

\bibitem{COVQ} F. Alajaji and N. Phamdo, ``Soft-decision COVQ for Rayleigh-fading channels,'' {\em IEEE Commun. Lett.}, vol.~2, pp.~162--164, June~1998.


\bibitem{arikan1} E.~Arikan and N.~Merhav, ``Guessing subject to distortion,''
{\em IEEE Trans. Inform. Theory}, vol.~44, pp.~1041--1056, May~1998.
                                                                                
\bibitem{arikan2} E.~Arikan and N.~Merhav, ``Joint source-channel coding
and guessing with application to sequential decoding,''
{\em IEEE Trans. Inform. Theory}, vol.~44, pp.~1756--1769, Sept.~1998.


\bibitem{Arimoto2} S.~Arimoto, ``'On the converse to the coding theorem for discrete memoryless
channels,'' {\em IEEE Trans. Inform. Theory}, vol.~19,
pp.~357--359, May~1973.

\bibitem{Arimoto} S.~Arimoto, ``Computation of random coding exponent
functions,'' {\em IEEE Trans. Inform. Theory}, vol.~22,
pp.~665--671, Nov.~1976.

\bibitem{Ayanoglu} E.~Ayano\v{g}lu and R.~Gray, ``The design of joint source and channel trellis waveform coders,'' {\em IEEE Trans. Inform. Theory}, vol.~33, pp.~855--865,
Nov.~1987.


\bibitem{balakirsky} V.~B.~Balakirsky, ``Joint source-channel coding
with variable length codes,'' {\em Probl. Inform. Transm.},
vol.~1, no.~37, pp.~10-23, Jan.--Mar.~2001.


\bibitem{barg} A.~Barg and A.~McGregor, ``Distance distribution of binary codes
and the error probability of decoding,'' {\em IEEE Trans. Inform. Theory},
vol.~51, pp.~4237-4246, Dec.~2005.


\bibitem{bert} D. P. Bertsekas, with A. Nedi\'c and A. E. Ozdagler, {\em Convex Analysis and Optimization}, Athena Scientific, Belmont, MA, 2003.

\bibitem{Blahut} R.~E.~Blahut, ``Hypothesis testing and information theory,'' {\em IEEE Trans. Inform. Theory}, vol.~IT-20, pp.~405--417, July~1974.

\bibitem{Brian} Brian D.~Bunday, {\em Basic Optimisation Methods}. London: Arnold,~1984.

\bibitem{Chen} P.-N.~Chen and F.~Alajaji, ``Optimistic Shannon coding theorems for arbitrary single-user
systems,'' {\em IEEE Trans. Inform. Theory}, vol.~45,
pp.~2623--2629, Nov.~1999.

\bibitem{Thomas} T.~M.~Cover and J.A.~Thomas, {\em Elements of
Information Theory}, New York: Wiley,~1991.

\bibitem{Csiszar1} I.~Csisz\'{a}r, ``Joint source-channel error exponent,'' {\em Probl. Contr. Inform. Theory}, vol.~9, pp.~315--328,~1980.

\bibitem{Csiszar2} I.~Csisz\'{a}r, ``On the error exponent of source-channel
transmission with a distortion threshold,'' {\em IEEE Trans.
Inform. Theory}, vol.~28, pp.~823--828, Nov.~1982.

\bibitem{Csiszar3} I.~Csisz\'{a}r and J.~K\"{o}rner, {\em Information Theory: Coding Theorems for Discrete Memoryless Systems}. New
York: Academic,~1981.

\bibitem{Dunham} J.~G.~Dunham and R.~M.~Gray, ``Joint source and noisy channel
trellis encoding,'' {\em IEEE Trans. Inform. Theory}, vol.~27,
pp.~516--519, July~1981.

\bibitem{Farvardin} N.~Farvardin, ``A study of vector quantization for noisy
channels,'' {\em IEEE Trans. Inform. Theory}, vol.~36, no.~4,
pp.~799--809, July~1990.

\bibitem{Fine} T.~Fine, ``Properties of an optimum digital system and
applications,'' {\em IEEE Trans. Inform. Theory}, vol.~10,
pp.~443--457, Oct.~1964.

\bibitem{Gallager} R.~G.~Gallager, {\em Information Theory and
Reliable Communication}, New York: Wiley,~1968.

\bibitem{Gastpar} M.~Gastpar, B.~Rimoldi and M.~Vetterli, ``To code, or not to code: lossy source-channel communication
revisited,'' {\em IEEE Trans. Inform. Theory}, vol.~49,
pp.~1147--1158, May~2003.

\bibitem{Gibson} J.~D.~Gibson and T.~R.~Fisher, ``Alphabet-constrained data
compression,'' {\em IEEE Trans. Inform. Theory}, vol.~28,
pp.~443--457, May~1982.

\bibitem{Gray} R.~M.~Gray and D.~S.~Ornstein, ``Sliding-block joint source/noisy-channel coding
theorems,'' {\em IEEE Trans. Inform. Theory}, vol.~22,
pp.~682--690, Nov.~1976.

\bibitem{Hagenauer} J.~Hagenauer, ``Source-controlled channel
decoding,'' {\em IEEE Trans. Commun.}, vol.~43, pp.~2449--2457,
Sep. 1995.

\bibitem{Han} T.~S.~Han, {\em Information-Spectrum Methods in Information
Theory}, Springer, 2003.

\bibitem{Hellman} M.~E.~Hellman, ``Convolutional source encoding,'' {\em IEEE Trans. Inform. Theory}, vol.~21, pp.~651--656, Nov. 1975.

\bibitem{Hochwald} B. Hochwald and K. Zeger, ``Tradeoff between source and channel coding,'' {\em IEEE Trans. Inform. Theory}, vol.~43, pp.~1412--1424, Sep.~1997.

\bibitem{Jelinek} F.~Jelinek, {\em Probabilistic Information Theory}, New
York, McGraw Hill,~1968.


\bibitem{koshelev} V.~N.~Koshelev, ``Direct sequential encoding and decoding
for discrete sources,'' {\em IEEE Trans. Inform. Theory,} vol.~19,
pp.~340--343, May~1973.


\bibitem{Kumazawa} H. Kumazawa, M. Kasahara, and T. Namekawa, ``A construction of vector quantizers for noisy channels,'' {\em Electron. Eng. Jpn.}, vol. 67-B, no.~4, pp.~39--47, 1984.

\bibitem{Kurtenbach} A.~Kurtenbach and P.~Wintz, ``Quantizing for noisy channels,''
{\em IEEE Transactions on Communication Technology}, vol.~COM-17,
pp.~ 291--302, Apr.~1969.

\bibitem{Lim} J. Lim and D.~L.~Neuhoff, ``Joint and tandem source-channel coding with complexity and delay constraints,'' {\em IEEE Trans. Commun.}, vol.~51, pp.~757--766,
May~2003.

\bibitem{Luenberger} D. G. Luenberger, {\em Optimization by Vector Space Methods}, Wiley, 1969.

\bibitem{Marton} K.~Marton, ``Error exponent for source coding with a fidelity criterion,'' {\em IEEE Trans. Inform. Theory}, vol.~IT-20, pp.~197--199, Mar.~1974.

\bibitem{Massey} J.~L.~Massey, ``Joint source and channel coding,'' in {\em
Communications and Random Process Theory,} J.~K.~Skwirzynski, ed.,
The Netherlands: Sijthoff and Nordhoff, pp.~279--293, 1978.

\bibitem{Miller} D. Miller and K. Rose, ``Combined source-channel vector quantization using deterministic annealing,'' {\em IEEE Trans. Commun.}, vol.~42, pp.~347--356,
Feb.-Apr. 1994.

\bibitem{Modestino} J.~W.~Modestino and D.~G.~Daut, ``Combined source-channel coding
of images,'' {\em IEEE Trans. Commun.}, vol.~27, pp.~1644-1659,
Nov.~1979.

\bibitem{Nam} Nam Phamdo and Fady Alajaji, ``Soft-decision demodulation design for COVQ over white, colored, and ISI Gaussian channels,'' {\em IEEE Trans.
Commun.}, vol.~46, No.~9, pp.~1499--1506, Sep.~2000.

\bibitem{rock} R. T. Rockafellar, {\em Conjugate Duality and Optimization}, SIAM, Philadelphia, 1974.

\bibitem{Royden} H. L. Royden, {\em Real Analysis}, Third Edition,
New York, 1988.

\bibitem{Sayood} K.~Sayood and J.~C.~Borkenhagen, ``Use of residual redundancy in the design of joint source/channel
coders,'' {\em IEEE Trans. Commun.}, vol.~39, pp.~838--846,
June~1991.

\bibitem{shannon} C.~E.~Shannon,``A mathematical theory of
communication,'' {\em Bell Syst. Tech. J.}, vol.~27, pp.~379--423
and pp.~623-656, Jul. and Oct. 1948.

\bibitem{Shamai} S.~Shamai, S.~Verd\'{u}, and R.~Zamir, ``Systematic lossy
source/channel coding,'' {\em IEEE Trans. Inform. Theory},
vol.~44, pp.~564--579, Mar.~1998.

\bibitem{Hadamard} M.~Skoglund and P.~Hedelin, ``Hadamard-based soft decoding for
vector quantization over noisy channels,'' {\em IEEE Trans.
Inform. Theory}, vol.~45, no.~2, pp.~515--532, Mar.~1999.

\bibitem{Skoglund} M. Skoglund, ``Soft decoding for vector quantization over noisy channels with
memory,'' {\em IEEE Trans. Inform. Theory}, vol.~45,
pp.~1293--1307, May 1999.

\bibitem{taricco} G. Taricco, ``On the capacity of the binary input Gaussian and Rayleigh fading
channels,'' {\em Eur. Trans. Telecommun.}, vol. 7, no. 2,
Mar.-Apr. 1996.

\bibitem{Vaisha} V.~A.~Vaishampayan and N.~Farvardin, ``Joint design of block source codes and modulation signal
sets,'' {\em IEEE Trans. Inform. Theory}, vol.~38, pp.~1230--1248,
July 1992.

\bibitem{Vembu} S.~Vembu, S.~Verd\'{u} and Y. Steinberg, ``The source-channel separation theorem
revisited,'' {\em IEEE Trans. Inform. Theory}, vol.~41,
pp.~44--54, Jan.~1995.

\bibitem{viterbi} A.~J.~Viterbi and J.~K.~Omura, {\em Principles of Digital Communication and
Coding}, McGraw-Hill, Inc., 1979.

\bibitem{Weissman} T. Weissman, E. Ordentlich, G. Seroussi, S.
Verd\'{u}, and M. J.~Weinberger, ``Universal discrete denoising:
known channel,'' {\em IEEE Trans. Inform. Theory}, vol.~51,
pp.~5--28, Jan. 2005.

\bibitem{Zeger} K.~A. Zeger and A.~Gersho, ``Pseudo-Gray coding,'' {\em IEEE
Trans. Commun.}, vol.~38, no.~ 12, pp.~2147--2158, Dec.~1990.

\bibitem{QBSC04} Y.~Zhong, F.~Alajaji, and L.~L.~Campbell, ``When is joint source-channel coding worthwhile: an information theoretic perspective,'' {\em Proc.\ 22nd Bienn.\ Symp.\ Commun.}, Canada, pp.~121--123, June~2004.

\bibitem{ISIT04} Y.~Zhong, F.~Alajaji, and L.~L.~Campbell, ``On the computation of the joint source-channel error exponent for memoryless
systems,'' {\em Proc. 2004 IEEE Int'l. Symp. Inform. Theory,}
p.~477, June-July~2004.

\bibitem{Zhu} G.-C.~Zhu, F.~Alajaji, J.~Bajcsy and P.~Mitran, ``Transmission of
non-uniform memoryless sources via non-systematic Turbo codes,''
{\em IEEE Trans. Commun.}, vol.~52, no.~8, pp.~1344--1354,
Aug.~2004.

\end{thebibliography}
\end{document}